\documentclass[11pt,a4paper]{article}
\pdfoutput=1
\usepackage{jheppub}
\usepackage{braket}

\usepackage{amsfonts}
\usepackage{amscd}
\usepackage{amssymb}
\usepackage{amsmath,bbm}
\usepackage{graphicx}
\usepackage{epsfig}
\usepackage{latexsym}
\usepackage{mathtools}
\usepackage{hyperref}
\usepackage{tikz-cd}
\usepackage[vcentermath]{youngtab}
    
\def \be  {\begin{equation}}
\def \ee  {\end{equation}}
\def \bea {\begin{equation}\begin{aligned}}
\def \eea {\end{aligned}\end{equation}}
\def \ba  {\begin{eqnarray}}
\def \ea  {\end{eqnarray}}
\def \bb  {\begin{thebibliography}}
\def \eb  {\end{thebibliography}}
\def \lab #1 {\label{#1}}

\newcommand\ep{\epsilon}

\newcommand\cA{\mathcal{A}}
\newcommand\cB{\mathcal{B}}

\newcommand\cH{\mathcal{H}}
\newcommand\cI{\mathcal{I}}

\newcommand\cM{\mathcal{M}}
\newcommand\cN{\mathcal{N}}
\newcommand\cO{\mathcal{O}}
\newcommand\cP{\mathcal{P}}
\newcommand\cQ{\mathcal{Q}}

\newcommand\cZ{\mathcal{Z}}
\newcommand\al{\alpha}

\newcommand\C{\mathbb{C}}

\newcommand\bR{\mathbb{R}}

\newcommand\bZ{\mathbb{Z}}
\newcommand\R{\mathbb{R}}

\newcommand\la{\langle}
\newcommand\ra{\rangle}

\definecolor{cardinal}{rgb}{0.6,0,0}
\definecolor{darkgreen}{rgb}{0,0.5,0}
\definecolor{golden}{rgb}{0.92, 0.7, 0}
\definecolor{midnight}{rgb}{0, 0, 0.5}
\definecolor{darkblue}{rgb}{0.2, 0, 0.8}

\begin{document}

\title{Boundaries, Vermas, and Factorisation}

\author[a]{Mathew Bullimore,}
\author[b]{Samuel Crew,}
\author[b]{Daniel Zhang}

\affiliation[a]{Department of Mathematical Sciences, Durham University, \\
 Durham, DH1 3LE, United Kingdom}

\affiliation[b]{Department of Applied Mathematics and Theoretical Physics, University of Cambridge, Cambridge, CB3 0WA, United Kingdom}

{\abstract{We revisit the factorisation of supersymmetric partition functions of 3d $\mathcal{N}=4$ gauge theories. The building blocks are hemisphere partition functions of a class of UV $\mathcal{N}=(2,2)$ boundary conditions that mimic the presence of isolated vacua at infinity in the presence of real mass and FI parameters. These building blocks can be unambiguously defined and computed using supersymmetric localisation. We show that certain limits of these hemisphere partition functions coincide with characters of lowest weight Verma modules over the quantised Higgs and Coulomb branch chiral rings. This leads to expressions for the superconformal index, twisted index and $S^3$ partition function in terms of such characters. On the way we uncover new connections between boundary 't Hooft anomalies, hemisphere partition functions and lowest weights of Verma modules.
}}

\maketitle

\setcounter{page}{1}


\section{Introduction}

Supersymmetric partition functions are useful tools to study interacting supersymmetric quantum field theories. In certain circumstances, these observables can be computed exactly using the method of supersymmetric localisation and this leads to a rich connection with geometric representation theory and enumerative geometry. 

For supersymmetric gauge theories in three dimensions with at least $\cN=2$ supersymmetry, partition functions on many supersymmetric backgrounds involving a compact space $\cM_3$ admit a factorization of the form
\be
\cZ_{\cM_3} = \sum_\al H_\al \widetilde H_\al \, ,
\ee
where the sum is over a finite set of vacua $\{\al\}$ and $H_\al$ is a partition function associated to the geometry $S^1 \times \{\text{hemisphere}\}$ or a twisted product $S^1 \times_q D^2$ with a boundary condition determined by the vacuum $\al$. 

This factorisation of supersymmetric partition functions originated in computations of the $S^3$ partition function~\cite{Pasquetti:2011fj} and has also been checked in many examples for the superconformal index~\cite{Hwang:2015wna,Hwang:2012jh,Dimofte:2011py} and $S^1\times S^2$ twisted index~\cite{Cabo-Bizet:2016ars,Crew:2020jyf}. Factorisation also plays an important role in the 3d-3d correspondence~\cite{Dimofte:2011ju,Dimofte:2014zga,Dimofte:2011py}. It can be derived using Higgs branch localisation~\cite{Benini:2013yva,Fujitsuka:2013fga} and from the gluing construction of~\cite{Dedushenko:2018aox,Dedushenko:2018tgx}.

The individual building blocks $H_\al$ of factorisation have a number of different interpretations in both physics and mathematics. A systematic approach is holomorphic blocks~\cite{Beem:2012mb}, which are defined in the IR as partition functions of massive theories on a twisted product $S^1 \times_q D^2$. This provides an elegant prescription to construct the building blocks $H_\al$ as solutions to certain difference equations but suffers from some ambiguities in the determination of classical and 1-loop contributions.

In this paper, we revisit the factorisation of supersymmetric partition functions from a UV perspective for gauge theories with $\cN=4$ supersymmetry. We will define the building blocks $H_\al$ as the hemisphere partition functions of a distinguished set of boundary conditions $\{\cB_\al\}$ preserving $\cN=(2,2)$ supersymmetry and labelled by isolated massive vacua $\al$
\bea 
	H_{\alpha} = \cZ_{\cB_\al}\,.
\eea 
These hemisphere partition functions can be computed exactly using supersymmetric localisation.

\begin{figure}[htp]
\centering
\tikzset{every picture/.style={line width=0.75pt}} 
\begin{tikzpicture}[x=0.4pt,y=0.4pt,yscale=-1,xscale=1]

\draw  [color={rgb, 255:red, 57; green, 112; blue, 177 }  ,draw opacity=1 ][line width=1.5]  (269.38,140.03) -- (269.38,295.5) -- (161.13,238.97) -- (161.13,83.5) -- cycle ;
\draw [line width=0.75]    (94,189.5) -- (215.25,189.5) ;
\draw [line width=0.75]  [dash pattern={on 0.84pt off 2.51pt}]  (456.25,189.5) -- (548.5,189.5) ;
\draw [shift={(551.5,189.5)}, rotate = 180] [fill={rgb, 255:red, 0; green, 0; blue, 0 }  ][line width=0.08]  [draw opacity=0] (8.93,-4.29) -- (0,0) -- (8.93,4.29) -- cycle    ;
\draw [line width=0.75]    (361,189.5) -- (456.25,189.5) ;

\draw (301,182) node [anchor=north west][inner sep=0.75pt]    {$\simeq $};
\draw (226,145) node [anchor=north west][inner sep=0.75pt]    {$\mathcal{B}_{\alpha }$};
\draw (555,182) node [anchor=north west][inner sep=0.75pt]    {$\alpha $};
\end{tikzpicture}
\caption{The distinguished set of boundary conditions $\cB_\al$ mimic the presence of an isolated vacuum $\al$ at infinite distance, at least for computations amenable to supersymmetric localisation.}
\label{fig:thimble-schematic}
\end{figure}
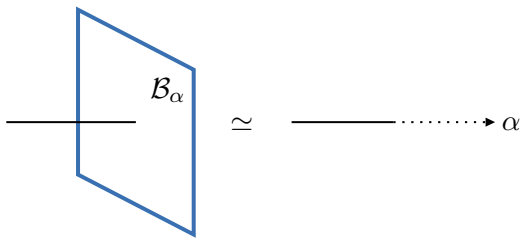

The boundary conditions $\cB_\al$ are designed to mimic isolated massive vacua $\al$ at infinite distance in the presence of generic real mass and FI parameters, at least for computations amenable to supersymmetric localisation. This is illustrated schematically in figure~\ref{fig:thimble-schematic}. Boundary conditions of this type were first studied in two dimensions for Landau-Ginzburg models and massive sigma models in~\cite{Hori:2000ck} and play an important part in 2d mirror symmetry. A systematic description in massive 2d theories was developed in~\cite{Gaiotto:2015zna,Gaiotto:2015aoa}. The importance of such boundary conditions in 3d $\cN=4$ theories was discussed in~\cite{Bullimore:2016nji}, which also gave an explicit UV construction in abelian gauge theories.

An important feature is that the set of boundary conditions $\{\cB_\al\}$  depend on the real mass and FI parameters. More precisely, they depend on a choice of chambers $\mathfrak{C}_H$, $\mathfrak{C}_C$ in the spaces of real mass and FI parameters. The walls separating chambers correspond to mass and FI parameters where the theory no longer has isolated vacua. As a consequence, the factorisation jumps across these walls in such a way that the partition function $Z_{\cM_3}$ is unchanged.

The hemisphere partition functions depend on four parameters,
\be
H_\al = H_\al(q,t,x,\xi) \, ,
\ee
where $q$, $t$ are fugacities dual to to combinations of isometries and R-symmetries while $x$, $\xi$ are fugacities dual to Higgs and Coulomb branch global symmetries. The hemisphere partition functions of the boundary conditions $\{\cB_\al\}$ in a given chamber $\mathfrak{C}_H$, $\mathfrak{C}_C$ are characterised by their common analytic properties in the fugacities $x$, $\xi$. They differ from the holomorphic blocks of 3d $\cN=4$ gauge theories presented in~\cite{Bullimore:2014awa,Zenkevich:2017ylb,Aprile:2018oau} in the classical and 1-loop contributions.

The hemisphere partition function can be related via the state-operator correspondence to a half superconformal index counting local operators at the origin of $\mathbb{R}^2 \times \mathbb{R}_{\ge 0}$. The relation between these objects is more accurately
\begin{equation}
    H_\al = e^{\phi_\al} \cI_\al
\end{equation}
where $\cI_\al$ is the half superconformal index of the boundary condition $\cB_\al$ and the pre-factor $e^{\phi_\al}$ is determined by boundary 't Hooft anomalies for global and R-symmetries.

We focus on two limits of the hemisphere partition function with enhanced supersymmetry. They correspond to limits of the half superconformal index that count boundary operators transforming as the scalar components of boundary chiral and twisted chiral multiplets respectively. They are defined respectively by
\bea
\mathcal{X}_\al^H(x) & := \lim_{t^{\frac{1}{2}}\to q^{-1/4}}H_\al(q,t,x,\xi) \\
\mathcal{X}_\al^C(\xi) & := \lim_{t^{\frac{1}{2}}\to q^{1/4}}H_\al(q,t,x,\xi) \, .
\eea
Although our notation indicates that these limits depend only on a single parameter, they retain a small additional dependence on the remaining parameters due to boundary mixed 't Hooft anomalies contributing to the pre-factor $e^{\phi_\al}$.

These boundary operators counted by this limit of the half superconformal index transform as modules for the quantised algebras $\cA_H$, $\cA_C$ of functions on the Higgs and Coulomb branch respectively \cite{Bullimore:2016nji}, as illustrated in figure~\ref{fig:intro-modules}. The quantisations are manifested by the $\Omega_A$- and $\Omega_B$-deformations respectively, studied in \cite{Yagi:2014toa, Bullimore:2015lsa, Bullimore:2016hdc, Beem:2018fng, Oh:2019bgz,Jeong:2019pzg}. Boundary conditions compatible with real mass and FI parameters in chambers $\mathfrak{C}_H$, $\mathfrak{C}_C$ generate modules that are lowest weight with respect to these chambers. In particular, boundary operators on the boundary conditions $B_\al$ generate lowest weight Verma modules $\cH^{(B)}_{\cB_\al}$, $\cH^{(A)}_{\cB_\al}$ for the algebras $\cA_H$, $\cA_C$ respectively.

\begin{figure}[htp]
\centering   
\tikzset{every picture/.style={line width=0.75pt}} 

\begin{tikzpicture}[x=0.48pt,y=0.48pt,yscale=-1,xscale=1]
\draw  [color={rgb, 255:red, 57; green, 112; blue, 177 }  ,draw opacity=1 ][line width=1.5]  (491,185.13) -- (491,356) -- (404,293.87) -- (404,123) -- cycle ;
\draw [line width=0.75]    (262.5,269) -- (444.25,231.5) ;
\draw  [color={rgb, 255:red, 74; green, 144; blue, 226 }  ,draw opacity=1 ][fill={rgb, 255:red, 74; green, 144; blue, 226 }  ,fill opacity=1 ][line width=1.5]  (444.25,231.5) .. controls (444.25,229.57) and (445.37,228) .. (446.75,228) .. controls (448.13,228) and (449.25,229.57) .. (449.25,231.5) .. controls (449.25,233.43) and (448.13,235) .. (446.75,235) .. controls (445.37,235) and (444.25,233.43) .. (444.25,231.5) -- cycle ;
\draw  [color={rgb, 255:red, 208; green, 2; blue, 27 }  ,draw opacity=1 ][fill={rgb, 255:red, 208; green, 2; blue, 27 }  ,fill opacity=1 ][line width=1.5]  (329.25,254.5) .. controls (329.25,252.57) and (330.37,251) .. (331.75,251) .. controls (333.13,251) and (334.25,252.57) .. (334.25,254.5) .. controls (334.25,256.43) and (333.13,258) .. (331.75,258) .. controls (330.37,258) and (329.25,256.43) .. (329.25,254.5) -- cycle ;
\draw  [draw opacity=0][line width=0.75]  (298.69,272.27) .. controls (297.71,288.34) and (294.18,300.24) .. (289.96,300.24) .. controls (284.99,300.24) and (280.96,283.68) .. (280.96,263.24) .. controls (280.96,242.81) and (284.99,226.24) .. (289.96,226.24) .. controls (293.8,226.24) and (297.07,236.1) .. (298.37,250) -- (289.96,263.24) -- cycle ; \draw  [line width=0.75]  (298.69,272.27) .. controls (297.71,288.34) and (294.18,300.24) .. (289.96,300.24) .. controls (284.99,300.24) and (280.96,283.68) .. (280.96,263.24) .. controls (280.96,242.81) and (284.99,226.24) .. (289.96,226.24) .. controls (293.8,226.24) and (297.07,236.1) .. (298.37,250) ;
\draw  [fill={rgb, 255:red, 0; green, 0; blue, 0 }  ,fill opacity=1 ][line width=1.5]  (298.89,254.36) -- (295.81,245.88) -- (299.89,245.39) -- cycle ;
\draw [color={rgb, 255:red, 208; green, 2; blue, 27 }  ,draw opacity=1 ][fill={rgb, 255:red, 129; green, 27; blue, 40 }  ,fill opacity=1 ][line width=1.5]    (334.25,254.5) -- (366.55,247.58) ;
\draw [shift={(370.46,246.74)}, rotate = 527.9100000000001] [fill={rgb, 255:red, 208; green, 2; blue, 27 }  ,fill opacity=1 ][line width=0.08]  [draw opacity=0] (11.61,-5.58) -- (0,0) -- (11.61,5.58) -- cycle    ;

\draw (450,182) node [anchor=north west][inner sep=0.75pt]  [font=\Large]  {$\mathcal{B}_{\alpha }$};
\draw (239,261) node [anchor=north west][inner sep=0.75pt]    {$x^{3}$};
\draw (427.25,241.5) node [anchor=north west][inner sep=0.75pt]    {$\mathcal{O}^{bdy}$};
\draw (284,304) node [anchor=north west][inner sep=0.75pt]    {$\epsilon $};
\draw (322.25,222.5) node [anchor=north west][inner sep=0.75pt]    {$\mathcal{O}^{bulk}$};

\end{tikzpicture}

\caption{Bulk operators in either omega background acting on boundary operators, defining a module for $\cA_H$ or $\cA_C$. The above represents $\cO^{bulk} \lvert \cO^{bdy}\ra$. }
\label{fig:intro-modules}
\end{figure}

These limits of the hemisphere partition function are then expected to reproduce the characters of the modules formed by boundary chiral or twisted chiral operators. Indeed, we show that these limits reproduce traces over Verma modules
\begin{equation}
\begin{split}
    \mathcal{X}_{\alpha}^H(x) & =  \text{Tr}_{\cH^{(B)}_{\cB_\al}}x^{J_H},\\
   \mathcal{X}_{\alpha}^C(\xi) & =  \text{Tr}_{\cH^{(C)}_{\cB_\al}} \xi^{J_C},
\end{split}    
\end{equation}
where $J_H$, $J_C$ denote complex moment map operators generating the Higgs and Coulomb branch symmetries. It is important here to work with the hemisphere partition function rather than half superconformal index: boundary 't Hooft anomalies encoded in $e^{\phi_\al}$ are crucial to reproduce the correct lowest weights of the Verma modules. We check this proposal explicitly for abelian gauge theories, where the boundary conditions $\cB_\al$ admit a description as exceptional Dirichlet boundary conditions.

Returning to factorisation, we explore the implications of this result for partition functions $\cZ_{\cM_3}$ on compact spaces. Following from the general structure of factorisation, we show that certain limits of the superconformal index, $S^1 \times S^2$ twisted index and $S^3$ partition function preserving additional supercharges can be expressed in terms of the characters of lowest weight Verma modules. In particular, 
\begin{equation}
\begin{gathered}
    \mathcal{Z}_{\text{SC}}^{B} = \sum_{\alpha} \mathcal{X}_\alpha^{H} (x) \mathcal{X}_{\alpha}^{H}(x^{-1})\,, \qquad
    \mathcal{Z}_{\text{SC}}^{A}  = \sum_{\alpha}\mathcal{X}_\alpha^{C} (\xi) \mathcal{X}_{\alpha}^{C}(\xi^{-1})\,, \\
    \mathcal{Z}_{\text{tw}}^{B} = \sum_{\alpha} \mathcal{X}_\alpha^{H} (x) \mathcal{X}_{\alpha}^{H}(x) \,,\qquad
    \mathcal{Z}_{\text{tw}}^{A}  = \sum_{\alpha} \mathcal{X}_\alpha^{C} (\xi) \mathcal{X}_{\alpha}^{C}(\xi) \\
    \mathcal{Z}_{S^3} =  \sum_{\alpha} \hat{\mathcal{X}}_\alpha^H (x)     \hat{\mathcal{X}}_{\alpha}^{C}(\xi) \,,
\end{gathered}    
\end{equation}
where $A$ and $B$ denote two different limits of the superconformal and twisted index preserving additional supercharges. In the factorisation of the $S^3$ partition function, the hatted characters involve an additional $\mathbb{Z}_2$ twist by the centre of the R-symmetry.\footnote{There are some additional phases that we omit in the introduction.} This reproduces the conjectured form of the $S^3$ partition function in~\cite{Gaiotto:2019mmf}  from the perspective of factorisation and extends it to the superconformal and twisted index. We illustrate these factorisations explicitly for supersymmetric QED with $N$ hypermultiplets.

The paper is organised as follows. In section  \ref{sec:bc} we discuss $\cN=(2,2)$ boundary conditions and the associated half superconformal index and hemisphere partition function. In section \ref{sec:thimbles} we consider boundary conditions which mimic an isolated massive vacua at infinity. In particular, we focus on abelian theories for which there exists an explicit UV construction of such boundary conditions as `exceptional Dirichlet', corresponding to thimbles. Finally in section \ref{sec-factorisation}  we discuss holomorphic factorisation of closed three-manifold partition functions in terms of our hemisphere partition functions associated to vacua. This directly yields various `IR formulae' for the superconformal index, twisted index and $S_b^3$ partition function in terms of characters of Verma modules. 

Appendices on boundary conditions and localisation on $S^1\times H^2$, the relation to the work \cite{Dimofte:2017tpi}, and the proof of our claims for general abelian theories are included.


\section{Boundary Conditions}
\label{sec:bc}

We consider boundary conditions in 3d $\cN=4$ gauge theories that flow to superconformal fixed points in the infrared and acquire isolated massive vacua in the presence of generic mass and FI parameter deformations. 

\subsection{Preliminaries}
\label{sec:prelim}

To introduce our notation, suppose the 3d $\cN=4$ theory has global symmetry $G_H \times G_C$ with a maximal torus $T_H \times T_C$ and Cartan subalgebra $\mathfrak{t}_H \oplus \mathfrak{t}_C$. We can then introduce real mass parameters $m \in \mathfrak{t}_H$ and FI parameters $t \in \mathfrak{t}_C$. 
We require that the theory has isolated massive vacua $v_\al$ for generic values of these parameters that preserve the maximal torus $T_H \times T_C$.

The generic condition means the mass and FI parameters lie in chambers $m\in \mathfrak{C}_H \subset \mathfrak{t}_H$ and $t\in \mathfrak{C}_C\subset \mathfrak{t}_C$. These chambers are cut out by co-dimension-1 walls where the tension of domains walls between vacua tends to zero. This tension is controlled by a certain central charge in the supersymmetry algebra
\be
Z_\al = \kappa_\al (m,t),
\ee
where
\be
\kappa_\al: \mathfrak{t}_H \times \mathfrak{t}_C \to \R
\label{eq:kappa}
\ee
is the effective $\cN=4$ supersymmetric mixed Chern-Simons coupling between $T_H$ and $T_C$ in the vacuum $v_\al$. The quantisation of Chern-Simons terms means that this lifts to a bilinear map $\kappa_\al : \Gamma_H \times \Gamma_C \to \mathbb{Z}$,
where $\Gamma_H$, $\Gamma_C$ denote co-character lattices. The walls separating chambers are loci where $Z_\al = Z_\beta$ for pairs of vacua.

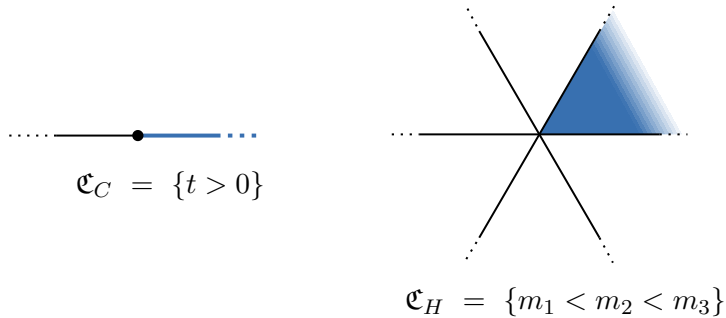
\begin{figure}[h]
\centering

\tikzset {_hcflgukgo/.code = {\pgfsetadditionalshadetransform{ \pgftransformshift{\pgfpoint{0 bp } { 0 bp }  }  \pgftransformrotate{-155 }  \pgftransformscale{2 }  }}}
\pgfdeclarehorizontalshading{_193tmw4bq}{150bp}{rgb(0bp)=(1,1,1);
	rgb(52.76785548244204bp)=(1,1,1);
	rgb(56.78431919642857bp)=(0.22,0.44,0.69);
	rgb(100bp)=(0.22,0.44,0.69)}
\tikzset{every picture/.style={line width=0.75pt}} 

\begin{tikzpicture}[x=0.45pt,y=0.45pt,yscale=-1,xscale=1]

\draw  [draw opacity=0][shading=_193tmw4bq,_hcflgukgo] (462,205) -- (562.59,33.67) -- (664.5,204.97) -- cycle ;
\draw [color={rgb, 255:red, 57; green, 112; blue, 177 }  ,draw opacity=1 ][line width=1.5]    (128,206) -- (152,206) -- (193,206) ;
\draw [line width=0.75]    (60,206) -- (125,206) ;
\draw  [fill={rgb, 255:red, 0; green, 0; blue, 0 }  ,fill opacity=1 ][line width=1.5]  (125,206) .. controls (125,204.34) and (126.34,203) .. (128,203) .. controls (129.66,203) and (131,204.34) .. (131,206) .. controls (131,207.66) and (129.66,209) .. (128,209) .. controls (126.34,209) and (125,207.66) .. (125,206) -- cycle ;
\draw [line width=0.75]  [dash pattern={on 0.84pt off 2.51pt}]  (21,206) -- (60,206) ;
\draw [color={rgb, 255:red, 57; green, 112; blue, 177 }  ,draw opacity=1 ][line width=1.5]  [dash pattern={on 1.69pt off 2.76pt}]  (193,206) -- (229,206) ;
\draw [line width=0.75]    (412,291.6) -- (443.21,237.54) -- (512,118.4) ;
\draw [line width=0.75]    (512,291.6) -- (412,118.4) ;
\draw [line width=0.75]    (562,205) -- (362,205) ;
\draw [line width=0.75]  [dash pattern={on 0.84pt off 2.51pt}]  (402.22,99.58) -- (412,118.4) ;
\draw [line width=0.75]  [dash pattern={on 0.84pt off 2.51pt}]  (512,291.6) -- (523.22,312.58) ;
\draw [line width=0.75]  [dash pattern={on 0.84pt off 2.51pt}]  (562,205) -- (585.22,205) ;
\draw [line width=0.75]  [dash pattern={on 0.84pt off 2.51pt}]  (339.22,205) -- (362,205) ;
\draw [line width=0.75]  [dash pattern={on 0.84pt off 2.51pt}]  (401.22,309.58) -- (412,291.6) ;
\draw [line width=0.75]  [dash pattern={on 0.84pt off 2.51pt}]  (512,118.4) -- (524.22,97.58) ;

\draw (74,233) node [anchor=north west][inner sep=0.75pt]    {$\mathfrak{C}_{C} \ =\ \{t >0\}$};
\draw (348,331) node [anchor=north west][inner sep=0.75pt]    {$\mathfrak{C}_{H} \ =\ \{m_{1} < m_{2} < m_{3}\}$};
\end{tikzpicture}
\caption{Chamber structure for supersymmetric QED with $N = 3$ fundamental hypermultiplets.}
\label{fig:chamber-example}
\end{figure}

A running example will be supersymmetric QED with $N$ fundamental hypermultiplets. In this case, we have $\mathfrak{t}_H = \mathbb{R}^{N-1}$ parametrised by real masses $(m_1,\ldots,m_N)$ with $\sum_j m_j = 0$ and $\mathfrak{t}_C = \mathbb{R}$ parametrised by a real FI parameter $t$. For generic values of these parameters, there are $N$ isolated massive vacua $v_i$ with central charge
\be
Z_i = m_i t \, .
\ee
The chambers are cut out by loci where $t = 0$ and $m_i = m_j$ for $i \neq j$. In section \ref{sec:thimbles}, we choose chambers $\mathfrak{C}_H = \{m_1<m_2<\cdots<m_N\}$ and $\mathfrak{C}_C = \{t>0\}$.
An example of this chamber structure is illustrated in figure~\ref{fig:chamber-example}.


\subsection{$\cN=(2,2)$ Boundary Conditions}
\label{subsec:(2,2)bc}

We consider boundary conditions preserving 2d $\cN=(2,2)$ superconformal symmetry. The boundary conditions support a global symmetry containing a subgroup of the bulk global symmetry $G_H \times G_C$ and any additional symmetries arising from boundary degrees of freedom. In this paper, we focus on boundary conditions preserving at least a maximal torus $T_H \times T_C$ of the bulk theory. 
 
For many interesting boundary conditions, the boundary R-symmetry $U(1)_V \times U(1)_A$ is identified with a maximal torus $U(1)_H \times U(1)_C$ of the bulk R-symmetry. However, it can also happen that $U(1)_H \times U(1)_C$ is spontaneously broken at the boundary but a linear combination involving boundary flavour symmetries is preserved, which we again denote by $U(1)_V \times U(1)_A$. The boundary conditions introduced in section~\ref{sec:thimbles} will be of the latter type.

The boundary global and R-symmetries are subject to boundary mixed `t Hooft anomalies. The possible boundary anomalies are as follows:
\begin{itemize}
\item A mixed anomaly between $U(1)_V$ and $U(1)_A$ with coefficient $\tilde{k}$. 
\item A mixed anomaly between $T_H$ and $U(1)_A$ with coefficient $k_A : \Gamma_H \to \mathbb{Z}$.
\item A mixed anomaly between $T_C$ and $U(1)_V$ with coefficient $k_V : \Gamma_C \to \mathbb{Z} $.
\item A mixed anomaly between $T_H$ and $T_C$ with coefficient $k: \Gamma_H \times \Gamma_C \to \mathbb{Z}$.
\end{itemize}
The last item is closely related to the bulk supersymmetric mixed Chern-Simons coupling and the boundary conditions introduced in section~\ref{sec:thimbles} will have exactly $k = \kappa_\al$. More broadly, boundary 't Hooft anomalies will play an important role throughout.

Let us briefly consider boundary conditions for a free hypermultiplet. A hypermultiplet contains contains two complex scalar fields $X$, $Y$ such that $(X,Y^\dagger)$ transforms as a doublet of $SU(2)_H$ R-symmetry while $(X,Y)$ transform as a doublet of $G_H = SU(2)$. The basic boundary conditions are
\bea
\cB_X & : \qquad \partial_\perp X |_\partial = 0 \qquad Y|_\partial  = 0 \, , \\
\cB_Y & : \qquad \partial _\perp Y |_\partial = 0 \qquad X|_\partial  = 0 \, ,
\label{eq:hyper-bc}
\eea
together with appropriate boundary conditions for the fermions. They break the global symmetry to $T_H = U(1)$ with a boundary mixed 't Hooft anomaly $k_A =+1$, $k_A =-1$ for $\cB_X$, $\cB_Y$. This is normalised such that the contribution from a boundary $\cN=(2,2)$ chiral multiplet of $U(1)$ charge $+1$ to the mixed anomaly is $2$. 


\subsection{Half Superconformal Index}
\label{sec:half-index}

The half superconformal index counts local operators supported on an $\cN=(2,2)$ boundary condition. For concreteness, we work on $\mathbb{R}^2 \times \mathbb{R}_{\geq 0}$ with coordinates $\{x^1,x^2,x^3\}$ where $x^3 \geq 0$. Then a superconformal $\cN=(2,2)$ boundary condition $\cB$ preserves a subalgebra of $\mathfrak{osp}(4|4,\mathbb{R})$ generated by the four supercharges $Q_+^{1\dot1}$, $Q_-^{1\dot2}$, $Q_-^{2\dot 1}$, $Q_+^{2\dot 2}$ and their conjugates in radial quantisation $S^+_{1\dot1}$, $S^-_{1\dot2}$, $S^-_{2\dot 1}$, $S^+_{2\dot 2}$.

We define the half superconformal index by
\be
\cI_\cB = \mathrm{Tr}_{\cH_\cB} (-1)^F  q^{J + \frac{R_V+R_A}{4}} t^{\frac{R_V-R_A}{2}} x^{F_H}\xi^{F_C} \, ,
\label{eq:sci-def}
\ee
where $J$ is the generator of rotations in the $x^{1,2}$-plane, $R_V$, $R_A$ are the generators of the boundary R-symmetry $U(1)_V \times U(1)_A$ and $F_H$, $F_C$ denotes the Cartan generator of the boundary flavour symmetry $T_H \times T_C$. The fermion number is chosen to be $(-1)^F = (-1)^{2J}$. Finally, $\cH_\cB$ denotes the space of states in radial quantisation annihilated by the pair of conjugate supercharges $Q^{1\dot 1}_+$ and $S_{1\dot 1}^+$ or equivalently their anti-commutator
\be
\{Q^{1\dot 1}_+,S_{1\dot 1}^+\}  = D - J - \frac{R_V +R_A}{2}  \, .
\ee
Unitarity bounds of the four supercharges preserved by the boundary condition imply that operators contributing to the index satisfy the inequality 
\be
J +\frac14(R_V+R_A) \geq 0\, ,
\ee
which is saturated only by the unit operator. The half superconformal index is therefore a formal Taylor series in $q^{1/4}$ starting with $1$, whose convergence requires $|q| <1$. These half indices can be computed as in \cite{Dimofte:2017tpi} and can be interpreted as a character of the boundary chiral algebra \cite{Costello:2020ndc}.

Here we have assumed that $U(1)_V \times U(1)_A$ is identified with a maximal torus of the bulk R-symmetry.
If there is mixing with boundary global symmetries then unitarity bounds are modified. In such cases, the half superconformal index may not start with $1$ and convergence may require additional constraints on the flavour fugacities. Examples of this phenomenon are discussed in section~\ref{sec:thimbles}.

For the basic hypermultiplet boundary conditions~\eqref{eq:hyper-bc}, the half superconformal index is given by
\bea
\cI_{\cB_X} & = \frac{(q^{\frac34}t^{-\frac12} x;q)_\infty}{(q^{\frac14}t^{\frac12} x;q)_\infty}  = 1 + q^{\frac14} t^{\frac12} x + \cdots \, , \\
\cI_{\cB_Y} & = \frac{(q^{\frac34}t^{-\frac12} x^{-1};q)_\infty}{(q^{\frac14}t^{\frac12} x^{-1};q)_\infty} = 1 + q^{\frac14} t^{\frac12} x^{-1} + \cdots \, .
\label{eq:hyper-pf}
\eea
Note that the leading contributions to the index beyond the unit operator are the boundary operators $X|_\partial$, $Y|_\partial$ supported on $\cB_X$, $\cB_Y$.

We are primarily interested in two limits $t^{\frac12} \to q^{\pm\frac14}$, where the remaining combinations of generators commute with additional supercharges. These limits require additional constraints on the flavour fugacities to maintain convergence, which is related to the response of boundary conditions to turning on bulk real mass and FI parameters.

\subsubsection{B-Limit}

The $B$-index is defined by
\be
\cI^{(B)}_{\cB} := \lim_{t^{\frac12} \to q^{-\frac14}} \cI_\cB = \mathrm{Tr}_{\cH^{(B)}_\cB}  x^{F_H} \, .
\label{eq:B-limit-index}
\ee
In the limit $t^{\frac12} \to q^{-\frac14}$, the generator $J+\frac{R_A}{2}$ conjugate to $q$ commutes with an additional supercharge $Q_-^{1\dot2}$. The index therefore receives contributions only from operators in the subspace $\cH^{(B)}_\cB \subset \cH_\cB$ annihilated by both supercharges $Q^{1\dot 1}_+$, $Q^{1\dot 2}_-$ and their conjugates in radial quantisation, or equivalently by the anti-commutators
\bea
\{Q^{1\dot 1}_+,S_{1\dot 1}^+\}  & = D - J - \frac{R_V}{2} -\frac{R_A}{2}  \, , \\
\{Q^{1\dot 2}_-,S_{1\dot 2}^-\}  & = D + J - \frac{R_V}{2} + \frac{R_A}{2}  \, .
\eea
Such operators transform as the scalar components of $\cN=(2,2)$ chiral multiplets and include the images of bulk Higgs branch operators under the bulk to boundary map. Their quantum numbers obey
\be
D = \frac{R_V}{2} \,, \quad J + \frac{R_A}{2} = 0
\ee
and therefore the index is independent of $q$. They are uncharged under $T_C$ so it is also independent of $\xi$. Finally, we can remove the $(-1)^F$ as such operators are bosons.

To maintain convergence, there must clearly be a constraint on $x$. We can regard this parameter as an element of the complexified maximal torus $T_H \otimes_\R \C$. If a boundary condition preserves $\cN=(2,2)$ supersymmetry in the presence of a real mass $m$, boundary operators contributing to the $B$-limit of the half superconformal index obey 
\be
\la m , F_H \ra \geq 0\, .
\label{eq:ineq-B}
\ee
The index will therefore converge if $-\log |x|$ lies in the same chamber as $m$. In summary:
\begin{itemize}
\item If a boundary condition is compatible with a real mass $m \in \mathfrak{C}_H$, the $B$-limit of the half superconformal index converges for $-\log|x| \in \mathfrak{C}_H$.
\end{itemize}

We illustrate this statement for a hypermultiplet. The $B$-limit of the half superconformal indices of the basic boundary conditions are
\bea
\cI^{(B)}_{\cB_X} & = 1+ x+x^2+ \cdots = \frac{1}{1-x} \, ,\\
\cI^{(B)}_{\cB_Y} & =  1 +x^{-1} +x^{-2} +\cdots = \frac{1}{1-x^{-1}} \, .
\eea
These expansions arise from monomials in the boundary Higgs branch operators $X|_\partial$ and $Y|_\partial$ respectively. The index of $\cB_X$ converges for $|x|<1$, while that of $\cB_Y$ converges for $|x|>1$. This is consistent with the fact that the $\cB_X$ is compatible with real mass parameter $m>0$, while $\cB_Y$ is compatible with $m<0$~\cite{Bullimore:2016nji}.

In section~\ref{sec:s3pf}, we will also encounter the closely related limit $t^{\frac12} \to e^{-\pi i} q^{-\frac14}$. Almost identical arguments hold except the differing sign leads to an additional factor of $(-1)^{R_V}$ in equation~\eqref{eq:B-limit-index}, such that the bottom components of chiral multiplets are counted with an additional sign depending on their vector R-charge.

\subsubsection{A-Limit}

The $A$-index is similarly defined by
\be
\cI^{(A)}_\cB := \lim_{t^{\frac12} \to q^{\frac14}} \cI = \mathrm{Tr}_{\cH^{(A)}_\cB}  \xi^{F_C} \, .
\label{eq:A-limit-index}
\ee
In the limit $t^{\frac12} \to q^{\frac14}$, the generator $J+\frac{R_V}{2}$ conjugate to $q$ now commutes with an additional supercharge $Q_-^{2\dot1}$. The index therefore receives contributions only from operators in the subspace $\cH^{(A)}_\cB \subset \cH_\cB$ annihilated by both supercharges $Q^{1\dot 1}_+$, $Q^{2\dot 1}_-$ and their conjugates in radial quantisation, or equivalently by the anti-commutators
\bea
\{Q^{1\dot 1}_+,S_{1\dot 1}^+\}  & = D - J - \frac{R_V}{2} -\frac{R_A}{2}  \, , \\
\{Q^{2\dot 1}_-,S_{2\dot 1}^-\}  & = D + J + \frac{R_V}{2} -\frac{R_A}{2} \, .
\eea
Such operators transform as the scalar component of $\cN=(2,2)$ twisted chiral multiplets and include the images of bulk Coulomb branch operators under the bulk to boundary map. The quantum numbers of such operators obey
\be
D = \frac{R_A}{2} \,, \quad J + \frac{R_V}{2} = 0
\ee
and therefore the index is independent of $q$. They are not charged under $T_H$ so it is also independent of $x$. Finally, we can again remove the $(-1)^F$ as such operators are bosons.

To maintain convergence, we now need a constraint on $\xi$. We can regard this parameter as an element of the complexified maximal torus $T_C \otimes_\R \C$. If a boundary condition preserves $\cN=(2,2)$ supersymmetry in the presence of a real FI parameter $t$, boundary operators contributing to the $A$-limit of the half superconformal index obey 
\be
\la t , F_C \ra \geq 0\, .
\label{eq:ineq-A}
\ee
The index will therefore converge if $-\log |\xi|$ lies in the same chamber as $t$. In summary:
\begin{itemize}
\item If a boundary condition is compatible with a real FI parameter $t \in \mathfrak{C}_C$, the $A$-limit of the half superconformal index converges for $-\log|\xi| \in \mathfrak{C}_C$.
\end{itemize}
For hypermultiplet boundary conditions,
\be
\cI^{(A)}_{\cB_X} = \cI^{(A)}_{\cB_Y} = 1 \, ,
\ee
which simply reflects the absence of bulk Coulomb branch operators that could supply twisted chiral operators at the boundary. This index is independent of $\xi$ so there is no issue with convergence in this case.

In section~\ref{sec:s3pf}, we will also encounter the closely related limit $t^{\frac12} \to e^{\pi i}q^{\frac14}$. Almost identical arguments hold except the differing sign leads to an additional factor of $(-1)^{R_A}$ in equation~\eqref{eq:A-limit-index}, such that the bottom components of twisted chiral multiplets are counted with an additional sign depending on their axial R-charge.


\subsection{Hemisphere Partition Function}\label{subsec:hemispherepartitionfunction}

The half superconformal index can be computed from a UV description by invoking the state-operator correspondence to relate it to a hemisphere partition function on $S^1 \times H^2$ and applying supersymmetric localisation. This essentially builds on similar computations for the bulk superconformal index, using either Coulomb branch or Higgs branch localisation. We give the details of this computation, and the form of boundary conditions on this geometry in appendix \ref{app:localisation}.

From one perspective, the $S^1 \times H^2$ background is a product
\be
ds^2 = d \tau^2 + r^2 (d\theta^2 + \sin^2\theta d\phi^2) \, ,
\ee
where $\tau \sim \tau +\beta r$ and $0 \leq \theta \leq \pi/2$ and the boundary condition $\cB$ supported at $\theta = \pi / 2$. The boundary conditions around $S^1$ are then twisted  according to the fugacities in the superconformal index~\eqref{eq:sci-def}. Another perspective is to replace the metric by an $S^1$-fibration over $H^2$ together with an appropriate background connection for the boundary global and R-symmetries around $S^1$. The fugacity $q$ is set to $e^{-2 \beta}$, see appendix \ref{app:localisation}.

The result of supersymmetric localisation leads to the computation of 1-loop determinants that require regularisation in a way compatible with the supersymmetry preserved. A consequence is that the hemisphere partition function $\cZ_\cB$ of an $\cN=(2,2)$ boundary condition is related to the superconformal index by a multiplicative factor,
\be
\cZ_\cB = e^{\phi_\cB} \cI_\cB \, ,
\ee
where $\phi_B$ is determined by the boundary mixed 't Hooft anomalies. In fact, this is true for $\mathcal{N}=2$ theories with $(0,2)$ boundary conditions, as we show in appendix \ref{app:localisation}. Specialising to the $\mathcal{N}=4$ case results in the only possible anomalies being those enumerated in section \ref{subsec:(2,2)bc}, with integer valued coefficients. An analysis of the 1-loop determinants then shows that
\be
\phi_\cB = \frac{1}{ 2 \log q} \sum_{ij}  \log y_i \cdot k_{ij} \cdot \log y_j \,,
\label{eq-pre-factor}
\ee
where the indices $i$, $j$ are summed over the Cartan generators of all boundary global and $R$-symmetries, $y_i$, $y_j$ denote the corresponding fugacities and the numbers $k_{ij}$ are the corresponding boundary mixed 't Hooft anomaly coefficients. Using our notation for the possible anomaly coefficients from section~\ref{subsec:(2,2)bc} this becomes
\bea
\phi_\cB & = \frac{1}{ \log q} \left[ \log \left(  q^{\frac14}t^{\frac12} \right) \cdot  \tilde{k} \cdot \log \left( q^{\frac14}t^{-\frac12} \right) \right] \\
& + \frac{1}{ \log q} \left[ \log \left(  q^{\frac14}t^{-\frac12} \right) \cdot  k_A \cdot \log x \right] \\
&  + \frac{1}{  \log q} \left[  \log \xi \cdot  k_V \cdot  \log \left( q^{\frac14}t^{\frac12} \right)\right] \\
& + \frac{1}{ \log q} \left[ \log \xi   \cdot k \cdot \log x \right] \, ,
\eea
where from our definition~\eqref{eq:sci-def}  of the half superconformal index the fugacities associated to $U(1)_V$ and $U(1)_A$ are $q^{\frac14}t^{\frac12}$ and $q^{\frac14}t^{-\frac12}$ respectively.

Let us illustrate this result for the basic boundary conditions~\eqref{eq:hyper-bc} for a hypermultiplet. Combining the results for Neumann and Dirichlet boundary conditions for 3d $\cN=2$ chiral multiplets found in \cite{Dimofte:2017tpi} we find
\bea
 \phi_{\cB_X} = + \frac{1}{\log q}  \log x\log(q^{\frac14}t^{-\frac12}) \, , \\ 
 \phi_{\cB_Y} = - \frac{1}{\log q}  \log x\log(q^{\frac14}t^{-\frac12}) \, ,
\label{eq:hyper-prefactor}
\eea
which reproduces the boundary mixed 't Hooft anomaly between the $U(1)$ global symmetry and $U(1)_A$ axial R-symmetry with coefficients $k_A = +1,-1$ for $\cB_X$, $\cB_Y$.

\subsubsection{B-Limit}

In the limit $t^{\frac12} \to q^{-\frac14}$, the fugacity conjugate to $U(1)_A$ becomes $q^{\frac12}$ while the fugacity conjugate to $U(1)_V$ becomes $1$. The overall factor relating the hemisphere partition function and the half superconformal index therefore no longer detects boundary 't Hooft anomalies involving $U(1)_V$. Explicitly, it becomes
\be
\phi^{(B)}_\cB  =   \frac{1}{2} k_A \cdot \log x  + \frac{\log x \cdot  k \cdot \log \xi}{\log q} \, ,
\ee
and exponentiating
\be
e^{\phi^{(B)}_\cB} = x^{\frac{k_A}{2} +  k \,\cdot \frac{\log \xi }{ \log q} } \, .
\label{eq-pre-factor-B}
\ee
Note that although the $B$-limit of the half superconformal index is independent of $\xi$, the hemisphere partition function may retain some dependence on $\log \xi$ through the boundary mixed anomaly between $T_C$ and $T_H$. We denote
\bea 
\lim_{t^{\frac12}\rightarrow q^{-\frac14}} \cZ_{\cB} \coloneqq \mathcal{X}^H_{\cB}\,.
\eea 
For the hypermultiplet, this limit is
\be
\mathcal{X}^H_{\cB_X} = \frac{x^{1/2}}{1-x}\, , \qquad \mathcal{X}^{H}_{\cB_Y} = \frac{x^{-1/2}}{1-x^{-1}} \, ,
\label{eq-hyper-hemi}
\ee
which encodes the anomaly coefficients $k_A = 1$ for $\cB_X$ and $k_A = - 1$ for $\cB_Y$. 

\subsubsection{A-Index}

In the limit $t^{\frac12} \to q^{\frac14}$, the fugacity conjugate to $U(1)_V$ becomes $q^{\frac12}$ while the fugacity conjugate to $U(1)_A$ becomes $1$. The overall factor relating the hemisphere partition function and the half superconformal index therefore no longer detects boundary 't Hooft anomalies involving $U(1)_A$. Explicitly, it becomes
\be
\phi^{(A)}_\cB  =   \frac{1}{2} k_V \cdot \log \xi  + \frac{\log \xi \cdot k \cdot \log x}{\log q} \, ,
\ee
and exponentiating
\be
e^{\phi^{(A)}_\cB} = {\xi}^{\frac{k_V}{2} +  k \,\cdot \frac{\log x }{ \log q} } \, .
\label{eq-pre-factor-A}
\ee
Note that although the $A$-limit of the half superconformal index is independent of $x$, the hemisphere partition function may retain some dependence on $\log x$ through the boundary mixed anomaly between $T_C$ and $T_H$. We denote
\bea 
\lim_{t^{\frac12}\rightarrow q^{\frac14}} \cZ_{\cB} \coloneqq \mathcal{X}^C_{\cB}\,.
\eea 
For the hypermultiplet, 
\be
\mathcal{X}^{C}_{\cB_X} = 1\, , \qquad \mathcal{X}^{C}_{\cB_Y} = 1 \, ,
\ee
as the only boundary mixed 't Hooft anomalies involve $U(1)_A$.


\subsection{Characters of Modules}
\label{sec:characters}

Let us return temporarily to the half superconformal index on $\mathbb{R}^2 \times \mathbb{R}_{\geq 0}$. We have considered two limits of the half superconformal index,
\bea
\cI^{(A)}_\cB = \mathrm{Tr}_{\cH^{(A)}_\cB}  \xi^{F_C} \, ,\\
\cI^{(B)}_\cB = \mathrm{Tr}_{\cH^{(B)}_\cB}   x^{F_H} \, ,
\label{eq:half-index-def-2}
\eea
where $\cH^{(B)}_\cB$ and $\cH^{(A)}_\cB$ denote respectively boundary operators that are the scalar components of $\cN=(2,2)$ chiral and twisted chiral multiplets. 

These setups admit deformations that can be described either as an omega background~\cite{Yagi:2014toa,Bullimore:2015lsa,Bullimore:2016hdc,Beem:2018fng} or passing to a `$Q+S$' type construction as in~\cite{Dedushenko:2016jxl,Dedushenko:2017avn,Dedushenko:2018icp, Oh:2019bgz, Jeong:2019pzg}. For concreteness, we will focus here on the omega background perspective.

There are two possible omega backgrounds $\Omega_A$, $\Omega_B$ in the $x^{1,2}$-plane. 
These deformations break superconformal symmetry but the boundary operators at the origin of the $x^{1,2}$-plane remain the same. However, bulk local operators are now constrained to the $x^3$-axis and generate non-commutative algebras $\cA_H$, $\cA_C$ that act on boundary operators. In this way, $\cH^{(A)}_\cB$, $\cH^{(B)}_\cB$ become modules for $\cA_H$, $\cA_C$, as described in~\cite{Bullimore:2016nji}. This is illustrated in figure~\ref{fig:intro-modules}.

The algebras $\cA_H$, $\cA_C$ are equivariant deformation quantisations of the Poisson algebras of functions on the Higgs and Coulomb branch respectively. They are determined by periods $t_\C \in \mathfrak{t}_C \otimes \C$ and $m_\C \in \mathfrak{t}_H \otimes \C$, which are complex mass and FI parameters. The algebras include operators $J_H$, $J_C$, whose commutators measure $T_H$, $T_C$ charge. For example, in $\cA_H$ we have
\be
[J_H , \cO_\gamma ] = \gamma\, \cO_\gamma
\ee
where $\cO_\gamma$ is a Higgs branch operator of charge $\gamma \in \Gamma^\vee_H$. This provides a grading of the non-commutative algebras $\cA_H$, $\cA_C$ by the character lattices $\Gamma_H^\vee$, $\Gamma_C^\vee$. Similarly, there is a weight decomposition of any module generated by a boundary condition preserving global symmetry $T_H$, $T_C$.

Now consider the operators
\be
J_m = m \cdot J_H \, , \qquad J_t = t \cdot J_C\, ,
\ee
where $m$ and $t$ are the real mass and FI parameters. An observation of~\cite{Bullimore:2016nji} is that boundary conditions compatible with real parameters $m$, $t$ determine modules that are lowest weight for the operators $J_m$, $J_t$, meaning their weights are bounded below. 

This property only depends on the chamber: if a module is lowest weight for $m\in\mathfrak{C}_H$, it is lowest weight for any other $m'\in \mathfrak{C}_H$ in the same chamber. Therefore, having fixed $\mathfrak{C}_H$, $\mathfrak{C}_C$, we simply refer to modules associated to compatible boundary conditions as lowest weight.

The modules $\cH^{(A)}_\cB$, $\cH^{(B)}_\cB$ will therefore have lowest weight states that we denote by $|\cB\ra^{(A)}$, $|\cB\ra^{(B)}$. If we were to add constants to the operators $J_H$, $J_C$ such that the lowest weight states have charge $0$, this would correspond to the charge measured by the generators $F_H$, $F_C$ appearing in the definition of the half superconformal index. The condition of lowest weight is then equivalent to the inequalities~\eqref{eq:ineq-B} and~\eqref{eq:ineq-A} and the characters of these modules coincide with the half superconformal indices in~\eqref{eq:half-index-def-2}. 

However, as we show for a general abelian theory in appendix \ref{app:abelian}, the charges of the lowest weight states measured by the operators $J_H$, $J_C$ are determined by boundary mixed 't Hooft anomalies:
\bea
J_H |\cB\ra^{(B)} & =  \left(\frac{1}{2} k_A + \frac{1}{\ep} t_\C \cdot k   \right)  |\cB\ra^{(B)} \, ,\\
J_C |\cB\ra^{(A)} & = \left( \frac{1}{2} k_V  + \frac{1}{\ep} m_\C  \cdot k  \right) |\cB\ra^{(A)} \, .
\eea
Let us now define the equivariant characters of these modules by
\bea\label{eq:vermamodulecharacters}
\mathcal{X}_{\cB}^H & = \text{Tr}_{\cH^{(B)}_\cB} x^{J_H} \, ,\\
\mathcal{X}_{\cB}^C & = \text{Tr}_{\cH^{(A)}_\cB} \xi^{J_C} \, .
\eea
Then the lowest weight states contribute the following multiplicative factors
\be
x^{\frac{1}{2} k_A + \frac{1}{\ep} t_\C \cdot k} \, , \qquad \xi^{\frac{1}{2} k_V  + \frac{1}{\ep} m_\C  \cdot k  } 
\ee
to these equivariant characters.

If we now compare to the multiplicative factor relating the hemisphere partition function to the half superconformal index in~\eqref{eq-pre-factor-B} and~\eqref{eq-pre-factor-A}, we can identify the hemisphere partition function with the character 
\bea
\lim_{t^{\frac{1}{2}}\rightarrow q^{-\frac{1}{4}}} \cZ_\cB & = \mathcal{X}_{\cB}^H,\\
\lim_{t^{\frac{1}{2}}\rightarrow q^{\frac{1}{4}}}   \cZ_\cB & = \mathcal{X}_{\cB}^C\,.
\eea
under the following identification of variables
\be
\ep \leftrightarrow -\log q ,\qquad m_\C \leftrightarrow -\log x ,\qquad t_\C \leftrightarrow - \log m_\C \, .
\ee
It would be desirable to give a more direct derivation of this correspondence by carefully understanding the map from the operator counting picture to the $S^1 \times H^2$ background used for supersymmetric localisation. Nevertheless, this relation will play an important role in the remainder of this paper.

\subsubsection{Example}

We briefly consider the $\Omega_B$ deformation of the free hypermultiplet. The quantised algebra $\cA_H$ is generated by the complex scalar fields $\hat X$, $\hat Y$ subject to $[\hat Y,\hat X] = \ep$. The basic boundary conditions correspond to the modules
\bea
\cH_{\cB_X}^{(B)} : \quad |n\ra = \hat X^n |0\ra \quad n \geq 0 \, , \\
\cH_{\cB_Y}^{(B)} : \quad |n\ra = \hat Y^n |0\ra \quad n \geq 0 \, ,
\eea
where for convenience we write $|0\ra := |\cB_X\ra^{(B)}$ or $|\cB_Y\ra^{(B)}$, which obeys $\hat Y |0\ra = 0$ and $\hat X |0\ra = 0$ respectively. 

The global symmetry $T_H =U(1)$ preserved by both boundary conditions is generated by the complex moment map
\be
J_H: = \frac{1}{\ep}\hat X\hat Y +\frac{1}{2} = \frac{1}{\ep}\hat Y \hat X - \frac{1}{2} \, .
\ee
such that
\bea
& \cH_{\cB_X}^{(B)} : \quad J_H | n \ra = +\left(n+\frac{1}{2}\right)  |n\ra \, , \\
& \cH_{\cB_Y}^{(B)} : \quad J_H | n \ra = -\left(n+\frac{1}{2}\right) |n\ra \, .
\eea
Note that the normal ordering of the moment map reproduces the expected shifts due to the boundary mixed anomaly $k_A = +1, -1$. We also see explicitly that $\cB_X$ is compatible with $m>0$ and lowest weight in the chamber $\mathfrak{C}_H = \{m>0\}$, while $\cB_Y$ is compatible with $m<0$ and lowest weight in opposite chamber $\mathfrak{C}_H =\{m<0\}$. 

The characters of these modules are
\bea
\mathcal{X}_{\cB_X}^H & = x^{1/2} \sum_{n\geq 0} x^n = \frac{x^{1/2}}{1-x} \, ,\\
\mathcal{X}_{\cB_Y}^H  & = x^{-1/2} \sum_{n\geq 0} x^{-n} = \frac{x^{-1/2}}{1-x^{-1}} \, ,
\eea
which converge to the function on the right when $|x| <1$ for $\cB_X$ and $|x|>1$ for $\cB_Y$. This is in perfect agreement with the hemisphere partition functions~\eqref{eq-hyper-hemi}.


\section{Thimble Boundary Conditions}
\label{sec:thimbles}

We now focus on a distinguished class of boundary conditions that mimic the presence of an isolated massive vacuum at infinity, at least for the purpose of computations preserving supersymmetry. This idea is illustrated figure \ref{fig:thimble-schematic}.

Boundary conditions of this type were first studied for 2d $\cN=(2,2)$ Landau-Ginzburg models and massive sigma models in~\cite{Hori:2000ck} and play an important part in 2d mirror symmetry. A systematic description in massive 2d $\cN=(2,2)$ theories has also been developed in~\cite{Gaiotto:2015zna,Gaiotto:2015aoa}. They were discussed for massive 3d $\cN=4$ theories in~\cite{Bullimore:2016nji} and constructed explicitly for abelian gauge theories. 

\subsection{General Idea}

First recall our restriction to 3d $\cN=4$ theories that have isolated massive vacua $v_\al$ in the presence of generic mass and FI parameter deformations. A choice of  generic real mass and FI parameters determines a pair of chambers $m \in \mathfrak{C}_H \subset \mathfrak{t}_H$ and $t\in\mathfrak{C}_C \subset \mathfrak{t}_C$ in which the theory remains massive.

The aim is to construct a collection of UV boundary conditions $\{\cB_\al\}$ in 1-1 correspondence with isolated massive vacua $v_\al$ that are simultaneously compatible with mass and FI parameters in the chambers $\mathfrak{C}_H$, $\mathfrak{C}_C$ and mimic the presence of an isolated massive vacuum $v_\al$ at infinity. The collection $\{\cB_\al\}$ depend on the chambers and may jump across walls in the space of mass and FI parameters.

Before turning to an example, we mention one generic feature of such boundary conditions. Since the boundary condition $\cB_\al$ is equivalent to the vacuum $v_\al$ at infinity, the mixed 't Hooft anomaly between $T_H$ and $T_C$ should coincide with the effective supersymmetric Chern-Simons coupling in the vacuum $v_\al$, namely 
\be
k(\cB_\alpha) = \kappa_\al \, ,
\ee
where $\kappa_\al : \Gamma_H \times \Gamma_C \to \mathbb{Z}$ is the bilinear map introduced in~\eqref{eq:kappa}. This will indeed be the case in our example. In the present work we focus on the example of supersymmetric QED, upcoming work \cite{Crew:2020psc} applies these ideas to a non-abelian theory with adjoint matter.

\subsection{Abelian Theories}

In abelian gauge theories, there is a proposal for constructing the collections $\{\cB_\al\}$ using `exceptional Dirichlet' boundary conditions~\cite{Bullimore:2016nji}. This involves a Dirichlet boundary condition for the $\cN=4$ vectormultiplet and a standard boundary condition for the hypermultiplets, deformed by non-vanishing expectation values such that a maximal torus $T_H \times T_C$ of the bulk global symmetry is preserved. We focus here on supersymmetric QED, leaving general abelian theories to appendix~\ref{app:abelian}.  

Let us then consider supersymmetric QED with gauge group $G = U(1)$ and $N$ fundamental hypermultiplets $(X_j,Y_j)$. The bulk global symmetries are $G_H = PSU(N)$ and $G_C = U(1)$ (enhanced to $SU(2)$ when $N=2$). Correspondingly, we can introduce real mass parameters $(m_1,\ldots,m_N)$ obeying $\sum_j m_j = 0$ and a real FI parameter $t$. 

The classical vacua are solutions of
\bea
& \sum_{j=1}^N |X_j|^2-|Y_j|^2 = t \, , \qquad && \sum_j X_j Y_j  = 0\, , \\
& (\sigma +m_j) X_j = 0 \, ,\qquad && \varphi X_j = 0\,, \\
& (\sigma -m_j) Y_j = 0 \, ,\qquad && \varphi Y_j = 0\,,
\eea
where $\sigma$ and $\varphi$ are the real and complex scalar fields in the vectormultiplet respectively. 

Assuming generic real mass and FI parameters, there are $N$ isolated massive vacua,
\be
v_i: \quad |X_j|^2-|Y_j|^2 = \begin{cases}
t & \mathrm{if} \quad j = i \\
0 & \mathrm{if} \quad j \neq i
\end{cases} \, ,
\qquad  X_j Y_j  = 0 \, ,
\qquad \sigma = - m_i \, ,
\qquad \varphi = 0\, ,
\ee
labelled by $i=1,\ldots,N$. The massive vacua have central charges 
\be
Z_i = \sum_{j=1}^N m_j \left(|X_j|^2-|Y_j|^2\right) \Big\rvert_{v_i} = m_i t
\ee
or equivalently mixed supersymmetric Chern-Simons term with components $\kappa_{i,j} = \delta_{ij}$.

In this case, generic parameters means concretely that $m_i \neq m_j$ for $i\neq j$ and $t \neq 0$. There are therefore $N!$ chambers $\mathfrak{C}_H \subset \mathfrak{t}_H$ specified by an ordering of the real masses and two chambers $\mathfrak{C}_C \subset \mathfrak{t}_H$ specified by the sign of $t$.

Henceforth, we fix
\be
\mathfrak{C}_H = \{m_1< m_2<\ldots<m_N\} \,, \qquad \mathfrak{C}_C = \{t>0\} \, .
\ee
We now consider exceptional Dirichlet boundary boundary conditions $\cB_i$ which behave as thimble boundary conditions in the presence of mass and FI parameters in these chambers. We refer to \cite{Bullimore:2016nji} for more details.\footnote{In the language of \cite{Bullimore:2016nji} we work with `right' boundary conditions, but our convention for the FI parameter is opposite.} The boundary condition $\cB_i$ imposes Dirichlet boundary conditions for the vector multiplet  with a non-vanishing expectation value
\be
\varphi|_\partial = \varphi_0
\ee
together with
\bea\label{eq:SQED-exceptionaldirichlet}
\partial_\perp Y_j  & = 0 ,\qquad X_j  = c\delta_{ij} \qquad && j\leq i \\
\partial_\perp X_j  & = 0 ,\qquad Y_j  = 0 \qquad && j>i \, ,
\eea
where $c \neq 0$.

The boundary conditions $\{\cB_i\}$ associated to the opposite chamber $\mathfrak{C}_H = \{t<0\}$ for the FI parameter are obtained by interchanging the boundary conditions for $X_j$ and $Y_j$ for all $j=1,\ldots,N$. Similarly, the boundary conditions associated to other chambers $\mathfrak{C}_H$ for the mass parameters are related by permutations of the hypermultiplets.

The boundary mixed 't Hooft anomalies can be computed following~\cite{Dimofte:2017tpi},
\bea\label{eq:anomalypolySQED}
\tilde{k}(\cB_i) & = 2i-N-1 , \\
k_V(\cB_i) & = 1\, ,\\
k_{A,j}(\cB_i) & = \begin{cases}
-1 & \text{if} \quad j < i \\
 2i-N-1 & \text{if} \quad j = i \\
+ 1 & \text{if} \quad j > i
\end{cases} \,, \\
k_j(\cB_i) & = \delta_{ij} \, ,
\eea
where $j =1\ldots,N$. The general abelian case is derived in appendix \ref{app:abelian}. As expected, the anomaly coefficient in the final line coincides with the components of the effective supersymmetric Chern-Simons term $\kappa_{i,j} = \delta_{ij}$ in the vacuum $v_i$.

\subsection{Half Superconformal Index}

We compute the half superconformal index of the boundary conditions $\cB_i$ in two steps. We first compute the half superconformal index of a Dirichlet boundary condition with $c = 0$ and then deform to $c \neq 0$. The second step involves a redefinition of the boundary symmetries and therefore we first review this process abstractly. This is similar in spirit to the construction of surface defects in~\cite{Gaiotto:2012xa}.

Suppose we have a Dirichlet boundary condition $\cB$ in a $U(1)$ gauge theory preserving a maximal torus $U(1)_V \times U(1)_A$ of the bulk R-symmetry and a distinguished boundary symmetry $U(1)_\partial$ arising from the bulk gauge symmetry. The half superconformal index of this boundary condition has the form
\be
\cI_\cB = \mathrm{Tr}_{\cH_\cB} (-1)^F  q^{J + \frac{R_V+R_A}{4}} t^{\frac{R_V-R_A}{2}} x^{F_H}\xi^{F_C} z^{F_g}\, ,
\ee
where $z$ and $F_g$ denote respectively the fugacity and generator of $U(1)_\partial$. Suppose we now initiate a boundary RG flow to a new superconformal boundary condition $\cB_c$ by turning on an expectation value $c$ for a hypermultiplet scalar of charge $+1$ under $U(1)_\partial$ and weight $Q_H$ under $T_H$. A hypermultiplet scalar also has $r_V = 1$ and therefore a linear combination of $U(1)_V$, $U(1)_\partial$, $T_H$ is spontaneously broken. However, the linear combinations
\bea
R'_V & := R_V - F_g \\
F'_H & := F_H - Q_HF_g
\label{eq:grading shift}
\eea
are preserved along the RG flow and become the boundary vector R-symmetry and Higgs branch flavour symmetry of boundary condition $\cB_c$. 

At the level of the half superconformal index, this is implemented by setting the weight of this field to unity, $q^{\frac14}t^{\frac12}x^{Q_H} z =1$ and eliminating $z$. Indeed, we find
\bea
\cI_\cB(z \to q^{-1/4}t^{-1/2}x^{-Q_F}) 
& = \mathrm{Tr}_{\cH_{\cB}} (-1)^F  q^{J + \frac{R'_V+R_A}{4}} t^{\frac{R'_V-R_A}{2}} x^{F'_H}\xi^{F_C} \\
& = \mathrm{Tr}_{\cH_{\cB_c}} (-1)^F  q^{J + \frac{R'_V+R_A}{4}} t^{\frac{R'_V-R_A}{2}} x^{F'_H}\xi^{F_C} \, .
\eea
In making this argument, we assume that any difference between $\cH_{\cB_c}$ and $\cH_{\cB}$ (with the gradings shifted by setting $z = q^{-\frac14}t^{-\frac12}x^{-Q_F}$) cancels out in the trace. This follows from the fact that $c$ is an exact deformation of the boundary action.

Let us now implement this procedure for exceptional Dirichlet boundary conditions. The first step is to evaluate the half superconformal index of the Dirichlet boundary condition with $c= 0$ in equation~\eqref{eq:SQED-exceptionaldirichlet} and preserves an additional boundary symmetry $U(1)_\partial$ with fugacity $z$.  This is given by 
\bea
\label{eq:SQEDdirichletpredeformation}
\sum_{m \in \mathbb{Z}} \left( \xi \left(q^{\frac14}t^{-\frac12}\right)^{2i-N} \right)^m\frac{(tq^{\frac{1}{2}};q)_{\infty}}{(q;q)_{\infty}} \prod_{j\leq i}   \frac{\left(q^{\frac{3}{4}+m} t^{-\frac{1}{2}}z^{-1}x_j^{-1};q\right)_{\infty}}{\left(q^{\frac{1}{4}+m}t^{\frac{1}{2}}z^{-1}x_j^{-1};q\right)_{\infty}}\prod_{j>i}
    \frac{\left(q^{\frac{3}{4}-m} t^{-\frac{1}{2}}z x_j;q\right)_{\infty}}{\left(q^{\frac{1}{4}-m}t^{\frac{1}{2}}z x_j;q\right)_{\infty}} \, .
\eea
where we have fugacities $\xi$ and $x_1,\ldots,x_N$ for $T_C$ and $T_H$ respectively and the $q$-Pochhammer symbols $(a,q)_\infty$ should be understood as expansions in $q$. The summation over $m\in \mathbb{Z}$ arises from boundary monopole operators. The power of $q^{\frac14}t^{-\frac12}$ multiplying $\xi$ is due a boundary mixed  't Hooft anomaly between $U(1)_A$ and $U(1)_\partial$.

The second step is introduce an expectation value $c\neq 0$ for $X_i$ and flow to the exceptional Dirichlet boundary condition $\cB_i$. As described above, this is implemented by setting $z = x_i^{-1}t^{-\frac{1}{2}}q^{-\frac{1}{4}}$. Performing this substitution in equation~\eqref{eq:SQEDdirichletpredeformation}, the half superconformal index is

\bea\label{eq:hemisphere-SQED-1loopandvortex}
    \cI_{\cB_i} &= \prod_{j=1}^{i-1} \frac{\left(q\frac{x_i}{x_j};q\right)_{\infty}}{\left(q^{\frac{1}{2}} t \frac{x_i}{x_j};q\right)_{\infty}} \prod_{j=i+1}^N \frac{\left(q^{\frac{1}{2}}t^{-1} \frac{x_j}{x_i};q\right)_{\infty}}{\left(\frac{x_j}{ x_i};q\right)_{\infty}}\\
    &\times  \sum_{m\geq0} \left(\left(q^{\frac{1}{4}}t^{-\frac{1}{2}}\right)^{N}\xi\right)^m \prod_{j=1}^N \frac{\left(q^{\frac{1}{2}}t\frac{x_i}{x_j};q\right)_m}{\left(q\frac{x_i}{x_j};q\right)_m} \, ,
\eea
where the summation now only extends over $m \in \mathbb{Z}_{\geq 0}$. The second line coincides with the vortex partition function for $\cN=4$ supersymmetric QED \cite{Dimofte:2010tz,Bullimore:2014awa,Zenkevich:2017ylb,Aprile:2018oau} and can be interpreted geometrically as a K-theoretic vertex function \cite{Aganagic:2016jmx,Aganagic:2017smx,Smirnov:2020lhm}.

Let us now consider limits of the half superconformal index preserving additional supercharges. First, in the $A$-limit $t^{\frac12} \to q^{\frac14}$, the contributions from ratios of $q$-Pochhammer symbols cancel out completely leaving 
\be
\cI^{(A)}_{\cB_i} =  \sum_{m\geq 0}\xi^m = \frac{1}{1-\xi} 
\ee
for all $i = 1,\ldots,N$. This converges when $|\xi| <0$, corresponding to the fact that the collection of exceptional Dirichlet boundary conditions $\{\cB_i\}$ are compatible with a real FI parameter in the chamber $\mathfrak{C}_C = \{ t >0\}$.  

Second, in the $B$-limit $t^{\frac12} \to q^{-\frac14}$, the contributions from $m>0$ vanish and the remaining contribution from $m=0$ converges to
\bea
\cI_{\cB_i}^{(B)} & = \prod_{j=1}^{i-1} \frac{1}{1-x_i/x_j}  \prod_{j=i+1}^{N} \frac{1}{1-x_j/x_i} 
\eea
provided that $|x_j| <|x_k|$ for $k<j$. This corresponds to the fact that the collection of exceptional Dirichlet boundary conditions $\{\cB_i\}$ are compatible with real mass parameters in the chamber $\mathcal{C}_H = \{ m_1<m_2<\cdots m_N\}$.

\subsection{Hemisphere Partition Function}\label{subsec:SQEDHPS}

We can now upgrade these computations to the hemisphere partition function. The ratios of $q$-Pochhammer symbols are replaced by regularised 1-loop determinants. The details are included in appendix \ref{app:localisation}. The result is an additional prefactor $e^{\phi_i}$ encoding the boundary mixed 't Hooft anomalies obtained by substituting (\ref{eq:anomalypolySQED}) into (\ref{eq-pre-factor}). Explicitly
\bea 
\cZ_{\cB_i} = \cZ^{\text{Cl}}_i  \cZ^{\text{1-loop}}_i  \cZ^{\text{Vortex}}_i
\eea
where:
\bea \label{eq-classicalHPSSQED}
 \cZ^{\text{Cl}}_i = e^{\phi_i}\,,
\eea 
\bea\label{eq-1-loopHPSSQED}
 \cZ^{\text{1-loop}}_i = \prod_{j=1}^{i-1} \frac{\left(q\frac{x_i}{x_j};q\right)_{\infty}}{\left(q^{\frac{1}{2}} t \frac{x_i}{x_j};q\right)_{\infty}} \prod_{j=i+1}^N \frac{\left(q^{\frac{1}{2}}t^{-1} \frac{x_j}{x_i};q\right)_{\infty}}{\left(\frac{x_j}{ x_i};q\right)_{\infty}}\,,
\eea 
\bea\label{eq-vortexHPSSQED}
\cZ^{\text{Vortex}}_i =  \sum_{m\geq0} \left(\left(q^{\frac{1}{4}}t^{-\frac{1}{2}}\right)^{N}\xi\right)^m \prod_{j=1}^N \frac{\left(q^{\frac{1}{2}}t\frac{x_i}{x_j};q\right)_m}{\left(q\frac{x_i}{x_j};q\right)_m} \,.
\eea 
The prefactor is given by
\bea\label{eq-prefactorSQEDHPS}
\phi_i = & \,(2i-N-1) \frac{\log\left(q^{\frac{1}{4}}t^{-\frac{1}{2}}\right)\log\left(q^{\frac{1}{4}}t^{\frac{1}{2}}\right)}{\log q} + \frac{\log{\xi}\log(x_i)}{\log q} + \frac{\log{\xi}\log\left(q^{\frac{1}{4}}t^{\frac{1}{2}}\right)}{\log q}\\
&\qquad+ \sum_{j < i} \frac{\log\left(q^{\frac{1}{4}}t^{-\frac{1}{2}}\right)\log(x_i/x_j)}{\log q}
+\sum_{j > i} \frac{\log\left(q^{\frac{1}{4}}t^{-\frac{1}{2}}\right)\log(x_j/x_i)}{\log q }\,.
\eea
We have limits:
\bea\label{eq-HPSHiggsVerma}
    \mathcal{X}_i^H = \lim_{t^{\frac12}\rightarrow q^{-\frac14}} \cZ_{\cB_i}  = e^{\frac{\log{\xi}\log(x_i)}{\log q}} 
    \prod_{j<i} \frac{\left(x_i /x_j\right)^{1/2}}{1-x_i / x_j}
    \prod_{j>i} \frac{\left(x_j/x_i\right)^{1/2}}{1-x_j/x_i} \,,
\eea
\bea\label{eq-HPSCoulombVerma}
    \mathcal{X}_i^C = \lim_{t^{\frac12}\rightarrow q^{\frac14}} \cZ_{\cB_i}  = 
     e^{\frac{\log{\xi}\log(x_i)}{\log q}} \frac{ \xi^{\frac{1}{2}}}{1-\xi} \,.
\eea

\subsection{Characters of Verma Modules}

The exceptional Dirichlet boundary conditions $\cB_i$ define lowest weight Verma modules for the quantised algebra of functions on the Coulomb branch and Higgs branch in the $\Omega_A$ and $\Omega_B$ deformations respectively. We now show that the $A$ and $B$-limits of the hemisphere partition function reproduces the characters of these representations.

\subsubsection{Higgs Branch}

The quantised Higgs branch chiral ring in supersymmetric QED can be constructed via quantum symplectic reduction. It is generated from $N$ commuting copies of the Heisenberg algebra
\be
[\hat{Y}_j,\hat{X}_j] = \epsilon\delta_{ij} \, , \quad j = 1,\ldots,N \, ,
\ee
restricting to gauge invariant combinations, and imposing the constraint
\be
\sum_{j=1}^N {:}\hat{X}_j \hat Y_j{:} = t_\C \, .
\ee
where the normal ordering is $ {:}\hat X_j \hat Y_j{:} = \hat X_j \hat Y_j + \frac{\varepsilon}{2} = \hat Y_j \hat X_j - \frac{\varepsilon}{2} $. These are the quantisations of the complex moment maps for the $U(1)$ subgroup of $T_H$ rotating the $j^{\text{th}}$ hypermultiplet. The complex FI parameter $t_\C$ determines the period of the deformation quantisation.

It is convenient to introduce gauge-invariant generators
\bea
h_j & = \, \hat X_j \hat Y_j - \hat X_{j+1}\hat Y_{j+1}  \,  ,\\
e_j & = \hat X_j \hat Y_{j+1} \quad j =1,\ldots,N-1\, , \\
f_j & = \hat X_{j+1} \hat Y_j  \quad j =1,\ldots,N-1 \, ,\\
\eea
such that
\bea
\left[ e_i, f_j \right] & = \ep \, \delta_{ij} h_j \, ,  \\
\left[ h_i, e_j \right] & =  + \ep \, A_{ij} e_j \, , \\ 
\left[ h_i, f_j \right] & = - \ep \, A_{ij} f_j \, , 
\eea
and
\bea
\text{ad}(e_i)^{1-A_{ij}} e_j & = 0 \,  ,\\
\text{ad}(f_i)^{1-A_{ij}} f_j & = 0 \, .
\eea
where $A_{ij}$ is the Cartan matrix of $\mathfrak{sl}_N$.  
The complex moment map equation then determines all of the Casimir elements of the enveloping algebra of $\mathfrak{sl}_N$ in terms of the period $t_\C$. We therefore find a central quotient of $U(\mathfrak{sl}_N)$.

More generally, it is convenient to introduce generators
\bea
e_{i,j} & = \hat X_i \hat Y_j \quad \text{for} \quad i < j\, ,\\
f_{i,j} & = \hat X_i \hat Y_j \quad \text{for} \quad i>j \, ,
\eea
such that for example, $e_{j,j+1} = e_j$ and $e_{j,j+2} = \frac{1}{\ep}[ e_j,e_{j+1}]$. We also note that the generator of the global symmetry $U(1)_m \subset T_H$ generated by real mass parameters $m_1,\ldots,m_N$ is
\bea
h_m 
&:= \sum_{j=1}^N m_j \, {:}\hat X_j \hat Y_j{:} \\
& =\sum_{j,\,k=1}^{N-1} (m_j - m_{j+1}) A^{-1}_{jk} h_j
\eea
such that
\bea
\left[h_m,e_{i,j} \right] & = \ep (m_i-m_j) e_{i,j}  \quad \text{for} \quad   i<j \, ,\\
\left[h_m,f_{i,j} \right] & = \ep (m_i-m_{j}) f_{i,j}  \quad \text{for} \quad  i>j \, . \\ 
\label{eq:comms}
\eea
This means that inside our chosen chamber $\mathfrak{C}_H=\{m_1<m_2< \cdots m_N\}$ for the real mass parameters, $e_{i,j}$ and $f_{i,j}$ are lowering and raising operators respectively for the weight associated to $h_m$.

Let us now consider the modules associated to the exceptional Dirichlet boundary conditions $\cB_i$ defined in equation~\eqref{eq:SQED-exceptionaldirichlet}. These modules are generated by acting on a vacuum state $|\cB_i\ra$ satisfying
\bea
\hat X_j \, | \cB_i \ra & = \delta_{ij} c |\cB_i\ra  \quad && \text{for}\quad j =1,\ldots,i  \, , \\
\hat Y_i \, |  \cB_i \ra & = 0 \quad && \text{for} \quad j = i+1,\ldots,N \, ,
\eea
where $c$ is a non-zero constant. In the action of gauge-invariant generators on the vacuum state, the constant $c$ can always be absorbed using the fact that the complex moment map equation annihilates this vacuum state. First, we find
\be
h_m |\cB_i \ra = \left[ \frac{\ep}{2} \left(\sum_{j>i} m_j - \sum_{j<i} m_j \right) + \left(t_\C - \frac{N-2i+1}{2} \ep \right) m_i  \right] |\cB_i \ra
\label{eq:h-action}
\ee
which encodes the boundary mixed 't Hooft anomalies for the global symmetry $U(1)_m$ in (\ref{eq:anomalypolySQED}) as claimed in section \ref{sec:characters}, after identifying $J_{H,i} = \frac{1}{\varepsilon}{:}\hat X_j \hat Y_j{:}$ and fugacities $x_i = e^{-m_i}$. In addition, the boundary state is annihilated by $e_{j,k}$ for all $j < k$. Finally, the operators not annihilating the boundary state are
\bea\label{eq:raisingoperatorsSQEDhiggs}
f_{i,j} \quad \text{for} \quad j < i \\
f_{k,i} \quad \text{for} \quad k>i 
\eea
and therefore their action on the boundary state generates a lowest weight Verma module in our chamber for the mass parameters.

We can now compute the character of this module using equation~\eqref{eq:h-action} and the commutators~\eqref{eq:comms} to find
\bea
  \text{Tr} \,e^{-\frac{h_m}{\ep}} 
  &=x_i^{\frac{t_{\mathbb{C}}}{\epsilon}} \prod_{j<i}\left(\frac{x_i}{x_j}\right)^{\frac{1}{2}} \prod_{k>i}\left(\frac{x_k}{x_i}\right)^{\frac{1}{2}}
  \prod_{j<i}\left(1+\frac{x_i}{x_j} +\frac{x_i^2}{x_j^2}+... \right)
  \prod_{k>i}\left(1+\frac{x_k}{x_i} +\frac{x_k^2}{x_i^2}+... \right)\\
  & = x_i^{\frac{t_{\mathbb{C}}}{\epsilon}} \prod_{j<i}\frac{\left(\frac{x_i}{x_j}\right)^{\frac{1}{2}}}{1-\frac{x_i}{x_j}} \prod_{k>i}\frac{\left(\frac{x_k}{x_i}\right)^{\frac{1}{2}}}{1-\frac{x_k}{x_i}} .
\eea
where the second line converges in our choice of chamber. This reproduces the $B$-limit (\ref{eq-HPSHiggsVerma}) of the hemisphere partition function.

\subsubsection{Coulomb Branch}

The quantised Coulomb branch chiral ring of supersymmetric QED is generated by the complex scalar $\varphi$ and the monopole operators $v^{\pm}$ subject to
\begin{equation}
\begin{split}
    & [\hat{\varphi}, \hat{v}_{\pm}] = \pm \epsilon \hat{v}_{\pm} \,,\\
    & \hat{v}_{+} \hat{v}_{-} = \prod_{i=1}^N\left(\varphi+m_{i,\mathbb{C}} - \frac{\epsilon}{2}\right) \,,\\
    & \hat{v}_{-} \hat{v}_{+} = \prod_{i=1}^N\left(\varphi+m_{i,\mathbb{C}} + \frac{\epsilon}{2}\right)  \, ,
\end{split}    
\end{equation}
which is a spherical rational Cherednik algebra. 

The topological global symmetry generated by a real FI parameter $t \in \mathbb{R}$ is generated by the operator $h_t = -t \hat \varphi$ such that
\be
[h_t , \hat v_\pm] = \mp \ep t   \hat v_\pm \, .
\ee
This means that in our chamber $\mathfrak{C}_H = \{t>0\}$, the monopole operator $\hat{v}_+$ is a lowering operator and $\hat v_-$ is a raising operator with respect to $h_t$. The minus sign in $h_t$ compared to $h_m$ comes from our convention for the FI parameter.

Let us now consider the modules for the quantised Coulomb branch algebra associated to the exceptional Dirichlet boundary conditions $\cB_i$. The modules are generated by boundary states $|\cB_i\ra$ that obey
\be
\left(\hat\varphi+m_{i,\C}+\frac{\ep}{2}\right) |\cB_i\ra = 0 ,\qquad \hat v_+ |\cB_i \ra = 0 \, .
\ee
Note that the expression in the brackets is the effective complex mass in the $\Omega_A$-deformation of the complex scalar $X_i$, which arises because $X_i$ receives a non-vanishing expectation value at the boundary. The second arises from an analysis of boundary monopole operators. The boundary condition therefore generates a lowest weight Verma module by acting with $\hat v_-$. The character of this module is
\begin{equation}
\begin{split}
    \text{Tr} \,e^{-\frac{h_t}{\varepsilon}} 
    &= \xi^{\frac{m_{i,\mathbb{C}}}{\epsilon}+\frac{1}{2}}(1+\xi+\xi^2+\ldots) \\
    &= \xi^{\frac{m_{i,\mathbb{C}}}{\epsilon}}\frac{\xi^{\frac{1}{2}}}{1-\xi} 
\end{split}
\end{equation}
which converges to the second line for $|\xi|<1$. This agrees with the result for the $A$-limit (\ref{eq-HPSCoulombVerma}) of the hemisphere partition function.

\section{Factorisation}\label{sec-factorisation}

We now consider the factorisation of 3d $\mathcal{N}=4$ partition functions on closed 3-manifolds in terms of hemisphere partition functions associated to vacua. As a corollary to our analysis for hemisphere partition functions, we show that 3-manifold partition functions can be factorised in terms of Verma module characters of $\cA_H$ and $\cA_C$.

\subsection{Preliminaries}

For theories with $\cN \geq 2$ supersymmetry, partition functions on many 3-manifolds $\cM_3$ admit a factorisation schematically of the form
\bea
    \mathcal{Z}_{\mathcal{M}_3} =  \sum_{\alpha}H_{\alpha}\tilde{H}_{\alpha}
    \label{eq:fact}
\eea 
where $\alpha$ correspond to isolated vacua. The `$\sim$' operation implements a transformation of fugacities  corresponding to the orientation reversal and element $g$ of $SL(2,\mathbb{Z})$ gluing the boundary tori $\partial(S^1\times H^2) = T^2$ in the Heegaard decomposition of $\mathcal{M}_3$:
\bea
    \mathcal{M}_3 = \left( S^1 \times H^2 \right) \cup_g \left( S^1 \times H^2 \right).
\eea
In this work we focus on factorisations of the $S^1 \times S^2$ superconformal and twisted indices, and the partition function on the squashed sphere or ellipsoid.

 Our proposal for theories with $\cN=4$ supersymmetry is to identify the components $H_{\alpha}$ with hemisphere partition function on $S^1\times H^2$ computed with the particular boundary condition $\mathcal{B}_{\alpha}$:
\bea
     H_{\alpha} = \mathcal{Z}_{\cB_\alpha}\,.
\eea 
This identification depends on a choice of chambers $\mathfrak{C}_H$, $\mathfrak{C}_C$ in the spaces of real mass and FI parameters and the blocks $H_\al$ may differ from the traditional holomorphic blocks in classical and 1-loop contributions. This gives a clean geometric interpretation of factorisation where each block is associated to a vacuum in a systematic way. 

It is then natural to examine factorisation in limits that preserve additional supercharges. This yields various formulae for such partition functions as sums over vacua of pairs of characters of Verma modules for $\cA_H$, $\cA_C$. Such a formula was proposed for the $S^3$ partition function in \cite{Gaiotto:2019mmf}. The present work shows that this arises naturally from the more general factorisation in equation~\eqref{eq:fact}.  We check this explicitly for a free hypermultiplet and supersymmetric QED.

Partial factorisations have been demonstrated explicitly using Coulomb branch localisation in a number of examples \cite{Pasquetti:2011fj,Hwang:2012jh,Hwang:2015wna,Dimofte:2011py,Cabo-Bizet:2016ars,Crew:2020jyf}. Higgs branch localisation offers a more direct approach where the path integral is localised to a sum over vortex contributions \cite{Benini:2013yva,Fujitsuka:2013fga}. We note in contrast the factorisation we propose is exact, in the sense that the perturbative pieces of $\cZ_{\mathcal{M}_3}$ are fully factorised into those of $\mathcal{Z}_{\cB_\alpha}$.

\subsection{Superconformal Index}

The superconformal index on $S^1 \times S^2$ is defined analogously to the half superconformal index introduced in section \ref{sec:half-index} and so our discussion here is brief. The superconformal index is defined by
\begin{equation}\label{eq:scindexdefinition}
    \cZ_{\text{SC}}= \text{Tr}_{\cH_{\text{SC}}}  (-1)^F q^{J_3 + \frac{R_V+R_A}{4}}t^{\frac{R_V-R_A}{2}}x^{F_H}\xi^{F_C} \, ,
\end{equation}
where $\cH_{\text{SC}}$ is the space of local operators annihilated by the pair of conjugate supercharges $Q_+^{1\dot1}$ and $S^+_{1\dot1}$, or equivalently states in radial quantisation. The index can be computed as a path integral on $S^1 \times S^2$ \cite{Kim:2009wb, Imamura:2011su, Kapustin:2011jm, Fujitsuka:2013fga, Benini:2013yva}. 

We propose an exact factorisation of the superconformal index into hemisphere partition functions for the distinguished boundary conditions $\cB_\al$ associated to vacua,
\begin{equation}\label{eq:SCindexfactorisation}
    \mathcal{Z}_{\text{SC}}(q,t,x,\xi) = \sum_{\alpha} \mathcal{Z}_{\cB_\alpha}(q,t,x,\xi) \mathcal{Z}_{\cB_\alpha}(\bar{q},\bar{t},\bar{x},\bar{\xi})\,,
\end{equation}
where
\begin{equation}\label{eq:SCindexgluing}
    \bar{q}=q^{-1},\enspace \bar{t}=t^{-1},\enspace \bar{x}=x^{-1},\enspace \bar{\xi}=\xi^{-1}
\end{equation}
is the transformation of variables implementing the splitting of $S^1 \times S^2$.

We are interested in limits of the superconformal index as $t^{\frac{1}{2}} \to q^{\pm \frac{1}{4}}$, where the remaining generators commute with additional supercharges. These limits were also studied in \cite{Razamat:2014pta}, where it was noted that the superconformal index reproduces the Hilbert series of the Higgs and Coulomb branch, and thus depend only on fugacities $x$ and $\xi$ respectively. We make a connection here to characters of Verma modules for $\cA_H$, $\cA_C$.

The arguments are the same as in section \ref{sec:bc}, and using the exact factorisation (\ref{eq:SCindexfactorisation}), in the limit we recover the equivariant Coulomb and Higgs branch Hilbert series
\begin{equation}\label{eq:SCindexvermacharacterlimit}
\begin{split}
    \left[\lim_{t^{\frac{1}{2}}\rightarrow q^{\frac{1}{4}}} \mathcal{Z}_{\text{SC}} \right] (\xi)& = \sum_{\alpha} \mathcal{X}^{C}_{\alpha}(q,x,\xi) \mathcal{X}^{C}_{\alpha}(q^{-1},x^{-1}, \xi^{-1})\,, \\
    \left[\lim_{t^{\frac{1}{2}}\rightarrow q^{-\frac{1}{4}}} \mathcal{Z}_{\text{SC}} \right] (x) & = \sum_{\alpha} \mathcal{X}^{H}_{\alpha}(q,x,\xi) \mathcal{X}^{H}_{\alpha}(q^{-1},x^{-1}, \xi^{-1})\,,
\end{split}        
\end{equation}
expressed as a sum of products of Verma module characters for $\cA_H$ and $\cA_C$ respectively. Note that although $\mathcal{X}_{\alpha}^C$ retain a residual $q$ and $x$ dependence due to the mixed $T_H\times T_C$ boundary 't Hooft anomaly, these contributions cancel in the gluing such that the limit $t^{\frac{1}{2}}\rightarrow q^{\frac{1}{4}}$ of the superconformal index depends only on $\xi$. Analogous statements hold in the other limit.

\subsubsection{Example: Hypermultiplet}

We briefly consider factorisation of the superconformal index of a free hypermultiplet. In the chamber $\mathfrak{C}_H = \{m>0\}$, the factorisation is in terms of the boundary condition $\cB_X$. In the absence of background flux, the superconformal index is\footnote{We use $\left\lVert \cdot \right\rVert_{\text{SC}}^2$ throughout this section to denote the gluing (\ref{eq:SCindexgluing}), and similar notation for the twisted index and ellipsoid partition function.}
\bea
	\cZ_{\text{SC}} = \frac{\left( q^{\frac{3}{4}}  x t^{-\frac{1}{2}}; q\right)_{\infty}}{\left( q^{\frac{1}{4}} x^{-1} t^{\frac{1}{2}};q\right)_{\infty}}
	\frac{\left( q^{\frac{3}{4}} x^{-1} t^{-\frac{1}{2}}; q\right)_{\infty}}{\left(q^{\frac{1}{4}}  x t^{\frac{1}{2}};q\right)_{\infty}}
	=   \left\lVert \cZ_{\cB_X}\right\rVert_{\text{SC}}^2\,,
\eea
where
\be
\cZ_{\cB_X} = e^{\frac{1}{\log q}  \log x\log(q^{\frac14}t^{-\frac12}) }\frac{(q^{\frac34}t^{-\frac12} x;q)_\infty}{(q^{\frac14}t^{\frac12} x;q)_\infty}
\label{eq:hyper-hemi}
\ee
is the full hemisphere partition function of $\cB_X$ and we have used the analytic continuation $(a;q)_{\infty} = \left(aq^{-1};q^{-1}\right)_{\infty}^{-1}$. Note that the contribution of boundary anomalies to the hemisphere partition function \eqref{eq:hyper-hemi} cancels out in the superconformal index.

The superconformal index in the $A$ limit $t^{\frac{1}{2}}\rightarrow q^{\frac{1}{4}}$ is $1$, reflecting the absence of a Coulomb branch. The superconformal index in the $B$-limit $t^{\frac{1}{2}}\rightarrow q^{-\frac{1}{4}}$ is
\bea 
	\lim_{t^{\frac{1}{2}}\rightarrow q^{-\frac{1}{4}}} \cZ_{\text{SC}}  = \mathcal{X}_{\cB_X}^H(x)  \mathcal{X}_{\cB_X}^H(x^{-1}) = -\frac{x}{(1-x)^2} \, ,
\eea 
which coincides with the equivariant Hilbert series of the Higgs branch $T^*\mathbb{C}$.

\subsubsection{Example: SQED}

The superconformal index of supersymmetric QED with $N$ hypermultiplets can be computed by localisation and was factorised into holomorphic blocks in \cite{Hwang:2012jh}. After an appropriate redefinition of parameters, shifting the fugacity $t$ to grade by the $\mathcal{N}=4$ superconformal R-charge and including the contribution of the $\mathcal{N}=2$ adjoint chiral multiplet we have:
\bea
     \mathcal{Z}_{\text{SC}} &=
     \sum_{\mathfrak{m \in \mathbb{Z}}}  \,\, \xi^{\mathfrak{m}}
     \left(\frac{q^{\frac{1}{2}}}{t}\right)^{\frac{N\lvert \mathfrak{m}\rvert}{2}}  \frac{\left(tq^{\frac{1}{2}};q\right)_{\infty}}{\left(t^{-1}q^{\frac{1}{2}};q\right)_{\infty}}   \\
     &\qquad\,\, \oint \frac{dz}{2\pi i z} \prod_{j=1}^N \frac{\left(z^{-1}x_j^{-1}t^{-\frac{1}{2}}q^{\frac{3}{4}+\frac{\lvert\mathfrak{m}\vert}{2}};q\right)_{\infty}}{\left(z x_j t^{\frac{1}{2}}q^{\frac{1}{4}+\frac{\lvert\mathfrak{m}\vert}{2}};q\right)_{\infty}}
     \frac{\left(z x_j t^{-\frac{1}{2}}q^{\frac{3}{4}+\frac{\lvert\mathfrak{m}\vert}{2}};q\right)_{\infty}}{\left(z^{-1} x_j^{-1} t^{\frac{1}{2}}q^{\frac{1}{4}+\frac{\lvert\mathfrak{m}\vert}{2}};q\right)_{\infty}}\\
     &= \sum_{i=1}^N \left\Vert \prod_{j\neq i} \frac{\left(q\frac{x_i}{x_j};q\right)_{\infty}}{\left(tq^{\frac{1}{2}} \frac{x_i}{x_j};q\right)_{\infty}}  \sum_{m\geq0} \left(\left(q^{\frac{1}{4}}t^{-\frac{1}{2}}\right)^{N}\xi\right)^m \prod_{j=1}^N \frac{\left(tq^{\frac{1}{2}}\frac{x_i}{x_j};q\right)_m}{\left(q\frac{x_i}{x_j};q\right)_m} \right\rVert_{\text{SC}}^2
\eea
where the contour encloses the poles
\bea
  z= x_j^{-1} t^{-\frac{1}{2}}q^{-\frac{1}{4}-\frac{\vert \mathfrak{m}\rvert}{2}-l}\quad j=1,\ldots,N\,,\quad l\in \mathbb{Z}_{\geq 0}\,.
\eea

The holomorphic block decomposition is not automatically written in terms of hemisphere partition functions of the boundary conditions $\cB_i$. In order to do so, we can rewrite the 1-loop contribution to $\cZ^{\text{1-loop}}_i$ of $\cZ_{\cB_i}$ given in equation \eqref{eq-1-loopHPSSQED} as
\begin{equation}\label{eq:SCindex-modified1loop}
    \cZ^{\text{1-loop}}_i =
    \prod_{j\neq i} \frac{\left(q\frac{x_i}{x_j};q\right)_{\infty}}{\left(tq^{\frac{1}{2}} \frac{x_i}{x_j};q\right)_{\infty}} \prod_{j=i+1}^N \frac{\theta\left(tq^{\frac{1}{2}} \frac{x_i}{x_j};q\right)}{\theta\left(q  \frac{x_i}{x_j};q\right)}\,,
\end{equation}
where we define $\theta(x;q) := (x;q)(qx^{-1};q)$. Then we note that the theta functions in \eqref{eq:SCindex-modified1loop} fuse trivially using the identity
\bea
\theta(aq^{\frac{m}{2}};q) \theta(a^{-1}q^{-\frac{m}{2}};q^{-1}) = 1\,,
\eea 
and also that the anomaly contribution to the hemisphere partition function in equation~\eqref{eq-classicalHPSSQED} satisfies $\| \mathcal{Z}^{\text{Cl}}_{i} \|_{\text{SC}}^2 = 1$. Combining these results we find
\begin{equation}
    \mathcal{Z}_{\text{SC}} = \sum_{i=1}^N \left \lVert \cZ_{\cB_i}  \right \rVert_{\text{SC}}^2\,, 
\end{equation}
as required. This computation for the superconformal index had the simple feature that the classical or anomaly contribution glues to $1$ and we could have worked with the half-superconformal index $\cI_{\cB_i}$. However, this will not be the case for the twisted index, where it plays a crucial role in recovering an exact factorisation. 

In the two limits with enhanced supersymmetry~\eqref{eq:SCindexvermacharacterlimit} we find
\bea\label{eqSCindexlimitSQED}
    \lim_{t^{\frac12}\rightarrow q^{\frac14}} \mathcal{Z}_{\text{SC}} & = \sum_{i=1}^N \mathcal{X}^{C}_i(q, x, \xi) \mathcal{X}^{C}_{i}(q^{-1}, x^{-1}, \xi^{-1})\\
    &= -\frac{N\xi}{(1-\xi)^2}\,,\\
    \lim_{t^{\frac12}\rightarrow q^{-\frac14}} \mathcal{Z}_{\text{SC}} & = \sum_{i=1}^N \mathcal{X}^{H}_{i}(q, x, \xi) \mathcal{X}^{H}_{i}(q^{-1}, x^{-1}, \xi^{-1}) \\
    &=  (-1)^{N-1}\sum_{i=1}^N \, \prod_{j\neq i} \frac{x_j/x_i}{(1-x_j/x_i)^2}\,,
\eea
which coincide with the equivariant Coulomb and Higgs branch Hilbert series for supersymmetric QED respectively, up to an overall sign. We note that as expected these depend only on $\xi$ and $x$ respectively.


\subsection{$S^1 \times S^2$ Twisted Index}

We next consider the twisted index of 3d $\cN=4$ supersymmetric gauge theories on $S^1 \times S^2$~\cite{Benini:2015noa, Closset:2016arn, Bullimore:2018jlp}. There are two versions of the twisted index depending on which R-symmetry is used to twist along $S^2$:
\begin{itemize}
\item The $A$-twisted index $\cZ_{\text{tw}}^A$ twists using $U(1)_H$.
\item The $B$-twisted index $\cZ_{\text{tw}}^B$ twists using $U(1)_C$. 
\end{itemize}
These two indices preserve a common pair of supercharges $Q_+^{1\dot1}$, $Q_-^{2\dot2}$ that commute with the combinations $J+\frac{R_V}{2}$ and $J+\frac{R_A}{2}$, and the anti-diagonal combination $R_V-R_A$. The twisted indices are then defined by
\bea
\label{eq:twistedindextrace}
    \cZ_{\text{tw}}^A & = \text{Tr}_{\cH_{S^2}^{A}} (-1)^F q^{J+\frac{R_V}{2}} t^{\frac{R_V-R_A}{2}} x^{F_H}\xi^{F_C} \, , \\
    \cZ_{\text{tw}}^B & = \text{Tr}_{\cH_{S^2}^{B}}  (-1)^F q^{J+\frac{R_A}{2}} t^{\frac{R_V-R_A}{2}} x^{F_H}\xi^{F_C}  \, ,
\eea
where $\cH_{S^2}^{A,B}$ denote respectively states in the $A$, $B$ twisted theory on $S^2$ that are annihilated by the supercharges $Q_+^{1\dot1}$ and $Q_-^{2\dot2}$.

It was shown in \cite{Bullimore:2018jlp} that the twisted indices are generating functions for a certain virtual Euler character of moduli spaces of twisted quasi-maps $\cQ$ from $S^2$ to $\mathcal{M}_H$. The twisted index can be computed by Coulomb branch localisation and factorised into holomorphic blocks \cite{Cabo-Bizet:2016ars,Crew:2020jyf}.\footnote{It would be would be interesting to verify this with a Higgs branch localisation scheme including the angular momentum deformation $q$.} Geometrically, this can be understood as a factorisation
\begin{equation}
     \cZ_{\text{tw}}  \simeq \sum_{\alpha}  \left\lVert \chi \left(\mathfrak{Q}^{(\alpha)}\right) \right\rVert_{\text{tw}}^2
\end{equation}
where $\chi(\mathfrak{Q}^{(\alpha)})$ denotes schematically a generating function for a virtual Euler character of the moduli space $\mathfrak{Q}^{(\alpha)}$ of based quasi-maps tending to the vacuum $\nu_{\alpha}$~\cite{Crew:2020jyf}. 

In this section we propose an exact factorisation of the twisted indices in terms of hemisphere partition functions of the distinguished boundary conditions $\cB_\al$. In this work we do not consider turning on background fluxes for the flavour symmetries. In order to express this factorisation, it is first convenient to introduce $A$- and $B$-shifted hemisphere partition functions defined by 
\begin{equation}
    \mathcal{Z}_{\cB_\alpha}(q,t,x,\xi) = \mathcal{Z}_{\cB_{\alpha}}^{A}(q,tq^{-\frac{1}{2}},x,\xi) = \mathcal{Z}_{\cB_{\alpha}}^{B}(q,tq^{\frac{1}{2}},x,\xi)\,.
\end{equation}
Note that more accurately we mean that, for example in passing to the $A$-shifted hemisphere partition function, we replace $t^{\frac12} \rightarrow t^{\frac12}q^{\frac14}$. We then propose:
\begin{equation}\label{eq:proposedtwistedfactorisation}
\begin{split}
    \cZ_{\text{tw}}^A (q,t,x,\xi) &= \sum_{\alpha} \mathcal{Z}_{\cB_{\alpha}}^{A}(q,t,x,\xi) \mathcal{Z}_{\cB_{\alpha}}^{A} (\bar{q},\bar{t},\bar{x},\bar{\xi})\,,\\
    \cZ_{\text{tw}}^B (q,t,x,\xi) &=  \sum_{\alpha} \mathcal{Z}_{\cB_{\alpha}}^{B}(q,t,x,\xi) \mathcal{Z}_{\cB_{\alpha}}^{B} (\bar{q},\bar{t},\bar{x},\bar{\xi}) \,,\\
\end{split}    
\end{equation}
where the gluing is
\begin{equation}
    \bar{q}=q^{-1},\enspace \bar{t}=t,\enspace \bar{x}=x, \enspace \bar{\xi} = \xi\,.
\end{equation}

We are again interested in the limit $t^{\frac12}\to1$ of the $A$, $B$ twisted indices, which preserves the four supercharges commuting with $J+\frac{R_V}{2}$, $J+\frac{R_A}{2}$. Supersymmetry implies $\cZ_{\text{tw}}^A$ and $\cZ_{\text{tw}}^B$ are independent of the fugacities $x$ and $\xi$ respectively and (in the absence of background flux) both are independent of $q$. Therefore
\bea
\label{eq:twistedindextrace2}
    \cZ_{\text{tw}}^A& = \text{Tr}_{\cH_{S^2}^{A}} (-1)^F   \xi^{F_C}  \, , \\
    \cZ_{\text{tw}}^B & = \text{Tr}_{\cH_{S^2}^{B}}  (-1)^F  x^{F_H}  \, ,
\eea
where $\cH_{A,B}^{S^2}$ now denotes respectively states in the $A$, $B$ twisted theory on $S^2$ annihilated by all four supercharges commuting with $J+\frac{R_V}{2}$, $J+\frac{R_A}{2}$.

These limits compute the partition function of the fully topologically twisted theory, or equivariant Rozansky-Witten invariant, on $S^1 \times S^2$. In this case, the topological state-operator map can be invoked to show that the index counts operators in the cohomology of the scalar supercharges 
\bea
Q_A & := Q^{1\dot1}_+ + Q^{2\dot1}_- \,,\\
Q_B & := Q^{1\dot1}_+ + Q^{1\dot2}_- \, .
\eea
In `good' and `ugly' theories in the sense of \cite{Gaiotto:2008ak}, this coincides with local operators in the Coulomb and Higgs branch chiral ring and therefore the twisted indices $\cZ_{\text{tw}}^A$ and $\cZ_{\text{tw}}^B$ are expected to again reproduce the equivariant Hilbert series of the Coulomb and Higgs branch respectively. For example, the integral representation of the $B$-twisted index reproduces the Molien integral for the Hilbert series of the Higgs branch \cite{Closset:2016arn}. The $t^{\frac{1}{2}}\rightarrow 1$ limits of the $A$ and $B$ twisted indices therefore coincide with the $t^{\frac{1}{2}} \rightarrow q^{\frac14}$ and $t^{\frac{1}{2}} \rightarrow q^{-\frac14}$ limits of the superconformal index respectively.

From our proposed factorisation~\eqref{eq:proposedtwistedfactorisation} we recover the formulae
\begin{equation}
\begin{split}
    \left[\lim_{t^{\frac{1}{2}}\rightarrow 1} \cZ_{\text{tw}}^A\right] (\xi) & = \sum_{\alpha} \mathcal{X}^{C}_{\alpha}(q, x,\xi) \mathcal{X}^{C}_{\alpha}(q^{-1}, x,\xi) \,, \\
    \left[\lim_{t^{\frac{1}{2}}\rightarrow 1} \cZ_{\text{tw}}^B \right] (x) & = \sum_{\alpha} \mathcal{X}^{H}_{\alpha}(q, x,\xi) \mathcal{X}^{H}_{\alpha}(q^{-1}, x,\xi). \\
\end{split}        
\end{equation}
This gives clean formulae for the Hilbert series of 3d $\mathcal{N}=4$ chiral rings in terms of Verma modules constructed out of boundary operators, with a different gluing to the corresponding expressions for the limits of the superconformal index \eqref{eq:SCindexvermacharacterlimit}. Again, the gluing is such that, for example in the $A$-twist, the $x$ and $q$ dependence (which is solely in the classical piece of the $\cA_C$ Verma characters) cancels.

\subsubsection{Example: Hypermultiplet}

We briefly consider the twisted indices of a free hypermultiplet. In the absence of background flux for the flavour symmetry, the $B$-twisted index is
\bea 
\cZ^{B}_{\text{tw}} = \frac{t^{\frac{1}{2}} }{\left(1-xt^{\frac{1}{2}}\right) \left(1-x^{-1}t^{\frac{1}{2}}\right)} 
= -\left\lVert \cZ_{\cB_X}^B \right\rVert_{\text{tw}}^2
\eea 
where
\be
\cZ_{\cB_X}^B = e^{\frac{1}{\log q}  \log x\log(q^{\frac12}t^{-\frac12}) }\frac{(q t^{-\frac12} x;q)_\infty}{(t^{\frac12} x;q)_\infty}\,,
\ee
using the same analytic continuation of the q-Pochhammer as for the superconformal index. 

The $t^{\frac{1}{2}}\to1$ limit preserving additional supersymmetry is
\bea 
\lim_{t^{\frac{1}{2}}\rightarrow 1} \cZ^{B}_{\text{tw}} =  - \mathcal{X}_{\cB_X}^H(x)  \mathcal{X}_{\cB_X}^H(x)  = -\frac{x}{(1-x)^2} \,,
\eea 
which coincides with the equivariant Hilbert series of the free hypermultiplet.  The $A$-twisted index is $1$ and is reproduced by the factorisation $\| Z^{A}_{\cB_X}\|^2_{\text{tw}} = 1$. The $t^{\frac{1}{2}}\to1$ limit is therefore trivial and compatible with the absence of a Coulomb branch.

\subsubsection{Example: SQED}

We now demonstrate this factorisation explicitly in supersymmetric QED in the absence of background fluxes for global symmetries.\footnote{Non-trivial background fluxes for global symmetries can be incorporated easily and factorisation is in terms of the same hemisphere partition functions but with $q$-shifts of the fugacities by the appropriate fluxes \cite{Crew:2020jyf}.} The twisted indices can be expressed as the following contour integrals
\bea
\cZ_{\text{tw}}^A
&= -\frac{1}{(t^{\frac{1}{2}}-t^{-\frac{1}{2}})}\sum_{m\in \mathbb{Z}} ((-1)^N\xi)^m\int_{\Gamma_A} \frac{dz}{2\pi i z} z^{Nm} \prod_{j=1}^N \frac{(q^{\frac{1-m}{2}}t^{\frac{1}{2}}z^{-1}x_j^{-1},q)_m}{(q^{\frac{1-m}{2}}t^{\frac{1}{2}} z x_j ,q)_m} \\
\cZ_{\text{tw}}^B
& = -(t^{\frac{1}{2}}-t^{-\frac{1}{2}})\sum_{m\in \mathbb{Z}} ((-1)^N\xi)^m\int_{\Gamma_B} \frac{dz}{2\pi i z} z^{Nm} t^N\prod_{j=1}^N \frac{(q^{1-\frac{m}{2}}t^{\frac{1}{2}}z^{-1}x_j^{-1},q)_{m-1}}{(q^{-\frac{m}{2}}t^{\frac{1}{2}}zx_j,q)_{m+1}}
\eea
where the contour $\Gamma_A$ surrounds the poles at $z = x_j^{-1}t^{-\frac{1}{2}}q^{\frac{1-m}{2}+k}$ for $k = 0 ,\ldots, m-1,\, j=1,\ldots,N$ and $\Gamma_B$ surrounds the poles at $z = x_j^{-1} t^{-\frac{1}{2}}q^{-\frac{m}{2}+k}$ for $k = 0 ,\ldots, m,\, j=1,\ldots,N$. 

We now demonstrate the factorisation of these twisted indices according to~\eqref{eq:proposedtwistedfactorisation}. The $B$-twisted index factorises naturally when evaluated on the aforementioned poles \cite{Crew:2020jyf}  as:
\bea\label{eq:SQED-Btwistedindex}
    \cZ_{\text{tw}}^B = &(-1)^N \sum_{i=1}^{N} t^{\frac{1}{2}(N-1)}  x_i^{N} \left\lVert \left[ \prod_{j\neq i}^{N} \frac{\left(q \frac{x_i }{x_j };q\right)_{\infty}}{\left(t\frac{x_i }{x_j };q\right)_{\infty}}\right]
    \sum_{m \geq 0 }\left({\left(\frac{q}{t}\right)}^{N/2} \xi \right)^{m} \left[\prod_{j=1}^N\frac{\left(t \frac{x_i }{x_j };q\right)_{m}}{\left(q\frac{x_i }{x_j };q\right)_{m}}  \right] \right\rVert_{\text{tw}}^2\\
    =& (-1)^N\sum_{i=1}^{N} t^{\frac{1}{2}(2i-N-1)} 
    \left(\prod_{j\leq i } \frac{x_i}{x_j}\right) 
    \left(\prod_{j >i } \frac{x_j}{x_i}\right) 
    \left\lVert \cZ^{B, \text{1-loop}}_i  \cZ^{B, \text{Vortex}}_i(q,t,x_i ,\xi) \right\rVert_{\text{tw}}^2 \\
    =&(-1)^N \left\lVert \cZ_{\cB_i}^B(q,t,x ,\xi) \right \rVert_{\text{tw}}^2.
\eea 
where for instance $ \mathcal{Z}^{\text{1-loop}}_i (q,t,x,\xi) = \mathcal{Z}_{i}^{B, \text{1-loop}}(q,tq^{\frac{1}{2}},x,\zeta)$. The twisted index as written in the first equality in (\ref{eq:SQED-Btwistedindex}) is partially factorised in terms of vortex partition functions or holomorphic blocks. In passing from the first to the second line, the $1$-loop piece has been re-organised as in \eqref{eq:SCindex-modified1loop} (but with the shift of $t$) and we have used the following identity 
\bea\label{eq:qthetafusionrule}
\theta_{q}(aq^{m/2};q)\theta_{q}(aq^{-m/2},q^{-1}) &= (-1)^{m-1}a^{1-m}\\
\eea
to fuse the theta functions. In passing to the last line, the remaining monomial is identified with the $\left\lVert \mathcal{Z}^{B, \text{Cl}}_i	 \right\rVert_{\text{tw}}^2$. We thus produce a full factorisation (up to an overall sign) in terms of hemisphere partition functions $\cB_i$ associated to vacua in a fixed chamber for the mass parameters.

Similarly, the $A$-twisted index can be fully factorised:
\bea 
    \cZ_{\text{tw}}^A &=  \sum_{i=1}^{N} t^{\frac{1}{2}(-N+1)} \xi 
    \left[\prod_{j\neq i}^{N} \frac{\left(q \frac{x_i }{x_j };q\right)_{-1}}{\left(tq\frac{x_i }{x_j };q\right)_{-1}}\right]
    \left\lVert \sum_{m \geq 0 }\left(t^{-N/2} \xi \right)^{m} \left[\prod_{j=1}^N\frac{\left(tq \frac{x_i }{x_j };q\right)_{m}}{\left(q\frac{x_i }{x_j };q\right)_{m}}  \right] \right\rVert_{\text{tw}}^2\\
    &= \left\lVert \cZ_{\cB_i}^A (q, t, x, \xi )\right \rVert_{\text{tw}}^2.
\eea

In the $t^{\frac12} \to 1$ limit with enhanced supersymmetry, the twisted indices become
\bea
    \lim_{t^{\frac12} \rightarrow 1} \cZ_{\text{tw}}^B
    &=  (-1)^N\sum_{i=1}^N\mathcal{X}^{H}_i(q, x, \xi) \mathcal{X}^{H}_i (q^{-1}, x, \xi) \\
    &= (-1)^N\sum_{i=1}^N \, \prod_{j\neq i} \frac{x_j/x_i}{(1-x_j/x_i)^2} \\
    \lim_{t^{\frac12} \rightarrow 1} \cZ_{\text{tw}}^A &=\sum_{i=1}^N\mathcal{X}^{C}_i (q, x, \xi)  \mathcal{X}^{C}_i (q^{-1}, x, \xi) \,,\\
    &= \frac{N\xi}{(1-\xi)^2}\,,
\eea
in agreement with the equivariant Hilbert series and results \eqref{eqSCindexlimitSQED} for the $A$-limit and $B$-limit of the superconformal index.

\subsection{$S_b^3$ Partition Function}
\label{sec:s3pf}

The final case we consider is the partition function on the squashed sphere or ellipsoid $S_b^3$ \cite{Hama:2011ea}.
In reference~\cite{Pasquetti:2011fj} it was shown that the supersymmetric localisation computation of such partition functions can be factorised into holomorphic blocks. We propose that the sphere partition functions of 3d $\cN=4$ theories, deformed by an axial mass $T$, admits the following factorisation into hemisphere partition functions
\bea
    \mathcal{Z}_{S_b^3} = \sum_{\alpha}  \cZ_{\cB_\alpha} (q, t, x, \xi) \cZ_{\cB_\alpha} (\bar{q}, \bar{t}, \bar{x}, \bar{\xi}) 
    \label{eq:S3factorisation}
\eea
up to an overall phase. The parameters are identified by
\bea
     q&=e^{-2\pi i b Q}\,,
     &\qquad  t&=e^{2\pi b T}  \,,
     &\qquad   x&= e^{-2\pi b m} \,,
     &\qquad \xi &= e^{-2\pi b \eta} \,,   \\
     \bar{q}&=e^{-\frac{2\pi i}{b}Q} \,,
     &\qquad \bar{t}&=e^{\frac{2\pi T}{b}} \,,
     &\qquad \bar{x} &= e^{-\frac{2\pi m}{b}} \,,
     &\qquad  \bar{\xi} &= e^{-\frac{2\pi \eta}{b}}
\eea
where $Q\coloneqq b+\frac{1}{b}$.  Note we have used $\eta$ in this section for the FI parameter in order to avoid confusion. In writing expressions from $S^3_b$ in terms of exponentiated parameters, rational powers are defined such that $q^{r}\coloneqq  e^{-2\pi i r b Q }$ where $r \in \mathbb{Q}$.

We now consider the limit of the axial mass
\be
T \rightarrow \pm \frac{i}{2}\left(b-1/b\right)
\ee
preserving additional supersymmetry~\cite{Razamat:2014pta}. The resulting partition function depends on $b$ in a trivial way: this parameter can either be absorbed into masses and FI parameters or sent to $1$ to recover the matrix model for the round $S^3$ partition function as studied in~\cite{Benvenuti:2011ga,Kapustin:2009kz}. It has been proposed that the $S^3$ partition function in this limit can be expressed as a sum over massive vacua $\alpha$ of products of twisted characters of Verma modules of $\cA_H$ and $\cA_C$ \cite{Gaiotto:2019mmf}. Explaining this proposal was one of the original motivations for the present work. 

Beginning from the general factorised form of the $S_b^3$ partition function \eqref{eq:S3factorisation}, we note first that the limit $T \rightarrow \frac{i}{2}\left(b-1/b\right)$ sends
\bea 
t^{\frac{1}{2}} \rightarrow  e^{-\pi i }q^{-\frac{1}{4}}\,,\qquad \bar{t}^{\frac{1}{2}} \rightarrow  e^{ \pi i }\bar{q}^{\frac{1}{4}}\,.
\eea 
It is important to keep track of the minus sign in the exponentials, since this corresponds to a choice of branch in the logarithms appearing in the anomaly contribution to the hemisphere partition function. Taking the limit we find the following expression,
\bea\label{eq:S3vermafusiongeneral}
\lim_{T\rightarrow \frac{i}{2} (b-1/b)} \mathcal{Z}_{S_b^3} = 
\sum_{\alpha}  \hat{\mathcal{X}}^{H}_\alpha(q,x,\xi) \hat{\mathcal{X}}^{C}_\alpha(\bar{q},\bar{x} ,\bar\xi)\,.
\eea
Here the twisted characters are defined by
\bea
\hat{\mathcal{X}}^{H}_\alpha (q,x,\xi)  & \coloneqq \lim_{t^{\frac{1}{2}} \rightarrow  e^{-\pi i }q^{-\frac{1}{4}}} \cZ_{\cB_{\alpha}}(q,t,x,\xi) = e^{\hat{\phi}_{\cB_{\alpha}}^{(B)}}\mathrm{Tr}_{\cH^{(B)}_{\cB_{\alpha}}} (-1)^{R_V}  x^{F_H}\,, \\
\hat{\mathcal{X}}^{C}_\alpha(q,x , \xi) & \coloneqq \lim_{ t^{\frac{1}{2}} \rightarrow  e^{ \pi i } q^{\frac{1}{4}}} \cZ_{\cB_{\alpha}}(q, t, x, \xi) =  e^{\hat{\phi}_{\cB_{\alpha}}^{(A)}} \mathrm{Tr}_{\cH^{(A)}_{\cB_{\alpha}}} (-1)^{R_A} \xi^{F_C}\,.
\eea
These differ from the characters used previously by a $\mathbb{Z}_2$ twist by the centre of the Higgs or Coulomb branch R-symmetry. This is implemented in the trace by the additional factors of $(-1)^{R_V}$ and $(-1)^{R_A}$ respectively and in the classical or boundary anomaly contributions, which become
\bea\label{eq:twistedcharprefactor}
e^{\hat{\phi}_{\cB_{\alpha}}^{(B)}} &= x^{\frac{k_A}{2} +  k \,\cdot \frac{\log \xi }{ \log q} } \, e^{-\frac{\pi i k_V \cdot \log \xi  }{\log q}}\, e^{ \frac{\pi i k_A \cdot \log x  }{\log q}}\, e^{-\frac{\pi i \tilde{k}}{2}} e^{\frac{\pi^2 \tilde{k}}{\log q}} \,,\\
e^{\hat{\phi}_{\cB_{\alpha}}^{(A)}} &= \xi^{\frac{k_V}{2} +  k \,\cdot \frac{\log x }{ \log q} } \, e^{\frac{\pi i k_V \cdot \log \xi  }{\log q}}\, e^{ -\frac{\pi i k_A \cdot \log x  }{\log q}}\, e^{-\frac{\pi i \tilde{k}}{2}} e^{\frac{\pi^2 \tilde{k}}{\log q}} \,.\\
\eea 
The anomaly coefficients are those of the boundary condition $\cB_{\alpha}$, but we omit the index $\alpha$ to avoid clutter. Note the presence of additional phases compared to $e^{\phi_{\cB_{\alpha}}^{(B)}}$, $e^{\phi_{\cB_{\alpha}}^{(A)}}$. 

To recover the proposal of \cite{Gaiotto:2019mmf}, we write equation~\eqref{eq:S3vermafusiongeneral} in terms of sphere parameters as follows,
\bea 
\lim_{T\rightarrow \frac{i}{2} (b-1/b)} \mathcal{Z}_{S_b^3} = 
\sum_{\alpha}  &\left(e^{-\frac{\pi i}{2}}\right)^{\tilde{k}} e^{-\pi b\, k_A \cdot m}\, e^{-\frac{\pi k_V \cdot \eta}{b} }  e^{2\pi i m \cdot k \cdot \log \eta}\\
&\,\,\mathrm{Tr}_{\cH^{(B)}_{\cB_{\alpha}}} \left[(-1)^{R_V}  \left(e^{-2\pi b m }\right)^{F_H} \right]\, \mathrm{Tr}_{\cH^{(A)}_{\cB_{\alpha}}} \left[ (-1)^{R_A} \left(e^{-\frac{2\pi \eta}{b} }\right)^{F_C}\right]
\eea 
after conjugating fugacities and gluing. 

Finally, we note that the alternative limit $T \rightarrow  -\frac{i}{2}\left(b-1/b\right)$ is obtained simply by exchanging $b\leftrightarrow 1/b$ and thus barred and unbarred fugacities.

\subsubsection{Example: Hypermultiplet}

For a free hypermultiplet the partition function may be factorised using well-known double sine function identities
\bea
	\cZ_{S_b^3} = \frac{s_b\left(-m+T/2+iQ/4\right)}{s_b\left(-m-T/2-iQ/4\right)} = \left\lVert \cZ_{\cB_X}\right\rVert_{S_b^3}^2\,.
\eea
Then in the twisted trace limit we have
\bea\label{eq:S3freehyper}
\lim_{T\rightarrow \frac{i}{2} (b-1/b)} \cZ_{S_b^3}= \hat{\mathcal{X}}^H_{\cB_X}(q,x) \hat{\mathcal{X}}^C_{\cB_X}(\bar{q},\bar{x})
= \frac{1}{2\text{cosh}\pi b x}  \,,
\eea
where
\bea
\hat{\mathcal{X}}^H_{\cB_X}(q,x)  = e^{\frac{\pi i  \log x }{\log q}}  \frac{x^{\frac{1}{2}}}{1+x}\,,\qquad  \hat{\mathcal{X}}^C_{\cB_X}(\bar{q},\bar{x})  = e^{-\frac{\pi i \log \bar{x}}{\log \bar{q}}}\,.
\eea
Note that the plus sign in the denominator of the twisted Higgs branch character arises from the additional weight $(-1)^{R_V}$ and the fact that the raising operator is the scalar field $\hat{X}$ with $R_V = 1$. The Coulomb branch Verma module is trivial and the Coulomb branch twisted character simply counts the identity operator, whose contribution is a phase due to the $\bZ_2$ twist by the centre of $R_A$.

\subsubsection{Example: SQED}

A partially factorised form of the $S_b^3$ partition function of supersymmetric QED is found by a computation in~\cite{Pasquetti:2011fj},
\bea
    \mathcal{Z}_{S_b^3} = 
    & \,\frac{1}{s_b(T)}\oint dz e^{2\pi i \eta z } \prod_{i=1}^N \frac{s_b(z-m_i+T/2+iQ/4)}{s_b(z-m_i-T/2-iQ/4)}\\
    = &\sum_{i=1}^N 
    e^{-2\pi i \left(\eta + \frac{N}{2} (T+iQ/2)\right)\left(-m_i+\frac{1}{2}(T-iQ/2\right)+\frac{\pi i}{2}T^2}\\
    &\left \lVert \prod_{j\neq i}\frac{\left(q\frac{x_i}{x_j};q\right)_{\infty}}{\left(tq^{1/2} \frac{x_i}{x_j};q\right)_{\infty}} 
    \sum_{m\geq 0} \left(\left(q^{\frac{1}{4}}t^{-\frac{1}{2}}\right)^{N}\xi\right)^m \prod_{j=1}^N \frac{\left(tq^{\frac{1}{2}}\frac{x_i}{x_j};q\right)_m}{\left(q\frac{x_i}{x_j};q\right)_m} \right \rVert_{S_b^3}^2\,,
\eea
where the contour surrounds simple poles of the numerator at
\bea
    z= m_i- T/2+ i Q/4 + i m b + i n/b,\qquad m, n\geq 0, \qquad i=1,\ldots,N\,,
\eea 
and in the second line
\be
     x_j= e^{-2\pi b m_j}\,, \qquad \bar{x}_j = e^{-\frac{2\pi m_j}{b}}\,.
\ee
Notice this factorisation corresponds to hemisphere partition functions for boundary conditions compatible with mass parameters in different chambers. 

In order to bring this expression into the factorised form of~\eqref{eq:S3factorisation}, it is necessary to re-arrange the classical and 1-loop contributions. For the hemisphere partition function $\cZ_{\cB_i}$ given in section \ref{subsec:SQEDHPS}, we rewrite the $1$-loop piece as in \eqref{eq:SCindex-modified1loop} and use the identity: 
\bea
    &\theta\left(e^{2\pi b m} q^{\frac{n}{2}}; q\right)
    \theta\left(e^{\frac{2\pi m}{b}} \bar{q}^{\frac{n}{2}}; \bar{q} \right) =  e^{-\frac{\pi i}{12}\left(b^2 + \frac{1}{b^2}\right)}e^{-i \pi \left(m + \frac{i}{2}(1-n)Q\right)^2}
\eea
(see e.g. \cite{Narukawa2003}) to glue the theta functions in \eqref{eq:SCindex-modified1loop}. Then identifying $\sum m_j=0$, we fuse under $b \leftrightarrow \frac{1}{b}$ to obtain an exact factorisation
\bea \label{eq:ellipsoidgluing1}
    \mathcal{Z}_{S_b^3} = \,\, e^{-\frac{\pi i}{8}Q^2}  \sum_{i=1}^{N} 
   \left\lVert \cZ_{\cB_i}(q,t, x, \xi) \right\rVert_{S_b^3}^2\,,
\eea
up to an overall phase which we now drop.

We now take the twisted character limit to find:
\bea
	\lim_{T\rightarrow \frac{i}{2} (b-1/b)} \mathcal{Z}_{S_b^3}  = \sum_{i=1}^N \hat{\mathcal{X}}^{H}_i(q,x,\xi) \hat{\mathcal{X}}^{C}_i (\bar{q}, \bar{x}, \bar{\xi}) \,,
\eea 
where the twisted Verma characters are
\bea\label{twistedcharactersSQED}
	\hat{\mathcal{X}}^{H} (q,x,\xi)  &= e^{\hat{\phi}_i^{(B)}} \prod_{j<i}\frac{1}{1-\frac{x_i}{x_j}} \prod_{j>i}\frac{1}{1-\frac{x_j}{x_i}} \,,\\
	\hat{\mathcal{X}}^{C} (q,x,\xi)  &=  e^{\hat{\phi}_i^{(A)}} \frac{1}{1-(-)^N \xi}\,,
\eea 
and the prefactors $e^{\hat{\phi}_i^{(B)}}$ and $e^{\hat{\phi}_i^{(A)}}$ are given by \eqref{eq:twistedcharprefactor} and the anomaly coefficients \eqref{eq:anomalypolySQED}. The $(-1)^N$ in the denominator of the twisted character of the Coulomb branch Verma module is because the raising operator is the monopole $\hat{v}_{-}$ with $R_A = N$.  The raising operators in the Higgs branch Verma module are gauge invariant combinations (\ref{eq:raisingoperatorsSQEDhiggs}) with $R_V = 2$, so there is no additional sign in the twisted character. Also note that the result for $N=1$ is consistent with a hypermultiplet~\eqref{eq:S3freehyper} under mirror symmetry.

The result of \cite{Gaiotto:2019mmf} is recovered explicitly by gluing the pre-factors in \eqref{twistedcharactersSQED}.  Writing everything in terms of the sphere parameters we have
\bea 
	\lim_{T\rightarrow \frac{i}{2} (b-1/b)} \mathcal{Z}_{S_b^3}  = \sum_{i=1}^N & \left(e^{-\frac{\pi i}{2}}\right)^{2i-N-1}
	e^{2\pi i m_i \eta} \frac{ e^{-\frac{\pi \eta}{b}} }{1-(-1)^N e^{-\frac{2\pi  \eta}{b}}} \\
	&\prod_{j<i} \frac{e^{-\pi b(m_i-m_j)}}{1-e^{-2\pi b (m_i-m_j)}} \prod_{j>i} \frac{e^{-\pi b(m_j-m_i)}}{1-e^{-2\pi b (m_j-m_i)}} \,. 
\eea

\acknowledgments

The authors would like to thank Nick Dorey and Tadashi Okazaki for collaboration at an early stage of the project. It is also a pleasure to thank Davide Gaiotto, Masazumi Honda, Nakarin Lohitsiri and David Tong for useful discussions. We especially thank Tudor Dimofte for helpful comments on a draft of the paper. The work of MB is supported by the EPSRC Early Career Fellowship EP/T004746/1 ``Supersymmetric Gauge Theory and Enumerative Geometry" and the STFC Research Grant ST/T000708/1 ``Particles, Fields and Spacetime". The work of SC and DZ is partially supported by STFC consolidated grants ST/P000681/1, ST/T000694/1.

\appendix

\section{Hemisphere Partition Functions}\label{app:localisation}

In this appendix we discuss the formulation of 3d $\mathcal{N}=2$ theories on $S^1 \times H^2$, where $H^2$ is a hemisphere with a $U(1)$ isometry, and the computation of their partition functions. We impose 2d $\cN = (0,2)$ boundary conditions on $S^1 \times \partial H^2 \simeq S^1 \times S^1 = T^2$. We show this coincides with the half superconformal index up to the Casimir energy, which is precisely the equivariant integral of the boundary 't Hooft anomaly. The 3d $\cN=4$ cases of interest with $\cN=(2,2)$ boundary conditions can then be obtained as a specialisation.

The case where the $\mathcal{N}=2$ vector multiplet is assigned a Neumann boundary condition on $T^2$ was analysed in \cite{Yoshida:2014ssa}. We propose an extension to cover the Dirichlet boundary condition for the vector multiplet, and find the partition function is expressed as a sum over fluxes corresponding to boundary monopole configurations for the vector multiplet. We stick mostly to abelian theories for simplicity, and will return to give a fuller picture of the non-abelian case and localisation in future work.

\subsection{Supersymmetry and the Index}\label{subsec:HPSdefinitions}

Rigid supersymmetry on $S^1\times S^2$ was considered in \cite{Kim:2009wb,Imamura:2011su} for the purposes of computing the superconformal index via Coulomb branch localisation. The computation of the superconformal index via Higgs branch localisation was performed in \cite{Benini:2013yva,Fujitsuka:2013fga}. As the metric on $H^2$ is identical to the one on $S^2$, the same conformal Killing spinors can be used. In this appendix, we follow the conventions of \cite{Yoshida:2014ssa}.  The metric on the $S^1\times H^2$ with radius $r$ is:
\bea\label{eq:hemispheremetric}
    ds^2 = r^2d\theta^2 + r^2\sin^2 \theta d\phi^2 + d\tau^2,
\eea
 where $0\leq \theta \leq \pi/2$, $\phi \sim \phi+ 2\pi$, $\tau \sim \tau + \beta r$. We also use subscripts $\mu \in \{1,2,3\}$ for coordinates  $\{\theta,\phi,\tau\}$, and $\{\hat{1},\hat{2},\hat{3}\}$ for components in a frame specified by the dreibein 
\bea 
    e^{\hat{1}} = rd\theta \,,\quad e^{\hat{2}} = r \sin\theta d\varphi \,,\quad e^{\hat{3}} = d\tau.
\eea 
The Killing spinor equations are
\bea
    \nabla_{\mu}\epsilon = \frac{1}{2r} \gamma_{\mu}\gamma_{3}\epsilon \,,\quad  \nabla_{\mu}\epsilon = -\frac{1}{2r} \gamma_{\mu}\gamma_{3}\bar{\epsilon},
\eea
and we choose solutions
\bea\label{eq:Killingspinors}
    \epsilon_{\alpha} = e^{\tau/2r} e^{i\varphi/2} \begin{pmatrix} \cos\theta/2\\ \sin\theta/2
    \end{pmatrix}\,,
    \quad  \bar{\epsilon}_{\alpha} = e^{-\tau/2r} e^{-i\varphi/2} \begin{pmatrix} \sin\theta/2\\ \cos\theta/2
    \end{pmatrix}.
\eea
The supersymmetry transformations of $\cN=2$ vector and chiral multiplets are given in section 2 of \cite{Yoshida:2014ssa}. The transformations generated by $\epsilon, \bar{\epsilon}$ on the boundary $T^2$ generate a $\cN=(0,2)$ supersymmetry under which boundary conditions for the bulk multiplets must be compatible.  We define these in the next subsection. The spinors are not periodic around $S^1$, and thus twisted boundary conditions must be imposed. This is precisely compatible with the hemisphere partition function, which is a path integral over fields on $S^1\times H^2$ with the twisted periodicities dictated by the fugacities:
\bea\label{eq:twistedperiodicity}
    \Phi(\tau+\beta r) =  e^{-\beta_1 R} e^{-(\beta_1-\beta_2)J_3}x^{-F_H}\xi^{-F_C} \Phi(\tau).
\eea
Note the Killing spinors (\ref{eq:Killingspinors}) obey these conditions, e.g. $\bar{\epsilon}$ has R-charge $+1$ and $J_3 = -1/2$. From standard arguments, the path integral gives a trace over states on $H^2$:
\bea\label{eq:hemisphere-trace}
    \mathcal{Z}_{S^1 \times H^2} &= \text{Tr}_{\mathcal{H}(H^2)} \left[ (-)^{F} e^{-\beta_1(D-R-J_3)} e^{-{\beta_2}(D+J_3)} x^{F_H}\xi^{F_C} \right]\\
    &= \text{Tr}_{\mathcal{H}(H^2)} \left[ (-)^{F}q^{J_3+R/2} x^{F_H}\xi^{F_C} \right].
\eea
which is independent of $\beta_1$. Here $\beta=\beta_1+\beta_2$, $F=2J_3$ is the fermion number, $q=e^{-2\beta_2}$ and $F_{H,C}$ the generators of matter/ topological flavour symmetry with fugacities $x,\xi$ respectively. The 1-loop determinants can be computed using these periodicities. Alternatively, noting that (\ref{eq:twistedperiodicity}) contains a gauge transformation for flavour and R-symmetries, one can equivalently turn on background flat connections for these symmetries \cite{Benini:2013yva}. In either case the twisted periodicity condition corresponding to the angular momentum $J_3$ is implemented by the coordinate identification (eliminating $\beta_1$)
\bea 
    (\tau,\varphi) \sim (\tau+\beta r, \varphi -  i (\beta - 2\beta_2)).
\eea
To evaluate the classical action properly one needs to take this identification into account. Redefining
\bea\label{eq:localisationshiftedcoordinates}
   \tilde{\tau} = \tau\,,\quad \tilde{\varphi}= \varphi + \frac{i (\beta - 2\beta_2)}{\beta r } \tau \qquad \Rightarrow \qquad (\tilde{\tau}, \tilde{\varphi}) \sim (\tilde{\tau}+\beta r, \tilde{\varphi})\,,
\eea
the classical actions can be evaluated by integrating separately over $\tilde{\tau},\tilde{\varphi}$. To recover the Casimir energy corresponding to the boundary 't Hooft anomaly, necessary for exact holomorphic factorisation, we will see we should set $\beta_2=\beta$ and the fugacity $q=e^{-2\beta}$ and will do so from here on out.

\subsection{Boundary Conditions}\label{app:boundaryconditions}

We specify a set of $\cN =(0,2)$ boundary conditions on $T^2$ for 3d $\cN=2$ multiplets. These differ from \cite{Yoshida:2014ssa} in that they involve a Dirichlet boundary condition for the vector multiplet. We restrict to an abelian gauge group $G = U(1)^k$ with Lie algebra $\mathfrak{g}$, for simplicity. We define the complexified covariant derivative $\mathcal{D} = D+\sigma = \nabla + iA +\sigma$ where $A$ and $\sigma$ act in the appropriate representation. 
\begin{itemize}
    \item For the $\mathcal{N}=2$ vector multiplet $(A_{\mu},\sigma, \lambda,\bar{\lambda},D)$, the Dirichlet boundary condition at $\theta = \pi/2$ is:
\begin{equation}\label{eq:hemisphere-boundaryconditions}
\begin{split}
    &A_{2,3} = a_{2,3},\\
    & \partial_{\hat{1}}\left( i F_{\hat{1}\hat{2}} +F_{\hat{1}\hat{3}}\right)=0,\\
    &D = 0,\\
    &D_{\hat{1}}\sigma = 0,
\end{split}
\qquad
\begin{split}
    &\lambda_1 + \lambda_2 = 0,\\
    &\bar{\lambda}_1 + \bar{\lambda}_2 = 0,\\
    &\partial_{1}(\lambda_1 - \lambda_2)= 0, \\
    &\partial_{1}(\bar{\lambda}_1 - \bar{\lambda}_2)= 0.
\end{split}
\end{equation}
Here $a_{2,3}$ is a constant flat connection on the boundary torus $T^2$. This breaks the gauge symmetry to a global symmetry $G_{\partial}$ at the boundary.

\item For $\mathcal{N}=2$ chiral multiplets $(\phi,
\bar{\phi}, \psi,\bar{\psi}, F,\bar{F})$ of R-charge $\Delta$, the Neumann boundary condition is:
\begin{equation}\label{eq:hemisphere-boundaryconditionsneumannchiral}
    \begin{split}
    (N):\quad
    \end{split}
    \begin{split}
    &\mathcal{D}_{\hat{1}}\phi = 0,\\
    &\mathcal{D}_{\hat{1}}\bar{\phi} = 0,\\
    &\psi_{1}+\psi_2 = 0,\\  
    &\bar{\psi}_{1}+\bar{\psi}_2 = 0,\\  
    \end{split}
    \qquad
    \begin{split}
    &F=0,\\
    &\bar{F}=0,\\
    &\mathcal{D}_{\hat{1}}(\psi_{1}-\psi_2) +(\lambda_1-\lambda_2)\cdot \phi= 0, \\
    &\mathcal{D}_{\hat{1}}(\bar{\psi}_{1}-\bar{\psi}_2) - \bar{\phi}\cdot (\bar{\lambda}_1-\bar{\lambda}_2)= 0. 
    \end{split}
\end{equation}
The basic Dirichlet boundary condition is:
\begin{equation}\label{eq:hemisphere-boundaryconditionsdirichletchiral}
    (D):\begin{split}
    &\phi = 0,\\ 
    &\bar{\phi} = 0,\\ 
    &\psi_{1}-\psi_2 = 0,\\
    &\bar{\psi}_{1}-\bar{\psi}_2 = 0,
    \end{split}
    \qquad
    \begin{split}
    &\mathcal{D}_{\hat{1}}\left(ie^{\frac{\tau}{r}}e^{i\varphi}\mathcal{D}_{\hat{1}}\phi + F\right)+\frac{1}{2}(\lambda_1-\lambda_2)\cdot (\psi_1+\psi_2)=0,\\
    &\mathcal{D}_{\hat{1}}\left(ie^{-\frac{\tau}{r}}e^{-i\varphi}\mathcal{D}_{\hat{1}}\bar{\phi} + \bar{F}\right)+\frac{1}{2} (\bar{\psi}_1+\bar{\psi}_2)\cdot(\bar{\lambda}_1-\bar{\lambda}_2)= 0,\\
    &\mathcal{D}_{\hat{1}}(\psi_1+\psi_2) = 0,\\
    &\mathcal{D}_{\hat{1}}(\bar{\psi}_1+\bar{\psi}_2) = 0.
    \end{split}
\end{equation}
\end{itemize}
These are related to 3d lifts of the boundary conditions in \cite{Herbst:2008jq,Hori:2013ika,Honda:2013uca}. For purposes of application to 3d $\mathcal{N}=4$ theories with boundary conditions associated to vacua, we would like to turn on non-zero values at the boundary for the scalars $\phi$ in chiral multiplets which acquire non-zero VEVs in the vacuum, analogously to the operator picture on $\mathbb{R}^2 \times \mathbb{R}_{\geq 0}$. Thus we would like to deform the basic Dirichlet boundary condition for such scalars to
\bea\label{eq:hemisphere-exceptionalDirichlet}
    (D_c): \quad \phi = c\,, \quad \bar{\phi} = \bar{c}\,, \quad c\neq0\,,
\eea
keeping the same boundary conditions for the remaining fields in the $\mathcal{N}=2$ chiral. To preserve supersymmetry in the right column of (\ref{eq:hemisphere-boundaryconditionsdirichletchiral})  we demand
\bea
    \rho(A_{\hat{2}})=0\,,\quad \rho(A_{\hat{3}})-\frac{i\Delta}{r} = 0,
\eea
 at $\theta=\pi/2$, where $\rho$ is the gauge group representation of the chiral. If we choose to realise the twisted boundary conditions for flavour symmetries around $S^1$ as holonomies for background vector multiplets, then the condition becomes:
\begin{equation}\label{eq:exceptionaldirichletconstraint}
    \rho(A_{\hat{3}})+\sum_{l}\rho_{l}(A^{l}_{\hat{3}})-\frac{i\Delta}{r} = 0,
\end{equation}
 where $\rho_l$ is the flavour representation. In computing the path integral, a hermiticity condition is imposed on the gauge fields. Thus a constant boundary value for $\phi$ can be turned on only for chirals of zero R-charge. To turn on a non-zero VEV, the background gauge fields must obey the constraint in (\ref{eq:exceptionaldirichletconstraint}), and thus the boundary condition breaks the combination of flavour symmetries (including $G_{\partial}$) dual to the charges of the chiral. The result for such chirals with arbitrary R-charges can be obtained by analytically continuing the final partition function by complexifying flavour fugacities \cite{Benini:2013yva}.

\subsection{Localisation}\label{sec:resultoflocalisation}

In this section we describe the localisation computation of the hemisphere partition function for general 3d $\mathcal{N}=2$ theories with Dirichlet boundary conditions for the vector multiplet.

\paragraph{BPS Locus} We may use the same localising actions as in \cite{Yoshida:2014ssa,Imamura:2011su} (the SYM and matter actions which are Q-exact), but restrict the saddle points to the ones compatible with the boundary condition. These saddle points coincide with the BPS locus. 
\begin{itemize}
    \item For the $\cN=2$ vector multiplets, the key feature of Dirichlet boundary conditions is that they are 
    compatible with slicing in half a Dirac monopole on $S^2$. The saddle points are
    \bea\label{eq:hemisphere-saddlepoints}
    A = a_3 d\tau + 2\mathfrak{m}B_\alpha dx^{\alpha}, \qquad \sigma = \mathfrak{m}/r,
    \eea
    where $\alpha = 1,2$, $\mathfrak{m} \in \text{Hom}(U(1),G) \simeq \mathbb{Z}^k$ and $a_3 \in \mathfrak{g}$ is a constant. Note the constant value of $A_\tau$ is fixed to its boundary value (\ref{eq:hemisphere-boundaryconditions}). Here $B_i$ is the monopole of unit flux on  $ S^2$
    \bea
        B = \frac{1}{2}\omega,
    \eea
    where $\omega$ is the spin connection on $S^2$. The factor of two difference between (\ref{eq:hemisphere-saddlepoints}) and (22) of \cite{Imamura:2011su} comes from the fact that for a $U(1)$ monopole on the hemisphere to have a well defined flux $\frac{1}{2\pi}\int_{H^2}F = \mathfrak{m} \in \mathbb{Z}$, it must have the functional form of a monopole of twice the magnetic charge on $S^2$. Explicitly we could write
    \bea\label{eq:hemisphere-monopole}
        A = a_3 d\tau + \mathfrak{m}(\kappa-\cos\theta) d\phi\,,\quad
        \kappa = \begin{cases}
        1 \qquad \text{for} & \,\, \theta \in [0,\pi/2-\epsilon)\,,\\
        0 \qquad \text{for} & \,\, \theta \in (\pi/2-2\epsilon,\pi/2]\,.
        \end{cases}
    \eea
    This is trivialised at the boundary and thus compatible with (\ref{eq:hemisphere-boundaryconditions}). Thus in the path integral we sum over monopole sectors $\mathfrak{m}$, mirroring the half index computation \cite{Dimofte:2017tpi}.
    
    \item For an $\cN=2$ chiral multiplet wit $\Delta\neq0$ the BPS locus sets all components of a chiral multiplet to 0. For $\Delta=0$, the scalar is set everywhere to the constant value it takes at the boundary (\ref{eq:hemisphere-exceptionalDirichlet}), see \cite{Benini:2013yva} for details.
\end{itemize}
As usual in localisation, we set:
\begin{equation}
        \Phi = \Phi^{(0)}+ \frac{\Phi'}{\sqrt{\delta}},
\end{equation}
where $\Phi^{(0)}$ are BPS configurations and $\Phi'$ fluctuations around the locus. Then:
\bea
    \mathcal{Z}_{S^1 \times H^2} & = \lim_{\delta \rightarrow \infty} \int \mathcal{D}\Phi e^{-S[\Phi]-\delta Q\cdot V[\Phi]} \\
    & = \prod_{\mathfrak{m} \in \mathbb{Z}^{k}}e^{-S_{\text{cl}}[\Phi^{(0)}]} Z_{\text{1-loop}}(q, z, x, \xi, \mathfrak{m}),
\eea
where $\Phi$ denotes the set of all fields, and $Q\cdot V[\Phi]$ are the localising actions given in \cite{Yoshida:2014ssa,Imamura:2011su}, and are just the 3d $\mathcal{N}=2$ Yang-Mills and matter actions. The path integral is over all configurations obeying boundary conditions in section \ref{subsec:HPSdefinitions}, and twisted periodicities defined by the trace (\ref{eq:twistedperiodicity}). Here $z = e^{-i\beta r a_3}$ is the fugacity for the gauge symmetry which is broken to a flavour symmetry by the boundary condition. $S_{\text{cl}}$ is the action evaluated on the BPS locus. We now describe each ingredient in turn.

\paragraph{Classical Contribution}  
To implement a grading by the topological symmetry, we turn on a BPS configuration for a background vector multiplet
\bea
    A^{(T)} = \eta d\tau\,,\quad \sigma^{(T)} = D^{(T)} = \lambda^{(T)} = \bar{\lambda}^{(T)} = 0,
\eea
in the mixed bulk-boundary Chern-Simons term for an abelian gauge group (for a non-abelian gauge group, the topological symmetry just couples to the centre)
    \begin{equation}\label{eq:mixedCSterm}
    \begin{split}
        S_{\text{mCS}} = & \frac{i}{4\pi} \int_{S^1 \times H^2} d^{3}x \Big[\epsilon^{\mu\nu\rho}\left(\partial_{\mu}A_{\nu}A_{\rho}^{(T)}+\partial_{\mu}A_{\nu}^{(T)}A_{\rho}\right) \\
        &\qquad\qquad\qquad\qquad+ 
        \sqrt{g} (-\bar{\lambda}^{(T)}\lambda -\bar{\lambda}\lambda^{(T)}  + 2 \sigma^{(T)}D + 2 \sigma D^{(T)})\Big]\\
        & -\frac{1}{4\pi} \int_{T^2}   d^2 x \sqrt{g^{(2)}}  \, \,\left[A_2A_2^{(T)} + A_3A_3^{(T)}-2 \sigma \sigma^{(T)}\right]\,.
    \end{split}
\end{equation}
The boundary terms involving $A$ and $\sigma$ are required for invariance under infinitesimal gauge/flavour transformations, and supersymmetry respectively. The evaluation of the term in the first line has the usual subtlety. Using coordinates $\tilde{\tau},\tilde{\varphi}$ in (\ref{eq:localisationshiftedcoordinates}), we write e.g.
\bea
    A = a_3 d\tilde{\tau}-\mathfrak{m}\cos \theta \left(d\tilde{\varphi}+\frac{i}{r}d\tilde{\tau}\right),
\eea
and extend to connections on a 4-manifold $D^2\times H^2$ where the $S^1$ factor is the boundary of a flat disk $D^2$ with $\rho \in [0,1]$:
\bea
    \hat{A} = a_3 \rho^2 d\tilde{\tau}-\mathfrak{m}\cos \theta \left(d\tilde{\varphi}+\frac{i}{r}\rho^2d\tilde{\tau}\right),\quad
    \hat{A}^{(T)} = \eta \rho^2 d\tilde{\tau}.
\eea
The action is defined to be the evaluation on the extension over $D^2\times H^2$
\bea
    \frac{i}{4\pi} \int_{S^1 \times H^2} A^{(T)}\wedge F + A \wedge F^{(T)} \quad\equiv\quad & \frac{i}{2\pi} \int_{D^2 \times H^2} \hat{F}^{(T)}\wedge \hat{F} = \mathfrak{m}(i\beta r \eta ).
\eea
Including the boundary contribution from (\ref{eq:mixedCSterm}):
\bea
    e^{S_{mCS}} |_{\text{BPS}} = e^{- \frac{\log\left( \xi\right)\log \left( z q^{\mathfrak{m}}\right)}{\log q}},
\eea
where we defined
\bea
z = e^{-i\beta r a_3 }\,,\qquad \xi = e^{-i\beta r \eta},
\eea
as the fugacities for the $G_{\partial}$ and topological symmetries which we use throughout. We have set $\beta_2=\beta$, so that when this is combined with the anomalous contributions of the vector and chiral multiplets the prefactor reproduces the anomaly polynomial, as we shall see in section \ref{app:regularisationanolypolynomials}. The contribution of a (diagonal) Chern-Simons term at level $k$ can be obtained by dropping the $(T)$ superscript and multiplying by $k/2$.

\paragraph{1-loop Determinants.} Here we give the 1-loop determinants, with the proof for the chiral multiplet with Dirichlet boundary conditions in the next subsection. The results are stated for a general gauge group. 
\begin{itemize}
    \item \textbf{The $\mathcal{N}=2$ chiral in Neumann $(N)$}.
    For an $\mathcal{N}=2$ chiral in representation $\rho$ of the gauge group, $\rho_f$ of the flavour group and $R$-charge $\Delta$
        \bea
           Z_{\text{1-loop}}^{(N)} =  e^{\mathcal{E}\left[-\log\left(q^{\frac{\Delta}{2}+\rho(\mathfrak{m})}z^{\rho}x^{\rho_f}\right)\right]} \left(q^{\frac{\Delta}{2}+\rho(\mathfrak{m})}z^{\rho}x^{\rho_f};q\right)^{-1}_{\infty}
        \eea
        where the function
        \begin{equation}
            \mathcal{E}\left[x\right] = \frac{\beta_2}{12} - \frac{x}{4}+ \frac{x^2}{8\beta_2} 
        \end{equation}
        arises from a zeta regularisation as in \cite{Yoshida:2014ssa}. The factor of two difference in the way the monopole charge enters compared to the $S^1 \times S^2$ index is due to monopoles on the $H^2$ having the same functional form as monopoles on $S^2$ with twice the flux. 
        
    \item \textbf{The $\mathcal{N}=2$ chiral in Dirichlet $(D)$}.
    Similarly to above we obtain
        \bea\label{eq:1-loopdetchiraldirichlet}
            Z_{\text{1-loop}}^{(D)} = e^{-\mathcal{E}\left[-\log\left(q^{1-\frac{\Delta}{2}-\rho(\mathfrak{m})}z^{-\rho}x^{-\rho_f}\right)\right]} \left(q^{1-\frac{\Delta}{2}-\rho(\mathfrak{m})}z^{-\rho}x^{-\rho_f};q\right)_{\infty}.
        \eea
    \item \textbf{The $\mathcal{N}=2$ vector multiplet in Dirichlet}.
    Note this is also the contribution of a Neumann chiral in the adjoint, with charge R-charge 2. 
        \bea 
            Z_{\text{1-loop}}^{\text{vector}}=\left[e^{\mathcal{E}\left[-\log(q)\right]}\left(q ;q\right)^{-1}_{\infty}\right]^{\text{rk}G}
            \prod_{\alpha} e^{\mathcal{E}\left[-\log(q^{1+\alpha(\mathfrak{m})}z^{\alpha})\right]}\left(q^{1+\alpha(\mathfrak{m})}z^{\alpha};q\right)^{-1}_{\infty}.
        \eea
\end{itemize}
To compute the partition function with some chirals with a deformed Dirichlet boundary condition,  the procedure can be described as computing with Dirichlet boundary conditions and then setting to 1 the product of fugacities dual to the charges of the chiral, as in (\ref{eq:exceptionaldirichletconstraint}). This is  analogous to the half-index computation for these boundary conditions \cite{Dimofte:2017tpi}.

\subsection{Details: Chiral Multiplet with Dirichlet B.C.}

In this section we derive the 1-loop determinant of the chiral multiplet with a basic Dirichlet boundary condition about the saddle points (\ref{eq:hemisphere-saddlepoints}). Contrary to \cite{Yoshida:2014ssa}, we do not expand in terms of monopole spherical harmonics as they do not form a complete eigenbasis on $H^2$ for the differential operators in the Gaussian integrals in the presence of a monopole - we do not require regularity at the `south pole'. Instead the determinant is derived by matching bosonic and fermionic eigenmodes, similarly to the 2d result in \cite{Sugishita:2013jca}. We abuse notation and also denote the fluctuating parts of the scalar and fermion as $(\phi,\psi)$. The differential operators appearing at quadratic order are, after substituting the BPS locus (\ref{eq:hemisphere-saddlepoints}):
\bea\label{eq:gaussianoperators}
    D_{\text{scalar}} &= \tilde{D}_{\text{scalar}} + \left[-D^3 D_3+\frac{\mathfrak{m}^2}{r^2}+\frac{1-2\Delta}{r} D_{3}+\frac{\Delta(1-\Delta)}{r^2}\right],\\
    D_{\text{fermion}} &= \tilde{D}_{\text{fermion}} + \left[ D_3-\frac{1-2\Delta}{2r}\right].
\eea
We have multiplied the fermionic operator appearing in the action by $\gamma_3$ due to the spinor product $\epsilon\cdot\psi = \epsilon_2\psi_1-\epsilon_1\psi_2$, and defined
\bea
 \tilde{D}_{\text{scalar}} = -D^{i}D_{i}\,, \quad \tilde{D}_{\text{fermion}} = \gamma^3\gamma^{i}D_{i}-\frac{\mathfrak{m}}{r}\gamma^{3},
\eea
for $i,j=1,2$, and $\mathfrak{m}$ acting implicitly in the appropriate representation. All covariant derivatives are with respect to the background (\ref{eq:hemisphere-saddlepoints}), for example on spinors:
\bea
    D_{\mu} = \partial_{\mu} + \frac{1}{2} i w_{\mu}\sigma_{3}+ i\mathfrak{m}w_{\mu},
\eea
where $w_{\mu} = (0,-\cos\theta,0)$ is the spin connection. The 1-loop determinant will be given by
\bea 
    Z_{\text{1-loop}}^{(D)} = \frac{\text{det}D_{\text{fermion}}}{\text{det}D_{\text{scalar}}}
\eea
after a suitable regularisation. The boundary condition $\phi|_{\theta=\frac{\pi}{2}} = \psi_1-\psi_2|_{\theta=\frac{\pi}{2}} = 0$ is imposed on the fluctuating modes. As expected, there are large cancellations between bosonic and fermionic eigenmodes. 

We work in the setting where the twisted periodicities in (\ref{eq:twistedperiodicity}) due to the flavour symmetries are cancelled by turning on holonomies for their background vector multiplets but retain twisted periodicities due to the R-symmetry and angular momentum. We therefore have $D_3 = \nabla_3 + ia_3 + i\sum_l a_3^l$, where $a_3^l$ are  flat connection(s) for flavour symmetry. This operator commutes with $\tilde{D}_{\text{scalar}}, \tilde{D}_{\text{fermion}}$ in (\ref{eq:gaussianoperators}), and so we diagonalise them simultaneously. For a field of R-charge $R$, we expand in terms of fields:
\bea 
    \mathcal{O}_{n,m}(\theta,\varphi,\tau) = e^{\frac{\tau}{\beta r} \left(2\pi i n -(R+m)\beta_1+m\beta_2\right)}\mathcal{O}_{m}(\theta,\varphi),
\eea
where
\bea 
    J_3 \mathcal{O}_{m} = \left(-i\partial_{\varphi}+ \kappa \mathfrak{m} \right) \mathcal{O}_{m} =  m \mathcal{O}_{m}.
\eea 
Then $D_3$ acts as:
\bea 
    \beta r D_3 \mathcal{O}_{n,m} = \left[2\pi i n  -(R+m)\beta_1+m\beta_2+ i\beta r \rho(a_3) + i\beta r \rho_l(a_3^l)\right] \mathcal{O}_{n,m}.
\eea

\paragraph{Paired Eigenmodes} We now exhibit the pairing of fermionic and bosonic eigenmodes. If $\psi$ is a fermionic eigenmode  obeying the boundary condition $\psi_1-\psi_2|_{\theta=\frac{\pi}{2}} = 0$ and satisfying
\bea 
    \tilde{D}_{\text{fermion}} \psi = \left(\gamma^{3}\gamma^{i}D_{i}-\frac{\mathfrak{m}}{r}\gamma^{3}\right) \psi = \nu \psi,
\eea
then we can construct
\bea\label{eq:fermiontobosoneigenomode}
\phi' = \bar{\epsilon} \psi
\eea 
which obeys
\bea    
    \tilde{D}_{\text{scalar}} \phi' = -g^{ij}_{(2)} D_i D_j \phi' = \left( \nu(\nu+1)-\frac{\mathfrak{m}^2 }{r^2}\right) \phi'\,.
\eea
Similarly, for a scalar eigenmode $\phi$ such that
\bea 
     \tilde{D}_{\text{scalar}} \phi = M^2 \phi,
\eea
one can construct two spinor eigenmodes
\bea\label{eq:bosontofermion}
    \psi^{(1,2)} = \gamma^i \epsilon D_{i} \phi + \frac{\mathfrak{m}}{r} \epsilon \phi - \nu \gamma^3 \epsilon \phi,
\eea
where $\nu$ is a solution to $\nu(\nu+1)-\frac{\mathfrak{m}^2}{r^2}=M^2$, i.e. if $\nu$ is a solution so is $-\nu-1$. It is easy to check that eigenvalues of $\phi$ and the pair $\psi^{(1,2)}$ cancel in the determinant, noting that  the Killings spinors (\ref{eq:Killingspinors}) satisfy
\bea
    \beta r D_3 \epsilon =  \frac{\beta}{2} \epsilon\,,\qquad  \beta r D_3 \bar{\epsilon} =  -\frac{\beta}{2} \bar{\epsilon}\,.
\eea
Also $\psi^{(1,2)}$ and $\phi'$ obey the appropriate boundary condition.

\paragraph{Unpaired Eigenmodes} The non-cancelling contributions to the 1-loop determinant are the ones which do not participate in the pairing above, that is when (\ref{eq:fermiontobosoneigenomode}) or (\ref{eq:bosontofermion}) are undefined. An unpaired scalar eigenmode is a $\phi$ such that:
\bea\label{eq:unpairedscalareigenomde}
    \gamma^i \epsilon D_{i} \phi + \frac{\mathfrak{m}}{r} \epsilon \phi - \nu \gamma^3 \epsilon \phi = 0.
\eea 
Contracting with $\bar{\epsilon}$ gives
\bea 
    \partial_{\varphi} \phi + i(\kappa \mathfrak{m}-r \nu) \phi = 0.
\eea 
Using the ansatz $\phi = f(\theta) e^{-i\left(\kappa \mathfrak{m}- r \nu\right)\varphi}$ (suppressing $\tau$ dependence for now) and contracting (\ref{eq:unpairedscalareigenomde}) with $\bar{\epsilon}\gamma_3$ we obtain:
\bea 
    \sin\theta \,\partial_{\theta} f + \mathfrak{m}f - r \nu\cos\theta f = 0.
\eea 
There are no non-trivial solutions obeying the boundary condition and thus no unpaired scalar eigenmodes.\footnote{Note that this means that if we have an eigenmode $\psi$ with eigenvalue $\nu$ which is paired, we may always construct the eigenmode $\tilde{\psi}$ with eigenmode $-\nu-1$ by using first the map (\ref{eq:fermiontobosoneigenomode}) to construct $\phi'$, and then (\ref{eq:bosontofermion}) to construct $\psi_1$ proportional to $\psi$, and $\psi_2 \coloneqq  \tilde{\psi}$.}

We now look for unpaired spinor eignmodes $\psi$. If $\bar{\epsilon} \psi =0$ then we may write $\psi = \bar{\epsilon}\Phi$ where $\Phi$ is a scalar of R-charge $\Delta-2$ (so that $\psi$ has R-charge $\Delta-1$). Using the Killing spinor equations
\bea 
    &\left(\gamma^{3}\gamma^{i}D_{i}-\frac{\mathfrak{m}}{r}\gamma^{3}\right)\left(\bar{\epsilon}\Phi\right)= \nu \psi, \\
   \Rightarrow\quad &\gamma^3\gamma^i\bar{\epsilon}D_i \Phi = \left(\nu+\frac{1}{r}\right)(\bar{\epsilon}\Phi) + \frac{\mathfrak{m}}{r} \gamma^3\bar{\epsilon}\Phi.
\eea 
Contracting with $\epsilon$ and $\epsilon \gamma^3$ gives
\bea 
    \sin\theta\partial_{\theta}\Phi = -(r\nu+1)\cos\theta + \mathfrak{m}\Phi,\\
    \left(\partial_{\varphi} +i\mathfrak{m}(\kappa-\cos\theta)\right)\Phi = i(r\nu+1)\Phi - i \mathfrak{m}\cos\theta.
\eea 
Using the ansatz $\Phi = f(\theta)e^{i(r\nu+1-\kappa \mathfrak{m})\varphi}$, we find solutions:
\bea 
    \Phi = \sin(\theta/2)^{\mathfrak{m}}\cos(\theta/2)^{-\mathfrak{m}} (\sin\theta)^j e^{i(-j-\kappa\mathfrak{ m })\varphi}e^{\frac{\tau }{\beta r}\left(2\pi i n -(\Delta-2-j)\beta_1-j\beta_2\right)},
 \eea 
where $j = - r\nu -1$ is an integer such that $j+\mathfrak{m} \geq 0$. The last requirement is for regularity at $\theta = 0$. This is less restrictive than also requiring regularity at $\theta = \pi$ as for the $S^1\times S^2$ index, which would require $j \geq |\mathfrak{m}|$. The $\tau$-dependent exponential ensures the twisted periodicity condition (\ref{eq:twistedperiodicity}). The unpaired fermionic eigenmodes are thus
\bea 
    \psi = &\,e^{\frac{\tau}{\beta r}\left(2\pi i n -((\Delta-1)-(j+1/2))\beta_1-(j+1/2)\beta_2\right)} e^{i(-j-1/2-\kappa\mathfrak{m})\varphi}\\ &\times \sin(\theta/2)^{\mathfrak{m}}\cos(\theta/2)^{-\mathfrak{m}} (\sin\theta)^j \begin{pmatrix} \sin(\theta/2) \\ \cos(\theta/2)\end{pmatrix}.
\eea
\paragraph{The 1-loop Determinant} We now have all the ingredients needed to write down the 1-loop determinant for the $N=2$ chiral multiplet with Dirichlet boundary condition. 
\bea 
    Z_{\text{1-loop}}^{(D)}=&\prod_{n\in \bZ} \prod_{j\geq -\rho(\mathfrak{m})} \left[\beta r \left(\frac{1-2\Delta}{2r}+\frac{j+1}{r} \right)-\beta r D_3\right] \\
    =&\prod_{n\in \bZ} \prod_{j\geq 0} \left[-2\pi i n - i\beta r \rho(a_3) + (2j+2-\Delta-2\rho(\mathfrak{m}))\beta_2 - i\beta r \rho_l(a_3^l) \right]\\
    =&e^{-\mathcal{E}[-i\beta r \rho(a_3)- i\beta r \rho_l(a_3^l)+(2-\Delta-2\rho(\mathfrak{m}))\beta_2]}\left(e^{i\beta r\rho(a_3)+ i\beta r \rho_l(a_3^l)}q^{1-\Delta/2-\rho(\mathfrak{m})};q\right)_{\infty}.
\eea 
The final line has been zeta function regularised as in \cite{Yoshida:2014ssa}. 

\subsection{Regularisation and Anomaly Polynomials} \label{app:regularisationanolypolynomials}

We now show that the results in section \ref{sec:resultoflocalisation} for the $S^1\times H^2$ partition function reproduces the formula (3.31) in \cite{Dimofte:2017tpi} for the half-index $\mathcal{I}$ counting local operators inserted at the origin of $\bR^2 \times \bR_{\geq 0}$, up to a prefactor encoding boundary 't Hooft anomalies:
\bea\label{eq:casimirappendix}
    \mathcal{Z} = e^{\phi} \mathcal{I}\,.
\eea
Here $\phi$ is the Casimir energy, and is consistent with the results of \cite{Bobev:2015kza}. This result holds for an $\mathcal{N}=2$ SCFT, with the $\mathcal{N}=4$ results of section \ref{sec:bc} following in an obvious way. We also stick to an abelian gauge group, the non-abelian generalisation can be found by ensuring consistency with the maximal torus of the gauge group. The q-Pochhammer contributions clearly match, so we need only consider the classical contribution and the $\mathcal{E}$ functions. Examining each in turn:
\begin{itemize}
    \item The coupling to the topological symmetry gives
    \bea 
        e^{-\frac{2\log\xi  \log zq^{\mathfrak{m}}}{2\log q}}.
    \eea 
    Rewriting the term in the exponential as:
    \bea\label{eq:topologicalsymmetrybilinearform}
    \frac{1}{2\log q} 
    \begin{pmatrix} \log zq^{\mathfrak{m}} &  \log\xi \end{pmatrix}
    \begin{pmatrix} 0 & -1 \\ -1 & 0 \end{pmatrix}
    \begin{pmatrix} \log zq^{\mathfrak{m}} \\  \log\xi \end{pmatrix}
    \eea
    This is the same bilinear form encoding the mixed boundary 't Hooft anomaly between the boundary gauge symmetry and the topological symmetry in the anomaly polynomial contribution
    \bea\label{eq:anomalypolytopological}
    - 2 \mathbf{f}\mathbf{f}_{\xi}\,,
    \eea
    where $\mathbf{f},\mathbf{f}_{\xi}$ are field strengths for the corresponding symmetries. Isolating the $\mathfrak{m}$ dependence in (\ref{eq:topologicalsymmetrybilinearform}) recovers $\xi^{-\mathfrak{m}}$ which appears in the half index formulae of \cite{Dimofte:2017tpi}. The $\mathfrak{m}$ independent part contributes to $\phi$ with $-\log\xi\log z / \log q$.
    
    \item For a chiral with $(N)$ boundary conditions, transforming with charge $\rho$ under an abelian gauge group and R-charge $\Delta$, the anomalous contribution is 
    \bea
        e^{\mathcal{E}\left[-\log\left(q^{\frac{\Delta}{2}+\rho(\mathfrak{m})}z^{\rho}\right)\right]} =  C e^{\frac{1}{2\log q } \left[\ -\frac{1}{2}\left(\rho \log zq^{\mathfrak{m}}+(\Delta-1)\log q^{\frac{1}{2}}\right)^2 \right]}\,,
    \eea
    where $C\coloneqq e^{-\frac{\beta}{24}}$.  Up to this constant, this matches the bilinear form encoding the contribution of the chiral to the boundary 't Hooft anomaly polynomial
    \bea\label{eq:anomalypolyneumannchiral}
        -\frac{1}{2}  \left(\rho \mathbf{f} +(\Delta-1)\mathbf{r} \right)^2
    \eea
    after replacing $\log z \rightarrow \mathbf{f}$ and $\log q^{\frac{1}{2}}\rightarrow \mathbf{r}$. Here $\mathbf{r}$ is the field strength of the R-symmetry. Again the $\mathfrak{m}$ dependence matches the half index formula, and we obtain an overall contribution to the prefactor $\phi$ of:
    \bea
        \frac{1}{2\log q } \left[\ -\frac{1}{2}\left(\rho \log z+(\Delta-1)\log q^{\frac{1}{2}}\right)^2 \right]\,.
    \eea 
    
    \item Similarly for a chiral with $(D)$ boundary conditions, we have:
    \bea
        e^{-\mathcal{E}\left[-\log\left(q^{1-\frac{\Delta}{2}-\rho(\mathfrak{m})}z^{-\rho}\right)\right]} = C^{-1} e^{\frac{1}{2\log q } \left[\ \frac{1}{2}\left(\rho \log zq^{\mathfrak{m}}+(\Delta-1)\log q^{\frac{1}{2}}\right)^2 \right]}.
    \eea
    which matches the contribution to the boundary 't Hooft anomaly polynomial
    \bea \label{eq:anomalypolydirichletchiral}
        \frac{1}{2}  \left(\rho \mathbf{f} +(\Delta-1)\mathbf{r} \right)^2\,.
    \eea
    
    \item A U(1) $\mathcal{N}=2$ vector multiplet contributes
    \bea
        e^{\mathcal{E}\left[-\log q \right]} =  C e^{\frac{1}{2\log q } \left[\ -\frac{1}{2}\left(\log q^{\frac{1}{2}}\right)^2 \right]}\,,
    \eea
    matching the corresponding boundary 't Hooft anomaly polynomial contribution
    \bea\label{eq:anomalypolyvector}
    -\frac{1}{2}\mathbf{r}^2\,.
    \eea
\end{itemize}
In summary, up to factors of $C$, we are left with a prefactor $\phi$ given precisely by:
\bea
\phi = \frac{1}{2\log q } \mathcal{P} (\log q^{\frac{1}{2}}, \log z, \log \xi)
\eea
where $\mathcal{P} (\mathbf{r},\mathbf{f},\mathbf{f}_{\xi})$ is the anomaly polynomial encoding the boundary 't Hooft anomaly, consisting of contributions (\ref{eq:anomalypolytopological}), (\ref{eq:anomalypolyneumannchiral}), (\ref{eq:anomalypolyneumannchiral}) and (\ref{eq:anomalypolyvector}). For a non-abelian theory, it is the equivariant integral of the polynomial \cite{Bobev:2015kza}. 

In an $\mathcal{N}=4$ theory, with $(2,2)$  boundary condition, the factors $C$ always cancel, and the cancellations of the $\mathcal{E}$ reflect that only the mixed anomalies listed in section \ref{subsec:(2,2)bc} can occur, thus proving equation (\ref{eq-pre-factor}).

\section{General Abelian Theories}
\label{app:abelian}

With the result of section \ref{app:regularisationanolypolynomials} in hand, we prove the claims in section \ref{sec:bc} for a general 3d $\mathcal{N}=4$ abelian theory. That is, we show that if $\mathcal{P}_{\cB_{\alpha}} (\mathbf{r}, \mathbf{t}, \mathbf{f}_{x},\mathbf{f}_{\xi})$ is the boundary 't Hooft anomaly polynomial for a boundary condition $\mathcal{B}_{\alpha}$, the lowest weights of the corresponding Higgs and Coulomb branch Verma modules are given by:
\bea 
    \lim_{t^{\frac{1}{2}}\rightarrow q^{-\frac{1}{4}}} \phi_{\cB_{\alpha}} &=
    \phi_{\cB_{\alpha}}^{(B)} = 
    \lim_{t^{\frac{1}{2}}\rightarrow q^{-\frac{1}{4}}} \frac{1}{2\log q}  \mathcal{P}_{\cB_{\alpha}} (\log q^{\frac{1}{2}}, \log t, \log x , \log \xi)  \\
     \lim_{t^{\frac{1}{2}}\rightarrow q^{\frac{1}{4}}} \phi_{\cB_{\alpha}} &=
     \phi_{\cB_{\alpha}}^{(A)} = 
     \lim_{t^{\frac{1}{2}}\rightarrow q^{\frac{1}{4}}} \frac{1}{2\log q}  \mathcal{P}_{\cB_{\alpha}} (\log q^{\frac{1}{2}}, \log t, \log x , \log \xi) .
\eea
Further,  the mixed anomaly coefficient $k$ between $T_H$ an $T_C$, is equal to the central charge $\kappa_{\alpha}$ where $\alpha$ is the vacuum for the abelian theory associated to boundary condition $\cB_{\alpha}$.

We briefly recap exceptional Dirichlet boundary conditions for abelian 3d $\mathcal{N}=4$ theories. See \cite{Bullimore:2016nji} for more details. Consider a gauge group $G = U(1)^r$, with $N$ hypermultiplets $(X_i,Y_i)$. The Higgs and Coulomb branch flavour symmetries are
\bea 
    G_H = U(1)^{N-r} \coloneqq  U(1)^{r'}\,,\qquad G_C=U(1)^r\,.
\eea
We denote by: 
\bea
Q=\{Q^{i}_a\}^{1\leq i \leq N}_{1\leq a\leq r}\,,\qquad q=\{q^{i}_\beta\}^{1\leq i \leq N}_{1\leq \beta\leq r'}
\eea
the  matrices of gauge and flavour charges respectively. An exceptional Dirichlet boundary condition is labelled by a subset $S \subset (1,...,N)$ such that the charge submatrix $Q^{(S)}$ is non-degenerate and a sign vector $\epsilon$ so that the boundary condition sets
\bea 
\mathcal{B}:\quad 
    \begin{cases}
        Y_i \rvert = c_i  & \epsilon_i  = + \\
        X_i \rvert = c_i  & \epsilon_i  = -
    \end{cases}
\quad (i\in S), \quad 
    \begin{cases}
        Y_j \rvert = 0  & \epsilon_j  = + \\
        X_j \rvert = 0  & \epsilon_j  = -
    \end{cases}
\quad (j\notin S)
\eea
where the $c_i$ are non-zero. The scalars fixed to non-zero values at the boundary are those with non-zero values on the vacuum $\nu_{\alpha}$. This boundary condition fully breaks the gauge symmetry and preserves the flavour symmetry at the boundary. It is a thimble boundary condition for a certain chamber of masses and FIs.

Let us define the submatrices and subvectors for later use:
\begin{equation}
\begin{split}
Q^{S} =& \{Q^{i}_a\}^{i\in S}_{1\leq a\leq r}\,,\\
Q' =& \{Q^j_a\}^{j\notin S}_{1\leq a\leq r}\,,
\end{split}
\qquad
\begin{split}
q^{S} =& \{q^{i}_{\beta}\}^{i\in S}_{1\leq \beta \leq N-r}\,,\\
q' =& \{q^j_{\beta}\}^{j\notin S}_{1\leq \beta \leq N-r}\,,
\end{split}
\qquad
\begin{split}
\varepsilon^{S} =& \{\varepsilon_i\}_{i\in  S}\,,\\
\varepsilon'  =&\{\varepsilon_j\}_{j\notin S}\,.
\end{split}
\end{equation}

\paragraph{Anomaly Polynomial}
To compute the anomaly polynomial we can first compute it for the boundary condition with zero values for $c_i$, and then deform to the anomaly polynomial for $\mathcal{B}$ by setting to 1 the sum of field strengths dual to the charges of the $\cN=2$ chirals labelled by $S$, whose scalars are set to $c_i$. We define field strengths $\mathbf{r}$, $\mathbf{t}$, $\mathbf{f}_a$, $\mathbf{f}_{x_\beta}$, $\mathbf{f}_{\xi_a}$ for $(R_V+R_A)/2, (R_V-R_A)/2, G_{\partial}, G_H$ and $G_C$ respectively. With $c_i=0$, the anomaly polynomial receives contributions \cite{Dimofte:2017tpi}:
\begin{itemize}
    \item From the $\mathcal{N}=4$ vector multiplet:
        \bea 
            -2\mathbf{f} \cdot  \mathbf{f}_{\xi} + r \left(\frac{1}{2}\mathbf{t}^2 -\frac{1}{2}\mathbf{r}^2 \right) 
            = -2\mathbf{f} \cdot  \mathbf{f}_{\xi}  -\frac{r}{2}\left(\mathbf{r}+\mathbf{t}\right)\left(\mathbf{r}-\mathbf{t}\right)\,.
        \eea 
    \item From the $j^{\text{th}}$ $\mathcal{N}=4$ hypermultiplet: 
        \bea
            &-\varepsilon_j\left[\frac{1}{2}\left(\mathbf{f}_aQ_a^j+\mathbf{f}_{x_\beta}q_{\beta}^j+\frac{1}{2}\mathbf{t}-\frac{1}{2}\mathbf{r}\right)^2 -\frac{1}{2}\left(-\mathbf{f}_aQ_a^j-\mathbf{f}_{x_\beta}q_{\beta}^j+\frac{1}{2}\mathbf{t}-\frac{1}{2}\mathbf{r}\right)^2    \right]\\
            &=\varepsilon_j (\mathbf{r}-\mathbf{t})(\mathbf{f}_aQ_a^j+\mathbf{f}_{x_\beta}q_{\beta}^j)
        \eea
        where we sum over $a$ and $\beta$ implicitly. So from all $N$ hypers:
        \bea 
            (\mathbf{r}-\mathbf{t})(\mathbf{f}\cdot Q \cdot \varepsilon +\mathbf{f}_{x} \cdot q \cdot \varepsilon)\,.
        \eea
\end{itemize}
So the total anomaly polynomial before deformation is
\bea 
\mathcal{P} = -2\mathbf{f} \cdot  \mathbf{f}_{\xi}
-\frac{r}{2}\left(\mathbf{r}+\mathbf{t}\right)\left(\mathbf{r}-\mathbf{t}\right)+
 (\mathbf{r}-\mathbf{t})(\mathbf{f}\cdot Q \cdot \varepsilon +\mathbf{f}_{x} \cdot q \cdot \varepsilon)\,.
\eea 
Now deforming to non-zero $c$, set for each $i\in S$
\bea
    \mathbf{f}_a Q_a^i +\mathbf{f}_{x_\beta}q_{\beta}^i - \frac{\varepsilon_i}{2} \left(\mathbf{r}+\mathbf{t}\right) = 0\,,
\eea
or since $Q^{(S)}$ is invertible:
\bea 
    \mathbf{f} = -\mathbf{f}_{x} \cdot q^S\cdot {Q^{S}}^{-1}+ \frac{1}{2}(\mathbf{r}+\mathbf{t})\epsilon^{S}\cdot {Q^{S}}^{-1}\,.
\eea    
Substituting into the undeformed $\cP$, we arrive at anomaly polynomial for $\cB_{\alpha}$
\bea 
\mathcal{P}_{\mathcal{B}} = & \,2\mathbf{f}_x \cdot q^S \cdot {Q^S}^{-1}\cdot \mathbf{f}_{\xi} - \epsilon^S\cdot {Q^{S}}^{-1}\cdot \mathbf{f}_{\xi} \, (\mathbf{r}+\mathbf{t})
\\
&+ (\mathbf{r}-\mathbf{t})\, \mathbf{f}_{x}\cdot \left(q'-q^S\cdot {Q^{S}}^{-1}\cdot Q' \right)\cdot\epsilon'\\
&+\frac{1}{2}(\mathbf{r}-\mathbf{t}) (\mathbf{r}+\mathbf{t}) \left(\epsilon^S \cdot {Q^S}^{-1}\cdot  Q \cdot \epsilon - r \right)\,.
\eea    
We may easily read off the coefficients defined in section \ref{subsec:(2,2)bc} for the various mixed anomalies:
\bea\label{eq:anomalycoeffgeneralabelian}
    \tilde{k} &= \epsilon^S \cdot {Q^S}^{-1}\cdot  Q \cdot \epsilon - r \,,\\
    k_V &= - \epsilon^S\cdot {Q^{S}}^{-1} \,,\\
    k_A &= \left(q'-q^S\cdot {Q^{S}}^{-1}\cdot Q' \right)\cdot\epsilon' \,,\\
    k &= q^S \cdot {Q^S}^{-1}\,.
\eea 

\paragraph{Central Charges}
The central charge $\kappa_{\alpha}$ is the bilinear pairing such that:\footnote{The minus sign difference in these generators is due to our convention for the FI parameter.} 
\begin{equation}
 \kappa_{\alpha}(m_{\mathbb{R}},t_{\mathbb{R}}) = 
\begin{cases}
h_m (\nu_{\alpha}) &= m_{\mathbb{R}}\cdot \mu_{H,\mathbb{R}}(\nu_{\alpha}) \\
h_t(\nu_{\alpha}) &= -t_{\mathbb{R}} \cdot \mu_{C,\mathbb{R}}(\nu_{\alpha})
\end{cases}\,.
\end{equation}
The bilinear pairing for a general abelian theory is derived in section 7.4.2 of \cite{Bullimore:2016nji}. We briefly recap it here. Define
\be
w_j := |X_j|^2-|Y_j|^2 \,, \qquad W_j := X_j Y_j.
\ee
\begin{itemize}
    \item On $\mathcal{M}_H$ we have $h_m (\nu_{\alpha}) = m_{\mathbb{R}}\cdot \mu_{H,\mathbb{R}}|_{\nu_{\alpha}} = m_{\mathbb{R}}\cdot q \cdot w |_{\nu_{\alpha}}$. Now $w_j = 0$ for all $j\notin S$ at the vacuum. The remaining $w_i$ for $i\in S$ are determined by the real moment map $Q\cdot w  = t_{\mathbb{R}}$. Then one can see immediately that:
    \bea
        h_m (\nu_{\alpha})  = m_{\mathbb{R}}\cdot q^{S} \cdot {Q^{S}}^{-1} \cdot t_{\mathbb{R}},
    \eea
    so $\kappa_{\alpha}=q^S(Q^S)^{-1}$ coincides with the value of the anomaly coefficient $k$ in (\ref{eq:anomalycoeffgeneralabelian}).
    
    \item Considering $\mathcal{M}_C$ yields the same answer. At the vacuum $h_t(\nu_{\alpha}) =- \sigma \cdot t_{\mathbb{R}} |_{\nu_{\alpha}}$. At $\nu_{\alpha}$ the effective real mass of the hypermultiplets must vanish for all $i\in S$: $M^i = \sigma\cdot Q^i + m_{\mathbb{R}}\cdot q^i = 0 $. Thus $\sigma|_{\nu_{\alpha}} = -m_\mathbb{R}\cdot q^{S}(Q^S)^{-1}$ and so $h_t(\nu_{\alpha}) = h_m(\nu_{\alpha})$. 
\end{itemize}

\paragraph{Lowest Weights}
We show now that the  anomaly coefficients (\ref{eq:anomalycoeffgeneralabelian}) coincide with the lowest weight characters of the Verma module defined by $\cB_{\alpha}$ as described in section \ref{sec:characters}.
\begin{itemize}
    \item On the Higgs branch recall that the action of $\hat{W}_j = {:}\hat{X}_j \hat{X}_j{:} $  for $j \notin S$ is given by
    \bea 
    \hat{W}_j |\cB_{\alpha}\ra  = \frac{\varepsilon}{2} \epsilon_j |\cB_{\alpha}\ra\,.
    \eea
    For $i\in S$ it is fixed by the relation $Q\cdot \hat{W} = t_{\mathbb{C}} $, and on the lowest weight state
\bea
    \hat{W}_{i}|\cB_{\alpha}\ra &= - \left( {Q^{S}}^{-1}\cdot  Q' \cdot \hat{W}' \right)_i |\cB_{\alpha}\ra + \left({Q^{S}}^{-1} \cdot t_{\mathbb{C}} \right)_{i} |\cB_{\alpha}\ra \\
    &= \left( - \frac{1}{2} \epsilon \left({Q^{S}}^{-1}\cdot Q' \cdot \varepsilon' \right)_i  + \left({Q^{S}}^{-1} \cdot t_{\mathbb{C}} \right)_{i} \right)|\cB_{\alpha}\ra.
\eea
The Verma character for $\cA_H$ is
\bea 
\text{Tr}  \left[ e^{-\frac{1}{\epsilon} m_{\bR}\cdot q\cdot \hat{W}}\right],
\eea 
where we identify fugacities for the flavour symmetry $x_i = e^{-m_{\mathbb{R},i}}$. One can straightforwardly compute the character of the lowest weight state as
\bea\label{eq:highestweighthiggs}
     x^{\frac{1}{2}\left(q'-q^S\cdot {Q^{S}}^{-1}\cdot Q' \right)\cdot\epsilon' + \frac{1}{\epsilon}q^{S} \cdot {Q^{S}}^{-1} \cdot t_{\mathbb{C}}}
\eea 
matching the values of $k_A$ and $k$ in (\ref{eq:anomalycoeffgeneralabelian}).

\item On the Coulomb branch the vacuum obeys $\left(\hat{M}^i_{\mathbb{C}} - \frac{1}{2}\varepsilon_i\epsilon \right)|\mathcal{B} \ra= 0$ for all $i\in S$ where the (quantised) effective complex masses are given by $\hat{M}^i_{\mathbb{C}} = \hat{\varphi}\cdot Q^i + m_{\mathbb{C}}\cdot q^i$. Thus
\bea
    \hat{\varphi} |\mathcal{B}\ra = \left(\frac{\epsilon}{2} \varepsilon^S\cdot {Q^{S}}^{-1}- m_{\mathbb{C}} \cdot q^{S} \cdot {Q^{S}}^{-1} \right)|\mathcal{B}\ra\,.
\eea
The character of the Coulomb Verma is
\bea
\text{Tr} \left[ e^{ \frac{1}{\epsilon} t_{\mathbb{R}}\cdot \hat{\varphi}}\right]\,,
\eea 
and thus the character of the lowest weight state is
\bea\label{eq:highestweightcoulomb}
   \xi^{-\frac{1}{2}\varepsilon^S\cdot {Q^{S}}^{-1} + \frac{1}{\epsilon} m_{\mathbb{C}}\cdot q^S \cdot {Q^{S}}^{-1} }\,,
\eea
matching the values of $k_V$ and $k$ in (\ref{eq:anomalycoeffgeneralabelian}).
\end{itemize}
We conclude that the lowest weights of the Higgs and Coulomb branch algebra modules defined by $\mathcal{B}$ are indeed given by the limits of the prefactor/Casimir energy $\phi_{\mathcal{B}}$, which itself coincides with the anomaly polynomial describing boundary 't Hooft anomalies determined by $\mathcal{B}$.

\bibliographystyle{JHEP}
\bibliography{blocks_and_vermas}

\providecommand{\href}[2]{#2}\begingroup\raggedright\begin{thebibliography}{10}

\bibitem{Pasquetti:2011fj}
S.~Pasquetti, {\it {Factorisation of N = 2 Theories on the Squashed 3-Sphere}},
   {\em JHEP} {\bf 04} (2012) 120, [\href{http://arxiv.org/abs/1111.6905}{{\tt
  arXiv:1111.6905}}].

\bibitem{Hwang:2015wna}
C.~Hwang and J.~Park, {\it {Factorization of the 3d superconformal index with
  an adjoint matter}},  {\em JHEP} {\bf 11} (2015) 028,
  [\href{http://arxiv.org/abs/1506.03951}{{\tt arXiv:1506.03951}}].

\bibitem{Hwang:2012jh}
C.~Hwang, H.-C. Kim, and J.~Park, {\it {Factorization of the 3d superconformal
  index}},  {\em JHEP} {\bf 08} (2014) 018,
  [\href{http://arxiv.org/abs/1211.6023}{{\tt arXiv:1211.6023}}].

\bibitem{Dimofte:2011py}
T.~Dimofte, D.~Gaiotto, and S.~Gukov, {\it {3-Manifolds and 3d Indices}},  {\em
  Adv. Theor. Math. Phys.} {\bf 17} (2013), no.~5 975--1076,
  [\href{http://arxiv.org/abs/1112.5179}{{\tt arXiv:1112.5179}}].

\bibitem{Cabo-Bizet:2016ars}
A.~Cabo-Bizet, {\it {Factorising the 3D Topologically Twisted Index}},  {\em
  JHEP} {\bf 04} (2017) 115, [\href{http://arxiv.org/abs/1606.06341}{{\tt
  arXiv:1606.06341}}].

\bibitem{Crew:2020jyf}
S.~Crew, N.~Dorey, and D.~Zhang, {\it {Factorisation of 3d $ \mathcal{N} $ = 4
  twisted indices and the geometry of vortex moduli space}},  {\em JHEP} {\bf
  08} (2020), no.~08 015, [\href{http://arxiv.org/abs/2002.04573}{{\tt
  arXiv:2002.04573}}].

\bibitem{Dimofte:2011ju}
T.~Dimofte, D.~Gaiotto, and S.~Gukov, {\it {Gauge Theories Labelled by
  Three-Manifolds}},  {\em Commun. Math. Phys.} {\bf 325} (2014) 367--419,
  [\href{http://arxiv.org/abs/1108.4389}{{\tt arXiv:1108.4389}}].

\bibitem{Dimofte:2014zga}
T.~Dimofte, {\it {Complex Chern\textendash{}Simons Theory at Level k via the
  3d\textendash{}3d Correspondence}},  {\em Commun. Math. Phys.} {\bf 339}
  (2015), no.~2 619--662, [\href{http://arxiv.org/abs/1409.0857}{{\tt
  arXiv:1409.0857}}].

\bibitem{Benini:2013yva}
F.~Benini and W.~Peelaers, {\it {Higgs branch localization in three
  dimensions}},  {\em JHEP} {\bf 05} (2014) 030,
  [\href{http://arxiv.org/abs/1312.6078}{{\tt arXiv:1312.6078}}].

\bibitem{Fujitsuka:2013fga}
M.~Fujitsuka, M.~Honda, and Y.~Yoshida, {\it {Higgs branch localization of 3d
  $\mathcal{N}=2$ theories}},  {\em PTEP} {\bf 2014} (2014), no.~12 123B02,
  [\href{http://arxiv.org/abs/1312.3627}{{\tt arXiv:1312.3627}}].

\bibitem{Dedushenko:2018aox}
M.~Dedushenko, {\it {Gluing. Part I. Integrals and symmetries}},  {\em JHEP}
  {\bf 04} (2020) 175, [\href{http://arxiv.org/abs/1807.04274}{{\tt
  arXiv:1807.04274}}].

\bibitem{Dedushenko:2018tgx}
M.~Dedushenko, {\it {Gluing II: Boundary Localization and Gluing Formulas}},
  \href{http://arxiv.org/abs/1807.04278}{{\tt arXiv:1807.04278}}.

\bibitem{Beem:2012mb}
C.~Beem, T.~Dimofte, and S.~Pasquetti, {\it {Holomorphic Blocks in Three
  Dimensions}},  {\em JHEP} {\bf 12} (2014) 177,
  [\href{http://arxiv.org/abs/1211.1986}{{\tt arXiv:1211.1986}}].

\bibitem{Hori:2000ck}
K.~Hori, A.~Iqbal, and C.~Vafa, {\it {D-branes and mirror symmetry}},
  \href{http://arxiv.org/abs/hep-th/0005247}{{\tt hep-th/0005247}}.

\bibitem{Gaiotto:2015zna}
D.~Gaiotto, G.~W. Moore, and E.~Witten, {\it {An Introduction To The Web-Based
  Formalism}},  \href{http://arxiv.org/abs/1506.04086}{{\tt arXiv:1506.04086}}.

\bibitem{Gaiotto:2015aoa}
D.~Gaiotto, G.~W. Moore, and E.~Witten, {\it {Algebra of the Infrared: String
  Field Theoretic Structures in Massive ${\cal N}=(2,2)$ Field Theory In Two
  Dimensions}},  \href{http://arxiv.org/abs/1506.04087}{{\tt
  arXiv:1506.04087}}.

\bibitem{Bullimore:2016nji}
M.~Bullimore, T.~Dimofte, D.~Gaiotto, and J.~Hilburn, {\it {Boundaries, Mirror
  Symmetry, and Symplectic Duality in 3d $\mathcal{N}=4$ Gauge Theory}},  {\em
  JHEP} {\bf 10} (2016) 108, [\href{http://arxiv.org/abs/1603.08382}{{\tt
  arXiv:1603.08382}}].

\bibitem{Bullimore:2014awa}
M.~Bullimore, H.-C. Kim, and P.~Koroteev, {\it {Defects and Quantum
  Seiberg-Witten Geometry}},  {\em JHEP} {\bf 05} (2015) 095,
  [\href{http://arxiv.org/abs/1412.6081}{{\tt arXiv:1412.6081}}].

\bibitem{Zenkevich:2017ylb}
A.~Nedelin, S.~Pasquetti, and Y.~Zenkevich, {\it {T[SU(N)] duality webs: mirror
  symmetry, spectral duality and gauge/CFT correspondences}},  {\em JHEP} {\bf
  02} (2019) 176, [\href{http://arxiv.org/abs/1712.08140}{{\tt
  arXiv:1712.08140}}].

\bibitem{Aprile:2018oau}
F.~Aprile, S.~Pasquetti, and Y.~Zenkevich, {\it {Flipping the head of
  $T[SU(N)]$: mirror symmetry, spectral duality and monopoles}},  {\em JHEP}
  {\bf 04} (2019) 138, [\href{http://arxiv.org/abs/1812.08142}{{\tt
  arXiv:1812.08142}}].

\bibitem{Yagi:2014toa}
J.~Yagi, {\it {$\Omega$-deformation and quantization}},  {\em JHEP} {\bf 08}
  (2014) 112, [\href{http://arxiv.org/abs/1405.6714}{{\tt arXiv:1405.6714}}].

\bibitem{Bullimore:2015lsa}
M.~Bullimore, T.~Dimofte, and D.~Gaiotto, {\it {The Coulomb Branch of 3d
  ${\mathcal{N}= 4}$ Theories}},  {\em Commun. Math. Phys.} {\bf 354} (2017),
  no.~2 671--751, [\href{http://arxiv.org/abs/1503.04817}{{\tt
  arXiv:1503.04817}}].

\bibitem{Bullimore:2016hdc}
M.~Bullimore, T.~Dimofte, D.~Gaiotto, J.~Hilburn, and H.-C. Kim, {\it {Vortices
  and Vermas}},  {\em Adv. Theor. Math. Phys.} {\bf 22} (2018) 803--917,
  [\href{http://arxiv.org/abs/1609.04406}{{\tt arXiv:1609.04406}}].

\bibitem{Beem:2018fng}
C.~Beem, D.~Ben-Zvi, M.~Bullimore, T.~Dimofte, and A.~Neitzke, {\it {Secondary
  products in supersymmetric field theory}},  {\em Annales Henri Poincare} {\bf
  21} (2020), no.~4 1235--1310, [\href{http://arxiv.org/abs/1809.00009}{{\tt
  arXiv:1809.00009}}].

\bibitem{Oh:2019bgz}
J.~Oh and J.~Yagi, {\it {Chiral algebras from $\Omega$-deformation}},  {\em
  JHEP} {\bf 08} (2019) 143, [\href{http://arxiv.org/abs/1903.11123}{{\tt
  arXiv:1903.11123}}].

\bibitem{Jeong:2019pzg}
S.~Jeong, {\it {SCFT/VOA correspondence via $\Omega$-deformation}},  {\em JHEP}
  {\bf 10} (2019) 171, [\href{http://arxiv.org/abs/1904.00927}{{\tt
  arXiv:1904.00927}}].

\bibitem{Gaiotto:2019mmf}
D.~Gaiotto and T.~Okazaki, {\it {Sphere correlation functions and Verma
  modules}},  {\em JHEP} {\bf 02} (2020) 133,
  [\href{http://arxiv.org/abs/1911.11126}{{\tt arXiv:1911.11126}}].

\bibitem{Dimofte:2017tpi}
T.~Dimofte, D.~Gaiotto, and N.~M. Paquette, {\it {Dual boundary conditions in
  3d SCFT}},  {\em JHEP} {\bf 05} (2018) 060,
  [\href{http://arxiv.org/abs/1712.07654}{{\tt arXiv:1712.07654}}].

\bibitem{Costello:2020ndc}
K.~Costello, T.~Dimofte, and D.~Gaiotto, {\it {Boundary Chiral Algebras and
  Holomorphic Twists}},  \href{http://arxiv.org/abs/2005.00083}{{\tt
  arXiv:2005.00083}}.

\bibitem{Dedushenko:2016jxl}
M.~Dedushenko, S.~S. Pufu, and R.~Yacoby, {\it {A one-dimensional theory for
  Higgs branch operators}},  {\em JHEP} {\bf 03} (2018) 138,
  [\href{http://arxiv.org/abs/1610.00740}{{\tt arXiv:1610.00740}}].

\bibitem{Dedushenko:2017avn}
M.~Dedushenko, Y.~Fan, S.~S. Pufu, and R.~Yacoby, {\it {Coulomb Branch
  Operators and Mirror Symmetry in Three Dimensions}},  {\em JHEP} {\bf 04}
  (2018) 037, [\href{http://arxiv.org/abs/1712.09384}{{\tt arXiv:1712.09384}}].

\bibitem{Dedushenko:2018icp}
M.~Dedushenko, Y.~Fan, S.~S. Pufu, and R.~Yacoby, {\it {Coulomb Branch
  Quantization and Abelianized Monopole Bubbling}},  {\em JHEP} {\bf 10} (2019)
  179, [\href{http://arxiv.org/abs/1812.08788}{{\tt arXiv:1812.08788}}].

\bibitem{Crew:2020psc}
S.~Crew, N.~Dorey, and D.~Zhang, {\it {Blocks and Vortices in the 3d ADHM
  Quiver Gauge Theory}},  \href{http://arxiv.org/abs/2010.09732}{{\tt
  arXiv:2010.09732}}.

\bibitem{Gaiotto:2012xa}
D.~Gaiotto, L.~Rastelli, and S.~S. Razamat, {\it {Bootstrapping the
  superconformal index with surface defects}},  {\em JHEP} {\bf 01} (2013) 022,
  [\href{http://arxiv.org/abs/1207.3577}{{\tt arXiv:1207.3577}}].

\bibitem{Dimofte:2010tz}
T.~Dimofte, S.~Gukov, and L.~Hollands, {\it {Vortex Counting and Lagrangian
  3-manifolds}},  {\em Lett. Math. Phys.} {\bf 98} (2011) 225--287,
  [\href{http://arxiv.org/abs/1006.0977}{{\tt arXiv:1006.0977}}].

\bibitem{Aganagic:2016jmx}
M.~Aganagic and A.~Okounkov, {\it {Elliptic stable envelopes}},
  \href{http://arxiv.org/abs/1604.00423}{{\tt arXiv:1604.00423}}.

\bibitem{Aganagic:2017smx}
M.~Aganagic, E.~Frenkel, and A.~Okounkov, {\it {Quantum $q$-Langlands
  Correspondence}},  {\em Trans. Moscow Math. Soc.} {\bf 79} (2018) 1--83,
  [\href{http://arxiv.org/abs/1701.03146}{{\tt arXiv:1701.03146}}].

\bibitem{Smirnov:2020lhm}
A.~Smirnov and Z.~Zhou, {\it {3d Mirror Symmetry and Quantum $K$-theory of
  Hypertoric Varieties}},  \href{http://arxiv.org/abs/2006.00118}{{\tt
  arXiv:2006.00118}}.

\bibitem{Kim:2009wb}
S.~Kim, {\it {The Complete superconformal index for N=6 Chern-Simons theory}},
  {\em Nucl. Phys. B} {\bf 821} (2009) 241--284,
  [\href{http://arxiv.org/abs/0903.4172}{{\tt arXiv:0903.4172}}]. [Erratum:
  Nucl.Phys.B 864, 884 (2012)].

\bibitem{Imamura:2011su}
Y.~Imamura and S.~Yokoyama, {\it {Index for three dimensional superconformal
  field theories with general R-charge assignments}},  {\em JHEP} {\bf 04}
  (2011) 007, [\href{http://arxiv.org/abs/1101.0557}{{\tt arXiv:1101.0557}}].

\bibitem{Kapustin:2011jm}
A.~Kapustin and B.~Willett, {\it {Generalized Superconformal Index for Three
  Dimensional Field Theories}},  \href{http://arxiv.org/abs/1106.2484}{{\tt
  arXiv:1106.2484}}.

\bibitem{Razamat:2014pta}
S.~S. Razamat and B.~Willett, {\it {Down the rabbit hole with theories of class
  $ \mathcal{S} $}},  {\em JHEP} {\bf 10} (2014) 099,
  [\href{http://arxiv.org/abs/1403.6107}{{\tt arXiv:1403.6107}}].

\bibitem{Benini:2015noa}
F.~Benini and A.~Zaffaroni, {\it {A topologically twisted index for
  three-dimensional supersymmetric theories}},  {\em JHEP} {\bf 07} (2015) 127,
  [\href{http://arxiv.org/abs/1504.03698}{{\tt arXiv:1504.03698}}].

\bibitem{Closset:2016arn}
C.~Closset and H.~Kim, {\it {Comments on twisted indices in 3d supersymmetric
  gauge theories}},  {\em JHEP} {\bf 08} (2016) 059,
  [\href{http://arxiv.org/abs/1605.06531}{{\tt arXiv:1605.06531}}].

\bibitem{Bullimore:2018jlp}
M.~Bullimore, A.~Ferrari, and H.~Kim, {\it {Twisted Indices of 3d ${\mathcal N}
  = 4$ Gauge Theories and Enumerative Geometry of Quasi-Maps}},  {\em JHEP}
  {\bf 07} (2019) 014, [\href{http://arxiv.org/abs/1812.05567}{{\tt
  arXiv:1812.05567}}].

\bibitem{Gaiotto:2008ak}
D.~Gaiotto and E.~Witten, {\it {S-Duality of Boundary Conditions In N=4 Super
  Yang-Mills Theory}},  {\em Adv. Theor. Math. Phys.} {\bf 13} (2009), no.~3
  721--896, [\href{http://arxiv.org/abs/0807.3720}{{\tt arXiv:0807.3720}}].

\bibitem{Hama:2011ea}
N.~Hama, K.~Hosomichi, and S.~Lee, {\it {SUSY Gauge Theories on Squashed
  Three-Spheres}},  {\em JHEP} {\bf 05} (2011) 014,
  [\href{http://arxiv.org/abs/1102.4716}{{\tt arXiv:1102.4716}}].

\bibitem{Benvenuti:2011ga}
S.~Benvenuti and S.~Pasquetti, {\it {3D-partition functions on the sphere:
  exact evaluation and mirror symmetry}},  {\em JHEP} {\bf 05} (2012) 099,
  [\href{http://arxiv.org/abs/1105.2551}{{\tt arXiv:1105.2551}}].

\bibitem{Kapustin:2009kz}
A.~Kapustin, B.~Willett, and I.~Yaakov, {\it {Exact Results for Wilson Loops in
  Superconformal Chern-Simons Theories with Matter}},  {\em JHEP} {\bf 03}
  (2010) 089, [\href{http://arxiv.org/abs/0909.4559}{{\tt arXiv:0909.4559}}].

\bibitem{Narukawa2003}
A.~{Narukawa}, {\it {The modular properties and the integral representations of
  the multiple elliptic gamma functions}},  {\em arXiv Mathematics e-prints}
  (June, 2003) math/0306164, [\href{http://arxiv.org/abs/math/0306164}{{\tt
  math/0306164}}].

\bibitem{Yoshida:2014ssa}
Y.~Yoshida and K.~Sugiyama, {\it {Localization of 3d $\mathcal{N}=2$
  Supersymmetric Theories on $S^1 \times D^2$}},
  \href{http://arxiv.org/abs/1409.6713}{{\tt arXiv:1409.6713}}.

\bibitem{Herbst:2008jq}
M.~Herbst, K.~Hori, and D.~Page, {\it {Phases Of N=2 Theories In 1+1 Dimensions
  With Boundary}},  \href{http://arxiv.org/abs/0803.2045}{{\tt
  arXiv:0803.2045}}.

\bibitem{Hori:2013ika}
K.~Hori and M.~Romo, {\it {Exact Results In Two-Dimensional (2,2)
  Supersymmetric Gauge Theories With Boundary}},
  \href{http://arxiv.org/abs/1308.2438}{{\tt arXiv:1308.2438}}.

\bibitem{Honda:2013uca}
D.~Honda and T.~Okuda, {\it {Exact results for boundaries and domain walls in
  2d supersymmetric theories}},  {\em JHEP} {\bf 09} (2015) 140,
  [\href{http://arxiv.org/abs/1308.2217}{{\tt arXiv:1308.2217}}].

\bibitem{Sugishita:2013jca}
S.~Sugishita and S.~Terashima, {\it {Exact Results in Supersymmetric Field
  Theories on Manifolds with Boundaries}},  {\em JHEP} {\bf 11} (2013) 021,
  [\href{http://arxiv.org/abs/1308.1973}{{\tt arXiv:1308.1973}}].

\bibitem{Bobev:2015kza}
N.~Bobev, M.~Bullimore, and H.-C. Kim, {\it {Supersymmetric Casimir Energy and
  the Anomaly Polynomial}},  {\em JHEP} {\bf 09} (2015) 142,
  [\href{http://arxiv.org/abs/1507.08553}{{\tt arXiv:1507.08553}}].

\end{thebibliography}\endgroup


\providecommand{\href}[2]{#2}\begingroup\raggedright\endgroup

\end{document}